\documentclass[aps, prl, twocolumn, notitlepage,groupedaddress,floatfix]{revtex4-2}

\usepackage{amsmath}
\usepackage{amsfonts}
\usepackage{amsthm}
\usepackage{amssymb}

\usepackage{bm}
\usepackage{physics}

\usepackage{graphicx}
\usepackage{xcolor}
\usepackage[caption=false]{subfig}

\usepackage[breaklinks=true,colorlinks,citecolor=blue,linkcolor=red,urlcolor=blue]{hyperref}

\newcommand{\ii}{\mathrm{i}}
\newcommand{\ee}{\mathrm{e}}
\newcommand{\ce}{\varepsilon}
\newcommand{\one}{\openone}

\newcommand{\fig}{Fig.}

\newcommand{\eq}{Eq.}

\newcommand{\rref}{Ref.}

\newcommand{\SM}{\cite{SM}}

\newcommand{\sket}[1]{\lvert #1 \rangle}

\newcommand{\snorm}[1]{\langle #1 \rvert #1 \rangle}
\newcommand{\sbraket}[2]{\langle #1 \rvert #2 \rangle}

\begin{document}
\title{Orthogonality catastrophe beyond bosonization from post-selection}
\author{Martino Stefanini}
\email{mstefa@uni-mainz.de}
\affiliation{Institut f\"ur Physik, Johannes Gutenberg-Universit\"at Mainz, D-55099 Mainz, Germany}
\author{Jamir Marino}
\affiliation{Institut f\"ur Physik, Johannes Gutenberg-Universit\"at Mainz, D-55099 Mainz, Germany}

\date{\today}

\begin{abstract}
We show that the dynamics induced by post-selected measurements can serve as a controlled route to access physical processes beyond the boundaries of Tomonaga-Luttinger liquid physics. 
We consider a one-dimensional fermionic wire whose dynamics results from a sequence of weak measurements of the fermionic density at a given site, interspersed with unitary hopping dynamics. This realizes a non-Hermitian variant of the celebrated instance of a local scatterer in a fermionic system and its ensuing orthogonality catastrophe. 
We observe a distinct crossover in the system's time evolution as a function of the fermion density. In the high-density regime, reminiscent of the Hermitian case, a bosonized version of the model properly describes the dynamics while, as we delve into the low-density regime, the validity of bosonization breaks down, giving rise to irreversible behavior. Notably, this crossover from reversible to irreversible dynamics is non-perturbative in the measurement rate and can manifest itself even with relatively shallow measurement rates, provided that the system's density remains below the crossover threshold. Our results render a conceptually transparent model for exploring non-perturbative effects beyond bosonization,  which could be used as a stepping stone to explore novel routes for the control of non-linear dynamics in low-dimensional quantum systems.
\end{abstract}
\maketitle

\paragraph{Introduction.---}
The Tomonaga-Luttinger liquid (TLL) stands as the universal, low-energy description for interacting one-dimensional (1D) quantum systems, renowned for its mathematical elegance and success in equilibrium scenarios \cite{Giamarchi,GogolinNersesyanTsvelik}. 
Its technical cornerstone, known as bosonization, allows to transform virtually any interacting system involving 1D fermions, bosons or spins into a massless free theory of collective bosonic excitations. 
Extending its applicability beyond equilibrium, especially in non-equilibrium settings, has been a topic of exploration in both isolated \cite{PRL_TLL_Quench_Cazalilla,IucciCazalillaQuenchTLL,CazalillaQuenchesLL,SmoothQuenchTLL,NoneqQuenchTLL_Mirlin,IucciCazalilla_Quench_SineGordon,Diehl_thermalization_TLL,EPJB_QuenchBeyondTLL,PhysRevB.94.085122} and open systems \cite{DoraThermalizationTLL,DoraDissipInducedTLL,DiehlTLLHeating,FromlPRB1,DoraVaporizationTLL}, as well as in the realm of non-Hermitian systems \cite{nhTLL-Affleck,nhTLL-PhysRevA.94.053615,nhTLL-PhysRevLett.124.136802,nhTLL-NatCommAshida,nhTLL-PhysRevB.106.235125,nhTLL-PhysRevB.107.045110,nhTLL-PhysRevB.108.035104,nhTLL-PhysRevB.108.085433} and measurement-induced dynamics  \cite{AltmanMeasurementsTLL,Buchhold_2021,MonitoredFermionsDissipation,NonHermTLL,sun2023new}.

While the TLL framework proves remarkably robust even far from equilibrium in certain phenomenological cases \cite{Collura_XXZ_TLL,Pollman_XXZ_TLL,MedenTLLQuench,MedenTLLSteadyState,DiehlTLLHeating,Diehl_thermalization_TLL}, it lacks a formal renormalization group argument to guarantee its validity given the potential excitation of higher-energy degrees of freedom. The exploration of dynamics beyond TLL at equilibrium has seen significant attention in recent decades \cite{ImambekovSchmidtGlazman}, yet understanding its validity in non-equilibrium conditions remains largely uncharted territory. This is partly due to the inherent technical challenges posed by the problem itself \cite{PhysRevLett.101.126802,GutmanBosonization,PhysRevLett.105.256802} and the absence of iconic models to guide such inquiries.

Nonetheless, delving into this line of inquiry holds both fundamental and practical significance. It serves as a potential testing ground for theories encompassing universal dynamics beyond traditional thermodynamics, while also offering the prospect of discovering effective models capable of simplifying the numerical resolution of highly correlated quantum dynamics.
\par In this Letter, we embark on an initial exploration in this direction by proposing a model of non-unitary dynamics where non-perturbative effects beyond bosonization manifest in a controlled fashion, and where at the same time the mechanisms for the breakdown of bosonization can be traced back to a transparent physical picture. The latter feature is highly nontrivial, since there are only a few cases \cite{ImambekovSchmidtGlazman,GutmanBosonization} in which the breakdown is physically well understood. We implement a non-Hermitian version of the orthogonality catastrophe (OC) using post-selected continuous measurements of the local fermionic density at a given site. The OC \cite{AndersonOC,Mahan,Giamarchi,GogolinNersesyanTsvelik} is a paradigmatic phenomenon in solid-state physics, with countless applications ranging from the physics of X-rays \cite{Mahan,SchotteAndSchotte} to the Kondo effect \cite{Hewson,SlaveBosons,PRB_Anderson_Kondo}, quantum dots \cite{PRB_EdgeSing_QuantumDots,PRB_FermiGasResponse} and ultracold atoms \cite{PhysRevA.97.033612,PhysRevX.2.041020,PhysRevA.89.053617}. The OC describes the strong sensitivity of gapless fermionic systems to local perturbations: a single scattering center introduced in a metal generates a new ground state which is orthogonal to the unperturbed one \cite{AndersonOC}, due to the excitation of a diverging number of particle-hole pairs with vanishing energy. The dynamical signature of the OC is an algebraic decay in time of the return amplitude after turning on of the local potential. This characteristic decay has been proven by non-perturbative means \cite{NozieresDeDominicis,CombescotNozieres}, but it is also predicted by a simple bosonized description \cite{SchotteAndSchotte}---an early example of a successful non-equilibrium application of bosonization.

In this study, we consider a model where a local imaginary scattering potential is abruptly introduced into a noninteracting 1D fermionic system.
While non-equilibrium versions of the OC problem have been explored in previous literature \cite{Tonielli,PRB_LupoSchiro,PRL_SchiroMitra,berdanier2019universal}, our approach is distinct. We employ post-selection of measurements, which plays a pivotal role in our ability to control the extent to which the dynamics deviate from an effective bosonized description.
%
Our primary focus is on the return amplitude $\mathcal{L}(t) \equiv \braket{\psi(0)}{\psi(t)}$, which serves as a global indicator of how closely the full state aligns with the description provided by bosonization. Our findings reveal a distinct algebraic decay in $\mathcal{L}(t)$, consistent with the Hermitian scenario. However, in the non-Hermitian setting, we observe an additional exponential decay not anticipated by bosonization. Unlike conventional occurrences of decoherence in OC~\cite{Tonielli,PRB_LupoSchiro,PRL_SchiroMitra}, this exponential decay signifies a reversible/irreversible dynamical crossover directly associated with the breakdown of bosonization. This feature stands as a unique hallmark of our post-selected implementation of the OC.
The presence of a decay rate, indicative of irreversible dynamics, strongly depends on the initial density of fermions: it is significant only in the low-density regime, where deviations from bosonization are notable, but tapers to zero as the background fermion density increases. This perspective is corroborated by our examination of the system's kinetic energy, which grows linearly over time. This growth hints at the existence of energy absorption processes that naturally drive deviations from the bosonization description whenever fermionic densities are low. 

We interpret the breakdown of bosonization at low densities as a consequence of the presence of quasi-bound states in the single-particle spectrum of the non-Hermitian Hamiltonian that governs the dynamics. These states occur far from the Fermi surface, and exert their influence through the imaginary part of their eigenvalues. They exist because of the curvature of the fermionic dispersion, and so they disappear as the dispersion is linearized when bosonizing the Hamiltonian. \\

\paragraph{Model.---}\label{sect: model}
We consider a 1D system of noninteracting, spinless fermions $c_j$ that can hop on the lattice sites of a chain according to the   Hamiltonian $H=-J\sum_{j=0}^{L-1}(c_{j+1}^\dag c_{j}+c_{j}^\dag c_{j+1})=\sum_q \ce_q c_q^\dag c_q$. The chain has $L$ sites with periodic boundary conditions (pbc) and unit lattice spacing $a=1$. We also use units such that $\hbar=1$. The momentum modes are $c_q=L^{-1/2}\sum_j \ee^{-\ii q j}c_j$, corresponding to energies $\ce_q\equiv -2J\cos q$.
\par We assume that the system is initialized in its ground state $\ket{FS}$ (at a certain filling $N_f$, namely density $\bar{n}=N_f/L$), and then evolves according to the nonlinear Schr\"odinger equation
\begin{align}\label{eq: nlse}
    \ii \dv{}{t}\ket{\psi(t)}=\Big[K+\ii\frac{\gamma}{2}\ev{n_0}{\psi(t)}\Big]\ket{\psi(t)}~,
\end{align}
where $\gamma$ is a constant that can be of either sign, $n_0\equiv c_{j_0}^\dag c_{j_0}$ is the number operator at the site $j_0$, and where we have introduced the non-Hermitian Hamiltonian 
\begin{equation}\label{eq: nh Ham}
    K\equiv H-\ii\frac{\gamma}{2}n_0~.
\end{equation}
The above Hamiltonian \cite{PhysRevE.60.114,Haque,SciPost_Schiro,FromlPRB1} describes scattering off a localized potential (i.e. an impurity) that has an imaginary strength. It can be seen as the simplest non-Hermitian generalization of the OC problem. Equation \eqref{eq: nlse} can be solved as $\ket{\psi(t)}=\sket{\tilde{\psi}(t)}/\snorm{\tilde{\psi}(t)}^{1/2}$, where $\sket{\tilde{\psi}(t)}=\ee^{-\ii K t}\ket{\psi(0)}$, so that the dynamics is simply the normalized version of the non-unitary evolution generated by $K$.  
\par There are several scenarios that lead to the dynamics \eqref{eq: nlse}, and all of them have in common some form of post-selection. We briefly present here a few of the simplest strategies, leaving a more detailed discussion to the Supplemental Material~\SM.  
\par The most conceptually straightforward scenario is that of continuous monitoring \cite{WisemanMilburn,NielsenChuang}: the system undergoes a series of weak measurements of the density at site $j_0$ at a rate $\gamma$, interspersed with Hamiltonian evolution, with a set of measurement operators that includes
\begin{equation}\label{eq: M}
    M(\delta t)=\mathcal{N}\ee^{-\frac{\gamma\delta t}{2}n_0}~,
\end{equation}
along with other operators that we do not need to specify here ($\mathcal{N}$ is a suitable normalization constant that plays no role). Such dynamics would yield stochastic state trajectories, conditioned on the random measurement outcomes, and \eq~\eqref{eq: nlse} is obtained by selecting only those trajectories in which the outcome is always the one corresponding to the action of \eqref{eq: M}. This procedure is known as the “no-click” limit \cite{AltmanMeasurementsTLL,SchiroNonHermPRB,SchiroNonHermSciPost,TurkeshiPhysRevB.103.224210}. Physically, the measurement operator \eqref{eq: M} for $\gamma>0$ may be taken to represent a measurement that has found no particles at $j_{0}$, since it favors states with $n_{0}=0$. Vice versa, $\gamma<0$ represents a measurement that finds one particle at $j_{0}$.
\par It is also possible to simulate \eq~\eqref{eq: nlse} with a dissipative dynamics governed by a Lindblad master equation, if the jump operator $\mathcal{J}$ \cite{DaleyQuantumTrajectories,WisemanMilburn} is such that the non-Hermitian part of the dissipative dynamics $\widetilde{H}=H-\tfrac{\ii}{2}\mathcal{J}^\dag\mathcal{J}$ reproduces \eq~\eqref{eq: nh Ham} \footnote{{Up to a constant shift $K-\widetilde{H}\propto\one$.}} and if the dynamics can be confined to a suitable subspace by a post-selection of experimental runs \cite{Rey_Hot_reactive_fermions,nhTLL-PhysRevA.94.053615,QubitExceptionalPoint}. A simple example is provided by a localized single particle loss $\mathcal{J}=\gamma^{1/2}c_{j_0}$ \cite{ThesisFroml,FromlPRL,FromlPRB1,FromlPRB2}. For each time $t$ of the dynamics, the total number of particles is measured and only the runs in which no fermions have been lost are kept. The observables measured within this set of runs correspond to the state $\sket{\tilde{\psi}(t)}$, and must be divided by its squared norm, which is simply the probability of not losing particles in the time $t$. This form of post-selection can be extended to any dissipative dynamics for which there is a way to recognize the effect of dissipative events, the so-called jumps \cite{DaleyQuantumTrajectories}. For instance, this procedure is possible for localized single particle gain $\mathcal{J}=\gamma_g^{1/2}c_{j_0}^\dag$ \cite{LocalizedFermionSource}, which corresponds to $\gamma=-\gamma_g<0$, while it is not viable for localized dephasing $\mathcal{J}=\gamma^{1/2}n_0$ \cite{Dolgirev, Tonielli}. It should be possible to implement the procedure above in near-term experiments (e.g. see \cite{QubitExceptionalPoint,response_PTEP}) for localized losses \cite{Esslinger_JPhys,Esslinger_PRL,Esslinger_PRL_23,Esslinger_PRX,Esslinger_PRA_theory}.  The two strategies outlined above (measurement- and dissipation-based) are closely related, but while the first requires monitoring a local observable at each instant of time, the second only requires global measurements at the end of the dynamics. 
\par A different dissipative strategy employs an immobile impurity particle that induces a localized two-body loss \cite{DipolarMolecules2BLoss, PRL_2BLoss_Zeno, DissipativeFermiHubbard,response_sign_reversal,Bose_Hubb_2b_loss,Science_2b_loss} in the system, i.e. $\mathcal{J}=\gamma^{1/2}d\, c_{j_0}$, where $d$ annihilates the impurity. Then, the effective Hamiltonian $\widetilde{H}=H-\ii\gamma/2\, n_0\, d^\dag d$ can be switched from $H$ to $K$ by simply injecting the impurity, $d^\dag d=1$. Finally, in analogy with the Hermitian scenario \cite{PhysRevX.2.041020,Science_Ramsey_interferometry}, one only needs to measure observables of the impurity to reconstruct the dynamics \eqref{eq: nlse}. This strategy does not require post-selection of the experimental runs.
\\

\paragraph{Numerical evidences.---}
\begin{figure}
    \centering
    \includegraphics[width=\linewidth]{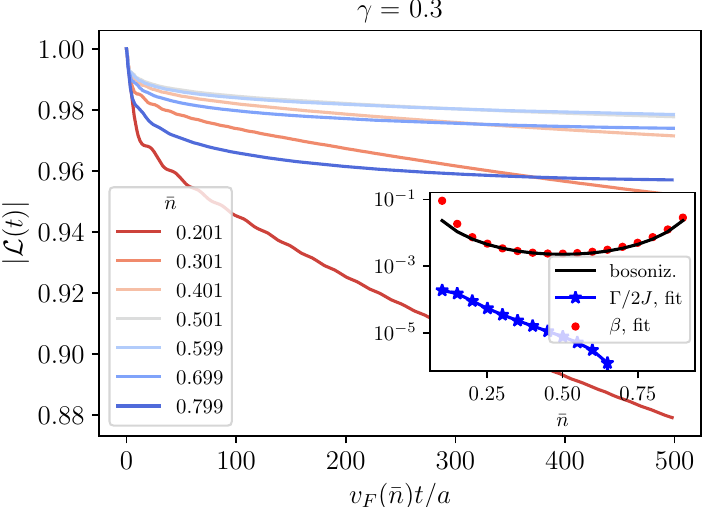}
    \caption{
    Absolute value of the return amplitude as a function of the rescaled time for $L=1000$, $J=0.5$, $\gamma=0.3$ and increasing density. The plot shows the absence of particle-hole symmetry as the low-density curves (in shades of red) decay faster than those at the conjugate densities $1-\bar{n}$ (shades of blue) because of an additional exponential envelope caused by the post-selected measurements. Inset: filling dependence of the two fitting parameters $\beta$ (OC exponent, red dots) and $\Gamma$ (exponential decay rate, blue stars), along with the bosonization prediction for $\beta$ (black line). The uncertainty on the fitted parameters is smaller than the size of the marker points.
    }
    \label{fig: loschmidt echo}
\end{figure}
We have computed the time evolution of the system using the algorithm described in \rref~\cite{CaoTilloyDeLuca}: we evolve the matrix of the single-particle wavefunctions $U_{ji}=[\phi_{p_i}(j)]^{i=1,\dots,N_f}_{j=1,\dots,L}$ with the single-particle versions of $H$ and $M$, and we keep the state normalized at each step.
\par We focus here on two main observables: the return amplitude and the total energy. The return amplitude is defined as $\mathcal{L}(t)\equiv \braket{\psi(0)}{\psi(t)}=\braket{FS}{\psi(t)}$, which is variously known as fidelity, Loschmidt echo, or impurity Green's function. As stated in the introduction, a power-law decay of the return amplitude is the hallmark of the OC \cite{Mahan,Giamarchi,GogolinNersesyanTsvelik}. 
\par The typical behavior of the return amplitude is shown in \fig~\ref{fig: loschmidt echo}. Our numerical computations show that after an initial transient, $\abs{\mathcal{L}(t)}$ displays two qualitatively different behaviors, and that we can go from one to the other by tuning the density. Above half filling, the return amplitude is well described by a power-law, while below it acquires an exponential decay. Notice that the lack of symmetry with respect to half-filling is to be expected, as the dynamics \eq~\eqref{eq: nlse} explicitly breaks particle-hole symmetry: loosely speaking, the exchange $c_j\leftrightarrow (-1)^j c_j^\dag$ leaves the Hamiltonian invariant while changing the sign of $\gamma$ \SM. As a consequence, we will show only data for $\gamma>0$, as those for $\gamma<0$ (for the case of post-selection for $n_{0}=1$) can be obtained from the correspondence $\bar{n}(\gamma<0)=1-\bar{n}(\gamma>0)$.   

\begin{figure}
    \centering
    \includegraphics[width=\linewidth]{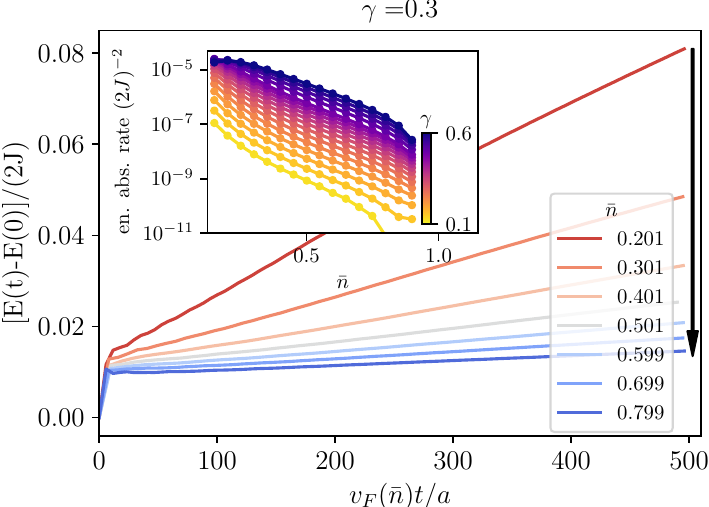}
    \caption{Energy absorption as a function of the rescaled time for $L=1000$, $J=0.5$, $\gamma=0.3$ and various densities. A monotonic behavior as a function of the density can be observed in the rate of energy absorption after a transient, which decreases upon increasing the density, as emphasized by the arrow on the right of the figure. Inset: density dependence of the energy absorption rate computed from a linear fit, at various values of $\gamma$. The rate decays exponentially with $\bar{n}$.}
    \label{fig: energy}
\end{figure}

\par We fit $\abs{\mathcal{L}(t)}$ assuming that after the transient it has the form $\abs{\mathcal{L}(t)}=A t^{-\beta}\ee^{-\Gamma t}$. The combined algebraic and exponential decay reflects the two processes that we expect to occur: the former is the OC behavior, while we interpret the latter as a sign of irreversible dynamics caused by the measurement process. This shape of $\abs{\mathcal{L}(t)}$ is valid for $0.1\lesssim\bar{n}\lesssim0.9$ and $\gamma\lesssim J$: outside this range of parameters, the return amplitude has a more complicated behavior.
\par The typical density dependence of the two parameters $\beta$ (the OC exponent) and $\Gamma$ (the decay rate) for a chain of $L=1000$ sites is shown in the inset of \fig~\ref{fig: loschmidt echo}. The two parameters behave very differently: while $\beta$ is essentially particle-hole symmetric, $\Gamma$ decreases monotonically (and exponentially) with the filling. We show in \SM~ that the behavior of the OC exponent $\beta$ can be predicted with bosonization, which yields $\beta_{b}(\bar{n})={\gamma^2}/{[2\pi v_F(\bar{n})]^2}$. This expression depends on the filling only through the Fermi velocity $v_F=2J\sin(\pi \bar{n})$, which is particle-hole symmetric, in accord with the numerics \footnote{The disagreement of the fitted exponent with respect to $\beta_b$ at larger $\gamma$ can be attributed to the local curvature of the dispersion around the Fermi energy, just as in the Hermitian case \cite{Giamarchi,GogolinNersesyanTsvelik}.}. However, the power-law behavior is all that bosonization predicts. Namely, the existence of a nonvanishing $\Gamma$ lies beyond bosonization. A numerical fit indicates (cf.~\SM) that for small $\gamma$ the decay rate behaves as $\ln\Gamma\approx a(\gamma)-b\bar{n}/\gamma^y$, where $a(\gamma)$ is approximately linear in $\gamma$, $b>0$ and $y\approx 1.4$. This result suggests that $\Gamma$ is non-perturbative in the measurement rate $\gamma$.
\par A second observable that we consider is the kinetic energy absorbed as a function of time, $E(t)\equiv \ev{H}{\psi(t)}$, whose evolution is shown in \fig~\ref{fig: energy} for a fixed $\gamma$ at various fillings. This quantity shows a steep transient, after which it begins to grow linearly in time. The interesting behavior is the dependence of the slope of the linear growth (namely, the energy absorption rate) on the filling. The inset in \fig~\ref{fig: energy} shows that the rate decreases exponentially with the fermion density. In the case of our model, bosonization would predict that the energy should saturate as a function of time, as in the Hermitian case. This predicted behavior depends on the excitation of only low-energy modes, which are those captured by the TLL Hamiltonian. The absorption of energy in the non-Hermitian scenario signals the broken time-reversal symmetry---or irreversibility---of the dynamics of the measured system. Indeed, the state will eventually converge to the many-body eigenstate of $K$ with the smallest imaginary part of the eigenvalue, although this happens at very late times~\cite{Dolgirev,FromlPRB1}, much larger than $L/v_F$.
\par We would like to point out that the value of the measurement rate $\gamma$ does not play an important role in our results, as in all the plots that we are showing it is smaller than the bandwidth $4J$. In this parameter regime, a larger $\gamma$ determines a quantitative change of the various effects, while the qualitative picture remains unchanged \footnote{Unless the density is too low or too high, so that the Fermi momentum crosses the “special momenta” $q^\ast$ reported in the following Section. There are also qualitative changes in the non-perturbative regime $\gamma>4J$ caused by the crossing of an exceptional point \cite{ThesisFroml,Haque}, not treated here.}. In the Hermitian OC scenario, the smallness of the impurity potential ensures the applicability of bosonization, but the power-law behavior is non-perturbative \cite{GogolinNersesyanTsvelik,Giamarchi,NozieresDeDominicis,CombescotNozieres}. Our calculations show that for the non-Hermitian version of the OC, bosonization is not applicable---yielding qualitatively wrong predictions---even in such perturbative regime.\\

\paragraph{Physical mechanism.---}
\begin{figure}
    \centering
    \includegraphics[width=\linewidth]{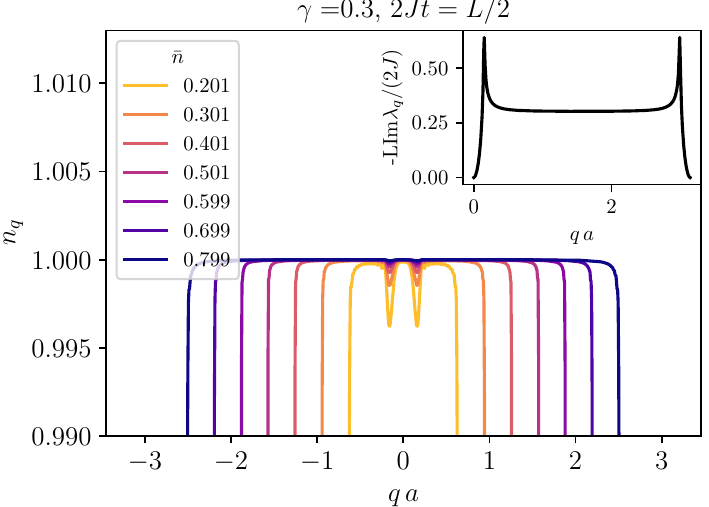}
    \caption{Occupation of the momentum states in the Brillouin zone for a system of $L=1000$ sites at the time $2J t=L/2$, with $2J=1$ and the measurement rate fixed at $\gamma/2J=0.3$. The plot reveals the extra depletion around the modes $q^\ast\approx \pm\arcsin{\gamma/4J}$, which cannot be accounted for by bosonization, and it shows how the depletion is deeper for smaller initial filling. The inset reports the imaginary part of the single-particle eigenvalues of the non-Hermitian Hamiltonian \eqref{eq: nh Ham}, displaying that the depleted modes correspond to a peak of the single-particle decay rates.}
    \label{fig: momentum occupancy}
\end{figure}
As observed in the numerics, the predictions of bosonization are violated if the density is low enough. We stress that our TLL predictions involve no approximations beyond the linearization of the dispersion relation, since we bosonize the model in such a way \cite{GogolinNersesyanTsvelik,Giamarchi,FermiEdgeKaneMatveevGlazman} that a quadratic Hamiltonian is obtained, to which no nonlinear terms need to be added within the usual theory of TLL. The validity of the linearization of the spectrum around the Fermi points typically relies only on the smallness of the impurity potential and gives reliable results in the Hermitian case \cite{Giamarchi,GogolinNersesyanTsvelik,SchotteAndSchotte}. So, what goes wrong in the bosonization procedure? The answer can be found by computing the single-particle spectrum of the Hamiltonian $K$. As shown in the inset of \fig~\ref{fig: momentum occupancy} (for more details, see~\SM), we find that the imaginary part of the eigenvalues (i.e. the decay rates of the eigenstates) is mostly flat near the middle of the band (momentum close to $\pi/2$), while it has two peaks near the band edges. This contrasts with the bosonized Hamiltonian, which predicts the same decay rate for all states \footnote{Indeed, it differs from a $\mathcal{P}\mathcal{T}$-symmetric \cite{ReviewNonHermitianUeda} Hamiltonian by an imaginary constant only, which is unobservable in the nonlinear dynamics \eq~\eqref{eq: nlse}.}. Therefore, bosonization is capable of describing the physics around the Fermi surface as long as the latter lies near the middle of the band (and this explains why it predicts the OC behavior correctly), but it cannot account for the nontrivial dynamics that occurs at the edges of the band. Indeed, this picture is confirmed by the time evolution of the occupation number of the momentum modes, shown in \fig~\ref{fig: momentum occupancy}: while the discontinuity of the Fermi surface remains intact, the decay peaks cause an extra depletion deep inside the Fermi sea. Moreover, it can be observed that the depletion of these modes is more pronounced for smaller densities, in agreement with the increasing deviation of the dynamics from the predictions of bosonization in that regime. The breakdown of bosonization that we are proposing is unusual, since it does not come from properties of the band dispersion (such as the curvature at the Fermi surface \cite{ImambekovSchmidtGlazman}), but rather it comes from the properties of the imaginary part of the spectrum---the eigenstates' lifetimes.
\par The existence of this faster-decaying modes has been already noticed in \rref~\cite{Haque}, in which they appear as the ones most strongly scattered by the imaginary potential. The momenta at which the peaks occur are $\pm q^\ast$ and $\pm(\pi-q^\ast)$, where $q^\ast$ is determined approximately \cite{Haque} from the requirement that the group velocity of the fermions is equal to the strength of the imaginary potential, $2J\sin q^\ast=\gamma/2$. This observation implies that these “special” modes are a consequence of the curvature of the band dispersion. Indeed, if we start with a perfectly linear fermion dispersion, these modes disappear, and we find again that the various observables are in excellent agreement with the results from bosonization. Thus, the breakdown of bosonization at low densities is a curvature effect, albeit different from the usual beyond-TLL physics \cite{ImambekovSchmidtGlazman}, since the latter involves only the curvature near the Fermi surface, whereas here we observe effects coming from the curvature at the band edges.
\par The observed decay properties within the Fermi surface explain why the system absorbs energy: the conservation of the number of particles implies that the fermions depleted from the low-momentum (and low-energy) special modes must populate higher-momentum states. The behavior of the return amplitude bears a similarity with the Hermitian OC in the presence of a bound state, where a non-analytic dependence of $\mathcal{L}(t)$ on the density is found \cite{Zagoskin_1997}---indeed, the special modes are quasi-bound states \cite{scaleFreeBoundStates}. A direct generalization \cite{SM} of the Hermitian OC results from \rref~\cite{CombescotNozieres} to the present model allows us to have an analytical understanding on how the density influences the un-renormalized return amplitude $\tilde{\mathcal{L}}(t)\equiv \sbraket{FS}{\tilde{\psi}(t)}=\ev{\ee^{-\ii K t}}{FS}$. This calculation clearly shows how the imaginary potential causes $\tilde{\mathcal{L}}(t)$ to acquire an exponential decay that depends on the properties of all occupied states and is therefore non-universal, and density-dependent. We infer again that any bosonized version of the model that focuses on the states around the Fermi surface only can not possibly capture this non-universal decay.
\paragraph{Conclusions.---}

From a fundamental point of view, our research unveils a novel mechanism for departures from bosonization. The mechanism is distinct from more traditional explanations rooted in the effects of band curvature of the dispersion relation~\cite{ImambekovSchmidtGlazman}, and in this regard it illustrates transparently the profound difference between unitary and dissipative systems when it comes to the breakdown of low-energy collective descriptions.

\par Upon initial examination~\SM, the Lindblad dynamics governed by localized noise dephasing, albeit featuring the same non-Hermitian Hamiltonian as this work, does not exhibit the same characteristic signatures of the measurement-induced dynamics that we found. This observation implies that quantum jumps may obscure the mechanism we have elucidated above. Unraveling these features and developing a simplified model for controlled breakdown of bosonization within Lindblad dynamics would represent a natural extension of our research.
\par Recently, measurement-induced phase transitions in low-dimensional fermionic systems have been captured using an effective bosonized description~\cite{Buchhold_2021}. Our findings could serve as a promising starting point for extending this analysis to interacting fermions or quantum spin chains undergoing monitored dynamics. At variance with~\cite{Buchhold_2021}, one could wonder whether the universality classes of such dynamical transitions may change when the Tomonaga-Luttinger liquid description becomes inapplicable and processes similar to those presented here come into play. Among all the possible ramifications of our work, this appears to us as one of the most promising. \\

\begin{acknowledgments}
    \paragraph{Acknowledgments.---} We wish to thank Z. Weinstein and S. Diehl for fruitful discussions and for providing comments on early versions of the manuscript. We also acknowledge the valuable observations provided by V. Meden. M.S. is grateful to R. J. Valencia Tortora and especially to O. Chelpanova for many useful discussions. This work has been supported by the Deutsche Forschungsgemeinschaft (DFG) through the grant HADEQUAM-MA7003/3-1. We gratefully acknowledge the computing time granted through the project “DysQCorr” on the Mogon II supercomputer of the Johannes Gutenberg University Mainz (\url{hpc.uni-mainz.de}), which is a member of the AHRP (Alliance for High Performance Computing in Rhineland Palatinate, \url{www.ahrp.info}), and the Gauss Alliance e.V.
\end{acknowledgments} 
\bibliography{biblioOC.bib}

\begin{thebibliography}{93}%
\makeatletter
\providecommand \@ifxundefined [1]{%
 \@ifx{#1\undefined}
}%
\providecommand \@ifnum [1]{%
 \ifnum #1\expandafter \@firstoftwo
 \else \expandafter \@secondoftwo
 \fi
}%
\providecommand \@ifx [1]{%
 \ifx #1\expandafter \@firstoftwo
 \else \expandafter \@secondoftwo
 \fi
}%
\providecommand \natexlab [1]{#1}%
\providecommand \enquote  [1]{``#1''}%
\providecommand \bibnamefont  [1]{#1}%
\providecommand \bibfnamefont [1]{#1}%
\providecommand \citenamefont [1]{#1}%
\providecommand \href@noop [0]{\@secondoftwo}%
\providecommand \href [0]{\begingroup \@sanitize@url \@href}%
\providecommand \@href[1]{\@@startlink{#1}\@@href}%
\providecommand \@@href[1]{\endgroup#1\@@endlink}%
\providecommand \@sanitize@url [0]{\catcode `\\12\catcode `\$12\catcode
  `\&12\catcode `\#12\catcode `\^12\catcode `\_12\catcode `\%12\relax}%
\providecommand \@@startlink[1]{}%
\providecommand \@@endlink[0]{}%
\providecommand \url  [0]{\begingroup\@sanitize@url \@url }%
\providecommand \@url [1]{\endgroup\@href {#1}{\urlprefix }}%
\providecommand \urlprefix  [0]{URL }%
\providecommand \Eprint [0]{\href }%
\providecommand \doibase [0]{https://doi.org/}%
\providecommand \selectlanguage [0]{\@gobble}%
\providecommand \bibinfo  [0]{\@secondoftwo}%
\providecommand \bibfield  [0]{\@secondoftwo}%
\providecommand \translation [1]{[#1]}%
\providecommand \BibitemOpen [0]{}%
\providecommand \bibitemStop [0]{}%
\providecommand \bibitemNoStop [0]{.\EOS\space}%
\providecommand \EOS [0]{\spacefactor3000\relax}%
\providecommand \BibitemShut  [1]{\csname bibitem#1\endcsname}%
\let\auto@bib@innerbib\@empty
\bibitem [{\citenamefont {Giamarchi}(2003)}]{Giamarchi}%
  \BibitemOpen
  \bibfield  {author} {\bibinfo {author} {\bibfnamefont {T.}~\bibnamefont
  {Giamarchi}},\ }\href@noop {} {\emph {\bibinfo {title} {Quantum Physics in
  One Dimension}}},\ \bibinfo {edition} {1st}\ ed.\ (\bibinfo  {publisher}
  {Clarendon Press},\ \bibinfo {address} {Oxford},\ \bibinfo {year}
  {2003})\BibitemShut {NoStop}%
\bibitem [{\citenamefont {Gogolin}\ \emph {et~al.}(1998)\citenamefont
  {Gogolin}, \citenamefont {Nersesyan},\ and\ \citenamefont
  {Tsvelik}}]{GogolinNersesyanTsvelik}%
  \BibitemOpen
  \bibfield  {author} {\bibinfo {author} {\bibfnamefont {A.~O.}\ \bibnamefont
  {Gogolin}}, \bibinfo {author} {\bibfnamefont {A.~A.}\ \bibnamefont
  {Nersesyan}},\ and\ \bibinfo {author} {\bibfnamefont {A.~M.}\ \bibnamefont
  {Tsvelik}},\ }\href@noop {} {\emph {\bibinfo {title} {Bosonization and
  Strongly Correlated Systems}}}\ (\bibinfo  {publisher} {Cambridge University
  Press},\ \bibinfo {address} {Cambridge},\ \bibinfo {year} {1998})\BibitemShut
  {NoStop}%
\bibitem [{\citenamefont {Cazalilla}(2006)}]{PRL_TLL_Quench_Cazalilla}%
  \BibitemOpen
  \bibfield  {author} {\bibinfo {author} {\bibfnamefont {M.~A.}\ \bibnamefont
  {Cazalilla}},\ }\bibfield  {title} {\bibinfo {title} {Effect of suddenly
  turning on interactions in the {Luttinger} model},\ }\href
  {https://doi.org/10.1103/PhysRevLett.97.156403} {\bibfield  {journal}
  {\bibinfo  {journal} {Phys. Rev. Lett.}\ }\textbf {\bibinfo {volume} {97}},\
  \bibinfo {pages} {156403} (\bibinfo {year} {2006})}\BibitemShut {NoStop}%
\bibitem [{\citenamefont {Iucci}\ and\ \citenamefont
  {Cazalilla}(2009)}]{IucciCazalillaQuenchTLL}%
  \BibitemOpen
  \bibfield  {author} {\bibinfo {author} {\bibfnamefont {A.}~\bibnamefont
  {Iucci}}\ and\ \bibinfo {author} {\bibfnamefont {M.~A.}\ \bibnamefont
  {Cazalilla}},\ }\bibfield  {title} {\bibinfo {title} {Quantum quench dynamics
  of the {Luttinger} model},\ }\href
  {https://doi.org/10.1103/PhysRevA.80.063619} {\bibfield  {journal} {\bibinfo
  {journal} {Phys. Rev. A}\ }\textbf {\bibinfo {volume} {80}},\ \bibinfo
  {pages} {063619} (\bibinfo {year} {2009})}\BibitemShut {NoStop}%
\bibitem [{\citenamefont {Cazalilla}\ and\ \citenamefont
  {Chung}(2016)}]{CazalillaQuenchesLL}%
  \BibitemOpen
  \bibfield  {author} {\bibinfo {author} {\bibfnamefont {M.~A.}\ \bibnamefont
  {Cazalilla}}\ and\ \bibinfo {author} {\bibfnamefont {M.-C.}\ \bibnamefont
  {Chung}},\ }\bibfield  {title} {\bibinfo {title} {Quantum quenches in the
  {Luttinger} model and its close relatives},\ }\href
  {https://doi.org/10.1088/1742-5468/2016/06/064004} {\bibfield  {journal}
  {\bibinfo  {journal} {J. Stat. Mech.}\ }\textbf {\bibinfo {volume} {2016}},\
  \bibinfo {pages} {064004} (\bibinfo {year} {2016})}\BibitemShut {NoStop}%
\bibitem [{\citenamefont {Dziarmaga}\ and\ \citenamefont
  {Tylutki}(2011)}]{SmoothQuenchTLL}%
  \BibitemOpen
  \bibfield  {author} {\bibinfo {author} {\bibfnamefont {J.}~\bibnamefont
  {Dziarmaga}}\ and\ \bibinfo {author} {\bibfnamefont {M.}~\bibnamefont
  {Tylutki}},\ }\bibfield  {title} {\bibinfo {title} {Excitation energy after a
  smooth quench in a {Luttinger} liquid},\ }\href
  {https://doi.org/10.1103/PhysRevB.84.214522} {\bibfield  {journal} {\bibinfo
  {journal} {Phys. Rev. B}\ }\textbf {\bibinfo {volume} {84}},\ \bibinfo
  {pages} {214522} (\bibinfo {year} {2011})}\BibitemShut {NoStop}%
\bibitem [{\citenamefont {Ngo~Dinh}\ \emph {et~al.}(2013)\citenamefont
  {Ngo~Dinh}, \citenamefont {Bagrets},\ and\ \citenamefont
  {Mirlin}}]{NoneqQuenchTLL_Mirlin}%
  \BibitemOpen
  \bibfield  {author} {\bibinfo {author} {\bibfnamefont {S.}~\bibnamefont
  {Ngo~Dinh}}, \bibinfo {author} {\bibfnamefont {D.~A.}\ \bibnamefont
  {Bagrets}},\ and\ \bibinfo {author} {\bibfnamefont {A.~D.}\ \bibnamefont
  {Mirlin}},\ }\bibfield  {title} {\bibinfo {title} {Interaction quench in
  nonequilibrium {Luttinger} liquids},\ }\href
  {https://doi.org/10.1103/PhysRevB.88.245405} {\bibfield  {journal} {\bibinfo
  {journal} {Phys. Rev. B}\ }\textbf {\bibinfo {volume} {88}},\ \bibinfo
  {pages} {245405} (\bibinfo {year} {2013})}\BibitemShut {NoStop}%
\bibitem [{\citenamefont {Iucci}\ and\ \citenamefont
  {Cazalilla}(2010)}]{IucciCazalilla_Quench_SineGordon}%
  \BibitemOpen
  \bibfield  {author} {\bibinfo {author} {\bibfnamefont {A.}~\bibnamefont
  {Iucci}}\ and\ \bibinfo {author} {\bibfnamefont {M.~A.}\ \bibnamefont
  {Cazalilla}},\ }\bibfield  {title} {\bibinfo {title} {Quantum quench dynamics
  of the sine-{Gordon} model in some solvable limits},\ }\href
  {https://doi.org/10.1088/1367-2630/12/5/055019} {\bibfield  {journal}
  {\bibinfo  {journal} {New Journal of Physics}\ }\textbf {\bibinfo {volume}
  {12}},\ \bibinfo {pages} {055019} (\bibinfo {year} {2010})}\BibitemShut
  {NoStop}%
\bibitem [{\citenamefont {Buchhold}\ \emph {et~al.}(2016)\citenamefont
  {Buchhold}, \citenamefont {Heyl},\ and\ \citenamefont
  {Diehl}}]{Diehl_thermalization_TLL}%
  \BibitemOpen
  \bibfield  {author} {\bibinfo {author} {\bibfnamefont {M.}~\bibnamefont
  {Buchhold}}, \bibinfo {author} {\bibfnamefont {M.}~\bibnamefont {Heyl}},\
  and\ \bibinfo {author} {\bibfnamefont {S.}~\bibnamefont {Diehl}},\ }\bibfield
   {title} {\bibinfo {title} {Prethermalization and thermalization of a
  quenched interacting {Luttinger} liquid},\ }\href
  {https://doi.org/10.1103/PhysRevA.94.013601} {\bibfield  {journal} {\bibinfo
  {journal} {Phys. Rev. A}\ }\textbf {\bibinfo {volume} {94}},\ \bibinfo
  {pages} {013601} (\bibinfo {year} {2016})}\BibitemShut {NoStop}%
\bibitem [{\citenamefont {Coira}\ \emph {et~al.}(2013)\citenamefont {Coira},
  \citenamefont {Becca},\ and\ \citenamefont {Parola}}]{EPJB_QuenchBeyondTLL}%
  \BibitemOpen
  \bibfield  {author} {\bibinfo {author} {\bibfnamefont {E.}~\bibnamefont
  {Coira}}, \bibinfo {author} {\bibfnamefont {F.}~\bibnamefont {Becca}},\ and\
  \bibinfo {author} {\bibfnamefont {A.}~\bibnamefont {Parola}},\ }\bibfield
  {title} {\bibinfo {title} {Quantum quenches in one-dimensional gapless
  systems},\ }\href {https://doi.org/10.1140/epjb/e2012-30978-y} {\bibfield
  {journal} {\bibinfo  {journal} {Eur. Phys. J. B}\ }\textbf {\bibinfo {volume}
  {86}},\ \bibinfo {pages} {55} (\bibinfo {year} {2013})}\BibitemShut {NoStop}%
\bibitem [{\citenamefont {Porta}\ \emph {et~al.}(2016)\citenamefont {Porta},
  \citenamefont {Gambetta}, \citenamefont {Cavaliere}, \citenamefont
  {Traverso~Ziani},\ and\ \citenamefont {Sassetti}}]{PhysRevB.94.085122}%
  \BibitemOpen
  \bibfield  {author} {\bibinfo {author} {\bibfnamefont {S.}~\bibnamefont
  {Porta}}, \bibinfo {author} {\bibfnamefont {F.~M.}\ \bibnamefont {Gambetta}},
  \bibinfo {author} {\bibfnamefont {F.}~\bibnamefont {Cavaliere}}, \bibinfo
  {author} {\bibfnamefont {N.}~\bibnamefont {Traverso~Ziani}},\ and\ \bibinfo
  {author} {\bibfnamefont {M.}~\bibnamefont {Sassetti}},\ }\bibfield  {title}
  {\bibinfo {title} {Out-of-equilibrium density dynamics of a quenched
  fermionic system},\ }\href {https://doi.org/10.1103/PhysRevB.94.085122}
  {\bibfield  {journal} {\bibinfo  {journal} {Phys. Rev. B}\ }\textbf {\bibinfo
  {volume} {94}},\ \bibinfo {pages} {085122} (\bibinfo {year}
  {2016})}\BibitemShut {NoStop}%
\bibitem [{\citenamefont {B\'acsi}\ and\ \citenamefont
  {D\'ora}(2023)}]{DoraThermalizationTLL}%
  \BibitemOpen
  \bibfield  {author} {\bibinfo {author} {\bibfnamefont {A.}~\bibnamefont
  {B\'acsi}}\ and\ \bibinfo {author} {\bibfnamefont {B.}~\bibnamefont
  {D\'ora}},\ }\bibfield  {title} {\bibinfo {title} {Lindbladian route towards
  thermalization of a {Luttinger} liquid},\ }\href
  {https://doi.org/10.1103/PhysRevB.107.125149} {\bibfield  {journal} {\bibinfo
   {journal} {Phys. Rev. B}\ }\textbf {\bibinfo {volume} {107}},\ \bibinfo
  {pages} {125149} (\bibinfo {year} {2023})}\BibitemShut {NoStop}%
\bibitem [{\citenamefont {B\'acsi}\ \emph
  {et~al.}(2020{\natexlab{a}})\citenamefont {B\'acsi}, \citenamefont {Moca},\
  and\ \citenamefont {D\'ora}}]{DoraDissipInducedTLL}%
  \BibitemOpen
  \bibfield  {author} {\bibinfo {author} {\bibfnamefont {A.}~\bibnamefont
  {B\'acsi}}, \bibinfo {author} {\bibfnamefont {C.~P.}\ \bibnamefont {Moca}},\
  and\ \bibinfo {author} {\bibfnamefont {B.}~\bibnamefont {D\'ora}},\
  }\bibfield  {title} {\bibinfo {title} {Dissipation-induced {Luttinger} liquid
  correlations in a one-dimensional fermi gas},\ }\href
  {https://doi.org/10.1103/PhysRevLett.124.136401} {\bibfield  {journal}
  {\bibinfo  {journal} {Phys. Rev. Lett.}\ }\textbf {\bibinfo {volume} {124}},\
  \bibinfo {pages} {136401} (\bibinfo {year} {2020}{\natexlab{a}})}\BibitemShut
  {NoStop}%
\bibitem [{\citenamefont {Buchhold}\ and\ \citenamefont
  {Diehl}(2015)}]{DiehlTLLHeating}%
  \BibitemOpen
  \bibfield  {author} {\bibinfo {author} {\bibfnamefont {M.}~\bibnamefont
  {Buchhold}}\ and\ \bibinfo {author} {\bibfnamefont {S.}~\bibnamefont
  {Diehl}},\ }\bibfield  {title} {\bibinfo {title} {Nonequilibrium universality
  in the heating dynamics of interacting {Luttinger} liquids},\ }\href
  {https://doi.org/10.1103/PhysRevA.92.013603} {\bibfield  {journal} {\bibinfo
  {journal} {Phys. Rev. A}\ }\textbf {\bibinfo {volume} {92}},\ \bibinfo
  {pages} {013603} (\bibinfo {year} {2015})}\BibitemShut {NoStop}%
\bibitem [{\citenamefont {Fr\"oml}\ \emph {et~al.}(2020)\citenamefont
  {Fr\"oml}, \citenamefont {Muckel}, \citenamefont {Kollath}, \citenamefont
  {Chiocchetta},\ and\ \citenamefont {Diehl}}]{FromlPRB1}%
  \BibitemOpen
  \bibfield  {author} {\bibinfo {author} {\bibfnamefont {H.}~\bibnamefont
  {Fr\"oml}}, \bibinfo {author} {\bibfnamefont {C.}~\bibnamefont {Muckel}},
  \bibinfo {author} {\bibfnamefont {C.}~\bibnamefont {Kollath}}, \bibinfo
  {author} {\bibfnamefont {A.}~\bibnamefont {Chiocchetta}},\ and\ \bibinfo
  {author} {\bibfnamefont {S.}~\bibnamefont {Diehl}},\ }\bibfield  {title}
  {\bibinfo {title} {Ultracold quantum wires with localized losses: Many-body
  quantum {Zeno} effect},\ }\href {https://doi.org/10.1103/PhysRevB.101.144301}
  {\bibfield  {journal} {\bibinfo  {journal} {Phys. Rev. B}\ }\textbf {\bibinfo
  {volume} {101}},\ \bibinfo {pages} {144301} (\bibinfo {year}
  {2020})}\BibitemShut {NoStop}%
\bibitem [{\citenamefont {B\'acsi}\ \emph
  {et~al.}(2020{\natexlab{b}})\citenamefont {B\'acsi}, \citenamefont {Moca},
  \citenamefont {Zar\'and},\ and\ \citenamefont
  {D\'ora}}]{DoraVaporizationTLL}%
  \BibitemOpen
  \bibfield  {author} {\bibinfo {author} {\bibfnamefont {A.}~\bibnamefont
  {B\'acsi}}, \bibinfo {author} {\bibfnamefont {C.~P.}\ \bibnamefont {Moca}},
  \bibinfo {author} {\bibfnamefont {G.}~\bibnamefont {Zar\'and}},\ and\
  \bibinfo {author} {\bibfnamefont {B.}~\bibnamefont {D\'ora}},\ }\bibfield
  {title} {\bibinfo {title} {Vaporization dynamics of a dissipative quantum
  liquid},\ }\href {https://doi.org/10.1103/PhysRevLett.125.266803} {\bibfield
  {journal} {\bibinfo  {journal} {Phys. Rev. Lett.}\ }\textbf {\bibinfo
  {volume} {125}},\ \bibinfo {pages} {266803} (\bibinfo {year}
  {2020}{\natexlab{b}})}\BibitemShut {NoStop}%
\bibitem [{\citenamefont {Affleck}\ \emph {et~al.}(2004)\citenamefont
  {Affleck}, \citenamefont {Hofstetter}, \citenamefont {Nelson},\ and\
  \citenamefont {Schollw\"ock}}]{nhTLL-Affleck}%
  \BibitemOpen
  \bibfield  {author} {\bibinfo {author} {\bibfnamefont {I.}~\bibnamefont
  {Affleck}}, \bibinfo {author} {\bibfnamefont {W.}~\bibnamefont {Hofstetter}},
  \bibinfo {author} {\bibfnamefont {D.~R.}\ \bibnamefont {Nelson}},\ and\
  \bibinfo {author} {\bibfnamefont {U.}~\bibnamefont {Schollw\"ock}},\
  }\bibfield  {title} {\bibinfo {title} {Non-{Hermitian Luttinger} liquids and
  flux line pinning in planar superconductors},\ }\href
  {https://doi.org/10.1088/1742-5468/2004/10/P10003} {\bibfield  {journal}
  {\bibinfo  {journal} {Journal of Statistical Mechanics: Theory and
  Experiment}\ }\textbf {\bibinfo {volume} {2004}},\ \bibinfo {pages} {P10003}
  (\bibinfo {year} {2004})}\BibitemShut {NoStop}%
\bibitem [{\citenamefont {Ashida}\ \emph {et~al.}(2016)\citenamefont {Ashida},
  \citenamefont {Furukawa},\ and\ \citenamefont
  {Ueda}}]{nhTLL-PhysRevA.94.053615}%
  \BibitemOpen
  \bibfield  {author} {\bibinfo {author} {\bibfnamefont {Y.}~\bibnamefont
  {Ashida}}, \bibinfo {author} {\bibfnamefont {S.}~\bibnamefont {Furukawa}},\
  and\ \bibinfo {author} {\bibfnamefont {M.}~\bibnamefont {Ueda}},\ }\bibfield
  {title} {\bibinfo {title} {Quantum critical behavior influenced by
  measurement backaction in ultracold gases},\ }\href
  {https://doi.org/10.1103/PhysRevA.94.053615} {\bibfield  {journal} {\bibinfo
  {journal} {Phys. Rev. A}\ }\textbf {\bibinfo {volume} {94}},\ \bibinfo
  {pages} {053615} (\bibinfo {year} {2016})}\BibitemShut {NoStop}%
\bibitem [{\citenamefont {D\'ora}\ and\ \citenamefont
  {Moca}(2020)}]{nhTLL-PhysRevLett.124.136802}%
  \BibitemOpen
  \bibfield  {author} {\bibinfo {author} {\bibfnamefont {B.}~\bibnamefont
  {D\'ora}}\ and\ \bibinfo {author} {\bibfnamefont {C.~P.}\ \bibnamefont
  {Moca}},\ }\bibfield  {title} {\bibinfo {title} {Quantum quench in
  $\mathcal{P}\mathcal{T}$-symmetric luttinger liquid},\ }\href
  {https://doi.org/10.1103/PhysRevLett.124.136802} {\bibfield  {journal}
  {\bibinfo  {journal} {Phys. Rev. Lett.}\ }\textbf {\bibinfo {volume} {124}},\
  \bibinfo {pages} {136802} (\bibinfo {year} {2020})}\BibitemShut {NoStop}%
\bibitem [{\citenamefont {Ashida}\ \emph {et~al.}(2017)\citenamefont {Ashida},
  \citenamefont {Furukawa},\ and\ \citenamefont {Ueda}}]{nhTLL-NatCommAshida}%
  \BibitemOpen
  \bibfield  {author} {\bibinfo {author} {\bibfnamefont {Y.}~\bibnamefont
  {Ashida}}, \bibinfo {author} {\bibfnamefont {S.}~\bibnamefont {Furukawa}},\
  and\ \bibinfo {author} {\bibfnamefont {M.}~\bibnamefont {Ueda}},\ }\bibfield
  {title} {\bibinfo {title} {Parity-time-symmetric quantum critical
  phenomena},\ }\href {https://doi.org/10.1038/ncomms15791} {\bibfield
  {journal} {\bibinfo  {journal} {Nature Communications}\ }\textbf {\bibinfo
  {volume} {8}},\ \bibinfo {pages} {15791} (\bibinfo {year}
  {2017})}\BibitemShut {NoStop}%
\bibitem [{\citenamefont {D\'ora}\ and\ \citenamefont
  {Moca}(2022)}]{nhTLL-PhysRevB.106.235125}%
  \BibitemOpen
  \bibfield  {author} {\bibinfo {author} {\bibfnamefont {B.}~\bibnamefont
  {D\'ora}}\ and\ \bibinfo {author} {\bibfnamefont {C.~P.}\ \bibnamefont
  {Moca}},\ }\bibfield  {title} {\bibinfo {title} {Full counting statistics in
  the many-body {Hatano-Nelson} model},\ }\href
  {https://doi.org/10.1103/PhysRevB.106.235125} {\bibfield  {journal} {\bibinfo
   {journal} {Phys. Rev. B}\ }\textbf {\bibinfo {volume} {106}},\ \bibinfo
  {pages} {235125} (\bibinfo {year} {2022})}\BibitemShut {NoStop}%
\bibitem [{\citenamefont {Yamamoto}\ and\ \citenamefont
  {Kawakami}(2023)}]{nhTLL-PhysRevB.107.045110}%
  \BibitemOpen
  \bibfield  {author} {\bibinfo {author} {\bibfnamefont {K.}~\bibnamefont
  {Yamamoto}}\ and\ \bibinfo {author} {\bibfnamefont {N.}~\bibnamefont
  {Kawakami}},\ }\bibfield  {title} {\bibinfo {title} {Universal description of
  dissipative {Tomonaga-Luttinger liquids} with $\mathrm{SU}(n)$ spin symmetry:
  Exact spectrum and critical exponents},\ }\href
  {https://doi.org/10.1103/PhysRevB.107.045110} {\bibfield  {journal} {\bibinfo
   {journal} {Phys. Rev. B}\ }\textbf {\bibinfo {volume} {107}},\ \bibinfo
  {pages} {045110} (\bibinfo {year} {2023})}\BibitemShut {NoStop}%
\bibitem [{\citenamefont {D\'ora}\ \emph {et~al.}(2023)\citenamefont {D\'ora},
  \citenamefont {Werner},\ and\ \citenamefont
  {Moca}}]{nhTLL-PhysRevB.108.035104}%
  \BibitemOpen
  \bibfield  {author} {\bibinfo {author} {\bibfnamefont {B.}~\bibnamefont
  {D\'ora}}, \bibinfo {author} {\bibfnamefont {M.~A.}\ \bibnamefont {Werner}},\
  and\ \bibinfo {author} {\bibfnamefont {C.~P.}\ \bibnamefont {Moca}},\
  }\bibfield  {title} {\bibinfo {title} {Quantum quench dynamics in the
  {Luttinger} liquid phase of the {Hatano-Nelson} model},\ }\href
  {https://doi.org/10.1103/PhysRevB.108.035104} {\bibfield  {journal} {\bibinfo
   {journal} {Phys. Rev. B}\ }\textbf {\bibinfo {volume} {108}},\ \bibinfo
  {pages} {035104} (\bibinfo {year} {2023})}\BibitemShut {NoStop}%
\bibitem [{\citenamefont {Dubey}\ \emph {et~al.}(2023)\citenamefont {Dubey},
  \citenamefont {Biswas},\ and\ \citenamefont
  {Kundu}}]{nhTLL-PhysRevB.108.085433}%
  \BibitemOpen
  \bibfield  {author} {\bibinfo {author} {\bibfnamefont {A.}~\bibnamefont
  {Dubey}}, \bibinfo {author} {\bibfnamefont {S.}~\bibnamefont {Biswas}},\ and\
  \bibinfo {author} {\bibfnamefont {A.}~\bibnamefont {Kundu}},\ }\bibfield
  {title} {\bibinfo {title} {Operator correlations in a quenched non-{Hermitian
  Luttinger} liquid},\ }\href {https://doi.org/10.1103/PhysRevB.108.085433}
  {\bibfield  {journal} {\bibinfo  {journal} {Phys. Rev. B}\ }\textbf {\bibinfo
  {volume} {108}},\ \bibinfo {pages} {085433} (\bibinfo {year}
  {2023})}\BibitemShut {NoStop}%
\bibitem [{\citenamefont {Garratt}\ \emph {et~al.}(2023)\citenamefont
  {Garratt}, \citenamefont {Weinstein},\ and\ \citenamefont
  {Altman}}]{AltmanMeasurementsTLL}%
  \BibitemOpen
  \bibfield  {author} {\bibinfo {author} {\bibfnamefont {S.~J.}\ \bibnamefont
  {Garratt}}, \bibinfo {author} {\bibfnamefont {Z.}~\bibnamefont {Weinstein}},\
  and\ \bibinfo {author} {\bibfnamefont {E.}~\bibnamefont {Altman}},\
  }\bibfield  {title} {\bibinfo {title} {Measurements conspire nonlocally to
  restructure critical quantum states},\ }\href
  {https://doi.org/10.1103/PhysRevX.13.021026} {\bibfield  {journal} {\bibinfo
  {journal} {Phys. Rev. X}\ }\textbf {\bibinfo {volume} {13}},\ \bibinfo
  {pages} {021026} (\bibinfo {year} {2023})}\BibitemShut {NoStop}%
\bibitem [{\citenamefont {Buchhold}\ \emph {et~al.}(2021)\citenamefont
  {Buchhold}, \citenamefont {Minoguchi}, \citenamefont {Altland},\ and\
  \citenamefont {Diehl}}]{Buchhold_2021}%
  \BibitemOpen
  \bibfield  {author} {\bibinfo {author} {\bibfnamefont {M.}~\bibnamefont
  {Buchhold}}, \bibinfo {author} {\bibfnamefont {Y.}~\bibnamefont {Minoguchi}},
  \bibinfo {author} {\bibfnamefont {A.}~\bibnamefont {Altland}},\ and\ \bibinfo
  {author} {\bibfnamefont {S.}~\bibnamefont {Diehl}},\ }\bibfield  {title}
  {\bibinfo {title} {Effective theory for the measurement-induced phase
  transition of dirac fermions},\ }\href
  {https://doi.org/10.1103%2Fphysrevx.11.041004} {\bibfield  {journal}
  {\bibinfo  {journal} {Physical Review X}\ }\textbf {\bibinfo {volume} {11}}
  (\bibinfo {year} {2021})}\BibitemShut {NoStop}%
\bibitem [{\citenamefont {Ladewig}\ \emph {et~al.}(2022)\citenamefont
  {Ladewig}, \citenamefont {Diehl},\ and\ \citenamefont
  {Buchhold}}]{MonitoredFermionsDissipation}%
  \BibitemOpen
  \bibfield  {author} {\bibinfo {author} {\bibfnamefont {B.}~\bibnamefont
  {Ladewig}}, \bibinfo {author} {\bibfnamefont {S.}~\bibnamefont {Diehl}},\
  and\ \bibinfo {author} {\bibfnamefont {M.}~\bibnamefont {Buchhold}},\
  }\bibfield  {title} {\bibinfo {title} {Monitored open fermion dynamics:
  Exploring the interplay of measurement, decoherence, and free {Hamiltonian}
  evolution},\ }\href {https://doi.org/10.1103/PhysRevResearch.4.033001}
  {\bibfield  {journal} {\bibinfo  {journal} {Phys. Rev. Res.}\ }\textbf
  {\bibinfo {volume} {4}},\ \bibinfo {pages} {033001} (\bibinfo {year}
  {2022})}\BibitemShut {NoStop}%
\bibitem [{\citenamefont {Yamamoto}\ \emph {et~al.}(2022)\citenamefont
  {Yamamoto}, \citenamefont {Nakagawa}, \citenamefont {Tezuka}, \citenamefont
  {Ueda},\ and\ \citenamefont {Kawakami}}]{NonHermTLL}%
  \BibitemOpen
  \bibfield  {author} {\bibinfo {author} {\bibfnamefont {K.}~\bibnamefont
  {Yamamoto}}, \bibinfo {author} {\bibfnamefont {M.}~\bibnamefont {Nakagawa}},
  \bibinfo {author} {\bibfnamefont {M.}~\bibnamefont {Tezuka}}, \bibinfo
  {author} {\bibfnamefont {M.}~\bibnamefont {Ueda}},\ and\ \bibinfo {author}
  {\bibfnamefont {N.}~\bibnamefont {Kawakami}},\ }\bibfield  {title} {\bibinfo
  {title} {Universal properties of dissipative {Tomonaga-Luttinger} liquids:
  Case study of a non-{Hermitian XXZ} spin chain},\ }\href
  {https://doi.org/10.1103/PhysRevB.105.205125} {\bibfield  {journal} {\bibinfo
   {journal} {Phys. Rev. B}\ }\textbf {\bibinfo {volume} {105}},\ \bibinfo
  {pages} {205125} (\bibinfo {year} {2022})}\BibitemShut {NoStop}%
\bibitem [{\citenamefont {Sun}\ \emph {et~al.}(2023)\citenamefont {Sun},
  \citenamefont {Yao},\ and\ \citenamefont {Jian}}]{sun2023new}%
  \BibitemOpen
  \bibfield  {author} {\bibinfo {author} {\bibfnamefont {X.}~\bibnamefont
  {Sun}}, \bibinfo {author} {\bibfnamefont {H.}~\bibnamefont {Yao}},\ and\
  \bibinfo {author} {\bibfnamefont {S.-K.}\ \bibnamefont {Jian}},\ }\href@noop
  {} {\bibinfo {title} {New critical states induced by measurement}} (\bibinfo
  {year} {2023}),\ \Eprint {https://arxiv.org/abs/2301.11337} {arXiv:2301.11337
  [quant-ph]} \BibitemShut {NoStop}%
\bibitem [{\citenamefont {Collura}\ \emph {et~al.}(2015)\citenamefont
  {Collura}, \citenamefont {Calabrese},\ and\ \citenamefont
  {Essler}}]{Collura_XXZ_TLL}%
  \BibitemOpen
  \bibfield  {author} {\bibinfo {author} {\bibfnamefont {M.}~\bibnamefont
  {Collura}}, \bibinfo {author} {\bibfnamefont {P.}~\bibnamefont {Calabrese}},\
  and\ \bibinfo {author} {\bibfnamefont {F.~H.~L.}\ \bibnamefont {Essler}},\
  }\bibfield  {title} {\bibinfo {title} {Quantum quench within the gapless
  phase of the $\text{spin}\ensuremath{-}\frac{1}{2}$ {Heisenberg XXZ }spin
  chain},\ }\href {https://doi.org/10.1103/PhysRevB.92.125131} {\bibfield
  {journal} {\bibinfo  {journal} {Phys. Rev. B}\ }\textbf {\bibinfo {volume}
  {92}},\ \bibinfo {pages} {125131} (\bibinfo {year} {2015})}\BibitemShut
  {NoStop}%
\bibitem [{\citenamefont {Pollmann}\ \emph {et~al.}(2013)\citenamefont
  {Pollmann}, \citenamefont {Haque},\ and\ \citenamefont
  {D\'ora}}]{Pollman_XXZ_TLL}%
  \BibitemOpen
  \bibfield  {author} {\bibinfo {author} {\bibfnamefont {F.}~\bibnamefont
  {Pollmann}}, \bibinfo {author} {\bibfnamefont {M.}~\bibnamefont {Haque}},\
  and\ \bibinfo {author} {\bibfnamefont {B.}~\bibnamefont {D\'ora}},\
  }\bibfield  {title} {\bibinfo {title} {Linear quantum quench in the
  {Heisenberg XXZ} chain: Time-dependent {Luttinger}-model description of a
  lattice system},\ }\href {https://doi.org/10.1103/PhysRevB.87.041109}
  {\bibfield  {journal} {\bibinfo  {journal} {Phys. Rev. B}\ }\textbf {\bibinfo
  {volume} {87}},\ \bibinfo {pages} {041109(R)} (\bibinfo {year}
  {2013})}\BibitemShut {NoStop}%
\bibitem [{\citenamefont {Karrasch}\ \emph {et~al.}(2012)\citenamefont
  {Karrasch}, \citenamefont {Rentrop}, \citenamefont {Schuricht},\ and\
  \citenamefont {Meden}}]{MedenTLLQuench}%
  \BibitemOpen
  \bibfield  {author} {\bibinfo {author} {\bibfnamefont {C.}~\bibnamefont
  {Karrasch}}, \bibinfo {author} {\bibfnamefont {J.}~\bibnamefont {Rentrop}},
  \bibinfo {author} {\bibfnamefont {D.}~\bibnamefont {Schuricht}},\ and\
  \bibinfo {author} {\bibfnamefont {V.}~\bibnamefont {Meden}},\ }\bibfield
  {title} {\bibinfo {title} {{Luttinger}-liquid universality in the time
  evolution after an interaction quench},\ }\href
  {https://doi.org/10.1103/PhysRevLett.109.126406} {\bibfield  {journal}
  {\bibinfo  {journal} {Phys. Rev. Lett.}\ }\textbf {\bibinfo {volume} {109}},\
  \bibinfo {pages} {126406} (\bibinfo {year} {2012})}\BibitemShut {NoStop}%
\bibitem [{\citenamefont {Kennes}\ and\ \citenamefont
  {Meden}(2013)}]{MedenTLLSteadyState}%
  \BibitemOpen
  \bibfield  {author} {\bibinfo {author} {\bibfnamefont {D.~M.}\ \bibnamefont
  {Kennes}}\ and\ \bibinfo {author} {\bibfnamefont {V.}~\bibnamefont {Meden}},\
  }\bibfield  {title} {\bibinfo {title} {{Luttinger} liquid properties of the
  steady state after a quantum quench},\ }\href
  {https://doi.org/10.1103/PhysRevB.88.165131} {\bibfield  {journal} {\bibinfo
  {journal} {Phys. Rev. B}\ }\textbf {\bibinfo {volume} {88}},\ \bibinfo
  {pages} {165131} (\bibinfo {year} {2013})}\BibitemShut {NoStop}%
\bibitem [{\citenamefont {Imambekov}\ \emph {et~al.}(2012)\citenamefont
  {Imambekov}, \citenamefont {Schmidt},\ and\ \citenamefont
  {Glazman}}]{ImambekovSchmidtGlazman}%
  \BibitemOpen
  \bibfield  {author} {\bibinfo {author} {\bibfnamefont {A.}~\bibnamefont
  {Imambekov}}, \bibinfo {author} {\bibfnamefont {T.~L.}\ \bibnamefont
  {Schmidt}},\ and\ \bibinfo {author} {\bibfnamefont {L.~I.}\ \bibnamefont
  {Glazman}},\ }\bibfield  {title} {\bibinfo {title} {One-dimensional quantum
  liquids: Beyond the {Luttinger} liquid paradigm},\ }\href
  {https://doi.org/10.1103/RevModPhys.84.1253} {\bibfield  {journal} {\bibinfo
  {journal} {Rev. Mod. Phys.}\ }\textbf {\bibinfo {volume} {84}},\ \bibinfo
  {pages} {1253} (\bibinfo {year} {2012})}\BibitemShut {NoStop}%
\bibitem [{\citenamefont {Gutman}\ \emph {et~al.}(2008)\citenamefont {Gutman},
  \citenamefont {Gefen},\ and\ \citenamefont
  {Mirlin}}]{PhysRevLett.101.126802}%
  \BibitemOpen
  \bibfield  {author} {\bibinfo {author} {\bibfnamefont {D.~B.}\ \bibnamefont
  {Gutman}}, \bibinfo {author} {\bibfnamefont {Y.}~\bibnamefont {Gefen}},\ and\
  \bibinfo {author} {\bibfnamefont {A.~D.}\ \bibnamefont {Mirlin}},\ }\bibfield
   {title} {\bibinfo {title} {Nonequilibrium {Luttinger} liquid: Zero-bias
  anomaly and dephasing},\ }\href
  {https://doi.org/10.1103/PhysRevLett.101.126802} {\bibfield  {journal}
  {\bibinfo  {journal} {Phys. Rev. Lett.}\ }\textbf {\bibinfo {volume} {101}},\
  \bibinfo {pages} {126802} (\bibinfo {year} {2008})}\BibitemShut {NoStop}%
\bibitem [{\citenamefont {Gutman}\ \emph
  {et~al.}(2010{\natexlab{a}})\citenamefont {Gutman}, \citenamefont {Gefen},\
  and\ \citenamefont {Mirlin}}]{GutmanBosonization}%
  \BibitemOpen
  \bibfield  {author} {\bibinfo {author} {\bibfnamefont {D.~B.}\ \bibnamefont
  {Gutman}}, \bibinfo {author} {\bibfnamefont {Y.}~\bibnamefont {Gefen}},\ and\
  \bibinfo {author} {\bibfnamefont {A.~D.}\ \bibnamefont {Mirlin}},\ }\bibfield
   {title} {\bibinfo {title} {Bosonization of one-dimensional fermions out of
  equilibrium},\ }\href {https://doi.org/10.1103/PhysRevB.81.085436} {\bibfield
   {journal} {\bibinfo  {journal} {Phys. Rev. B}\ }\textbf {\bibinfo {volume}
  {81}},\ \bibinfo {pages} {085436} (\bibinfo {year}
  {2010}{\natexlab{a}})}\BibitemShut {NoStop}%
\bibitem [{\citenamefont {Gutman}\ \emph
  {et~al.}(2010{\natexlab{b}})\citenamefont {Gutman}, \citenamefont {Gefen},\
  and\ \citenamefont {Mirlin}}]{PhysRevLett.105.256802}%
  \BibitemOpen
  \bibfield  {author} {\bibinfo {author} {\bibfnamefont {D.~B.}\ \bibnamefont
  {Gutman}}, \bibinfo {author} {\bibfnamefont {Y.}~\bibnamefont {Gefen}},\ and\
  \bibinfo {author} {\bibfnamefont {A.~D.}\ \bibnamefont {Mirlin}},\ }\bibfield
   {title} {\bibinfo {title} {Full counting statistics of a {Luttinger} liquid
  conductor},\ }\href {https://doi.org/10.1103/PhysRevLett.105.256802}
  {\bibfield  {journal} {\bibinfo  {journal} {Phys. Rev. Lett.}\ }\textbf
  {\bibinfo {volume} {105}},\ \bibinfo {pages} {256802} (\bibinfo {year}
  {2010}{\natexlab{b}})}\BibitemShut {NoStop}%
\bibitem [{\citenamefont {Anderson}(1967)}]{AndersonOC}%
  \BibitemOpen
  \bibfield  {author} {\bibinfo {author} {\bibfnamefont {P.~W.}\ \bibnamefont
  {Anderson}},\ }\bibfield  {title} {\bibinfo {title} {Infrared catastrophe in
  fermi gases with local scattering potentials},\ }\href
  {https://doi.org/10.1103/PhysRevLett.18.1049} {\bibfield  {journal} {\bibinfo
   {journal} {Phys. Rev. Lett.}\ }\textbf {\bibinfo {volume} {18}},\ \bibinfo
  {pages} {1049} (\bibinfo {year} {1967})}\BibitemShut {NoStop}%
\bibitem [{\citenamefont {Mahan}(1993)}]{Mahan}%
  \BibitemOpen
  \bibfield  {author} {\bibinfo {author} {\bibfnamefont {G.~D.}\ \bibnamefont
  {Mahan}},\ }\href@noop {} {\emph {\bibinfo {title} {Many-Particle
  Physics}}},\ \bibinfo {edition} {2nd}\ ed.\ (\bibinfo  {publisher} {Plenum
  press},\ \bibinfo {address} {New York},\ \bibinfo {year} {1993})\BibitemShut
  {NoStop}%
\bibitem [{\citenamefont {Schotte}\ and\ \citenamefont
  {Schotte}(1969)}]{SchotteAndSchotte}%
  \BibitemOpen
  \bibfield  {author} {\bibinfo {author} {\bibfnamefont {K.~D.}\ \bibnamefont
  {Schotte}}\ and\ \bibinfo {author} {\bibfnamefont {U.}~\bibnamefont
  {Schotte}},\ }\bibfield  {title} {\bibinfo {title} {Tomonaga's model and the
  threshold singularity of {X-ray} spectra of metals},\ }\href@noop {}
  {\bibfield  {journal} {\bibinfo  {journal} {Phys. Rev.}\ }\textbf {\bibinfo
  {volume} {182}},\ \bibinfo {pages} {479} (\bibinfo {year}
  {1969})}\BibitemShut {NoStop}%
\bibitem [{\citenamefont {Hewson}(1993)}]{Hewson}%
  \BibitemOpen
  \bibfield  {author} {\bibinfo {author} {\bibfnamefont {A.~C.}\ \bibnamefont
  {Hewson}},\ }\href {https://doi.org/10.1017/CBO9780511470752} {\emph
  {\bibinfo {title} {The Kondo Problem to Heavy Fermions}}},\ Cambridge Studies
  in Magnetism\ (\bibinfo  {publisher} {Cambridge University Press},\ \bibinfo
  {year} {1993})\BibitemShut {NoStop}%
\bibitem [{\citenamefont {Coleman}(1984)}]{SlaveBosons}%
  \BibitemOpen
  \bibfield  {author} {\bibinfo {author} {\bibfnamefont {P.}~\bibnamefont
  {Coleman}},\ }\bibfield  {title} {\bibinfo {title} {New approach to the
  mixed-valence problem},\ }\href {https://doi.org/10.1103/PhysRevB.29.3035}
  {\bibfield  {journal} {\bibinfo  {journal} {Phys. Rev. B}\ }\textbf {\bibinfo
  {volume} {29}},\ \bibinfo {pages} {3035} (\bibinfo {year}
  {1984})}\BibitemShut {NoStop}%
\bibitem [{\citenamefont {Yuval}\ and\ \citenamefont
  {Anderson}(1970)}]{PRB_Anderson_Kondo}%
  \BibitemOpen
  \bibfield  {author} {\bibinfo {author} {\bibfnamefont {G.}~\bibnamefont
  {Yuval}}\ and\ \bibinfo {author} {\bibfnamefont {P.~W.}\ \bibnamefont
  {Anderson}},\ }\bibfield  {title} {\bibinfo {title} {Exact results for the
  {Kondo} problem: One-body theory and extension to finite temperature},\
  }\href {https://doi.org/10.1103/PhysRevB.1.1522} {\bibfield  {journal}
  {\bibinfo  {journal} {Phys. Rev. B}\ }\textbf {\bibinfo {volume} {1}},\
  \bibinfo {pages} {1522} (\bibinfo {year} {1970})}\BibitemShut {NoStop}%
\bibitem [{\citenamefont {Heyl}\ and\ \citenamefont
  {Kehrein}(2012)}]{PRB_EdgeSing_QuantumDots}%
  \BibitemOpen
  \bibfield  {author} {\bibinfo {author} {\bibfnamefont {M.}~\bibnamefont
  {Heyl}}\ and\ \bibinfo {author} {\bibfnamefont {S.}~\bibnamefont {Kehrein}},\
  }\bibfield  {title} {\bibinfo {title} {{X-ray} edge singularity in optical
  spectra of quantum dots},\ }\href
  {https://doi.org/10.1103/PhysRevB.85.155413} {\bibfield  {journal} {\bibinfo
  {journal} {Phys. Rev. B}\ }\textbf {\bibinfo {volume} {85}},\ \bibinfo
  {pages} {155413} (\bibinfo {year} {2012})}\BibitemShut {NoStop}%
\bibitem [{\citenamefont {d'Ambrumenil}\ and\ \citenamefont
  {Muzykantskii}(2005)}]{PRB_FermiGasResponse}%
  \BibitemOpen
  \bibfield  {author} {\bibinfo {author} {\bibfnamefont {N.}~\bibnamefont
  {d'Ambrumenil}}\ and\ \bibinfo {author} {\bibfnamefont {B.}~\bibnamefont
  {Muzykantskii}},\ }\bibfield  {title} {\bibinfo {title} {Fermi gas response
  to time-dependent perturbations},\ }\href
  {https://doi.org/10.1103/PhysRevB.71.045326} {\bibfield  {journal} {\bibinfo
  {journal} {Phys. Rev. B}\ }\textbf {\bibinfo {volume} {71}},\ \bibinfo
  {pages} {045326} (\bibinfo {year} {2005})}\BibitemShut {NoStop}%
\bibitem [{\citenamefont {Grusdt}\ \emph {et~al.}(2018)\citenamefont {Grusdt},
  \citenamefont {Seetharam}, \citenamefont {Shchadilova},\ and\ \citenamefont
  {Demler}}]{PhysRevA.97.033612}%
  \BibitemOpen
  \bibfield  {author} {\bibinfo {author} {\bibfnamefont {F.}~\bibnamefont
  {Grusdt}}, \bibinfo {author} {\bibfnamefont {K.}~\bibnamefont {Seetharam}},
  \bibinfo {author} {\bibfnamefont {Y.}~\bibnamefont {Shchadilova}},\ and\
  \bibinfo {author} {\bibfnamefont {E.}~\bibnamefont {Demler}},\ }\bibfield
  {title} {\bibinfo {title} {Strong-coupling {Bose} polarons out of
  equilibrium: Dynamical renormalization-group approach},\ }\href
  {https://doi.org/10.1103/PhysRevA.97.033612} {\bibfield  {journal} {\bibinfo
  {journal} {Phys. Rev. A}\ }\textbf {\bibinfo {volume} {97}},\ \bibinfo
  {pages} {033612} (\bibinfo {year} {2018})}\BibitemShut {NoStop}%
\bibitem [{\citenamefont {Knap}\ \emph {et~al.}(2012)\citenamefont {Knap},
  \citenamefont {Shashi}, \citenamefont {Nishida}, \citenamefont {Imambekov},
  \citenamefont {Abanin},\ and\ \citenamefont {Demler}}]{PhysRevX.2.041020}%
  \BibitemOpen
  \bibfield  {author} {\bibinfo {author} {\bibfnamefont {M.}~\bibnamefont
  {Knap}}, \bibinfo {author} {\bibfnamefont {A.}~\bibnamefont {Shashi}},
  \bibinfo {author} {\bibfnamefont {Y.}~\bibnamefont {Nishida}}, \bibinfo
  {author} {\bibfnamefont {A.}~\bibnamefont {Imambekov}}, \bibinfo {author}
  {\bibfnamefont {D.~A.}\ \bibnamefont {Abanin}},\ and\ \bibinfo {author}
  {\bibfnamefont {E.}~\bibnamefont {Demler}},\ }\bibfield  {title} {\bibinfo
  {title} {Time-dependent impurity in ultracold fermions: Orthogonality
  catastrophe and beyond},\ }\href {https://doi.org/10.1103/PhysRevX.2.041020}
  {\bibfield  {journal} {\bibinfo  {journal} {Phys. Rev. X}\ }\textbf {\bibinfo
  {volume} {2}},\ \bibinfo {pages} {041020} (\bibinfo {year}
  {2012})}\BibitemShut {NoStop}%
\bibitem [{\citenamefont {Shashi}\ \emph {et~al.}(2014)\citenamefont {Shashi},
  \citenamefont {Grusdt}, \citenamefont {Abanin},\ and\ \citenamefont
  {Demler}}]{PhysRevA.89.053617}%
  \BibitemOpen
  \bibfield  {author} {\bibinfo {author} {\bibfnamefont {A.}~\bibnamefont
  {Shashi}}, \bibinfo {author} {\bibfnamefont {F.}~\bibnamefont {Grusdt}},
  \bibinfo {author} {\bibfnamefont {D.~A.}\ \bibnamefont {Abanin}},\ and\
  \bibinfo {author} {\bibfnamefont {E.}~\bibnamefont {Demler}},\ }\bibfield
  {title} {\bibinfo {title} {Radio-frequency spectroscopy of polarons in
  ultracold {Bose} gases},\ }\href {https://doi.org/10.1103/PhysRevA.89.053617}
  {\bibfield  {journal} {\bibinfo  {journal} {Phys. Rev. A}\ }\textbf {\bibinfo
  {volume} {89}},\ \bibinfo {pages} {053617} (\bibinfo {year}
  {2014})}\BibitemShut {NoStop}%
\bibitem [{\citenamefont {Nozi\`eres}\ and\ \citenamefont
  {De~Dominicis}(1969)}]{NozieresDeDominicis}%
  \BibitemOpen
  \bibfield  {author} {\bibinfo {author} {\bibfnamefont {P.}~\bibnamefont
  {Nozi\`eres}}\ and\ \bibinfo {author} {\bibfnamefont {C.~T.}\ \bibnamefont
  {De~Dominicis}},\ }\bibfield  {title} {\bibinfo {title} {Singularities in the
  {X-ray} absorption and emission of metals. iii. one-body theory exact
  solution},\ }\href {https://doi.org/10.1103/PhysRev.178.1097} {\bibfield
  {journal} {\bibinfo  {journal} {Phys. Rev.}\ }\textbf {\bibinfo {volume}
  {178}},\ \bibinfo {pages} {1097} (\bibinfo {year} {1969})}\BibitemShut
  {NoStop}%
\bibitem [{\citenamefont {Combescot}\ and\ \citenamefont
  {Nozi\`eres}(1971)}]{CombescotNozieres}%
  \BibitemOpen
  \bibfield  {author} {\bibinfo {author} {\bibfnamefont {M.}~\bibnamefont
  {Combescot}}\ and\ \bibinfo {author} {\bibfnamefont {P.}~\bibnamefont
  {Nozi\`eres}},\ }\bibfield  {title} {\bibinfo {title} {Infrared catastrophy
  and excitons in the {X-ray} spectra of metals},\ }\href@noop {} {\bibfield
  {journal} {\bibinfo  {journal} {J. Phys. (France)}\ }\textbf {\bibinfo
  {volume} {32}},\ \bibinfo {pages} {913} (\bibinfo {year} {1971})}\BibitemShut
  {NoStop}%
\bibitem [{\citenamefont {Tonielli}\ \emph {et~al.}(2019)\citenamefont
  {Tonielli}, \citenamefont {Fazio}, \citenamefont {Diehl},\ and\ \citenamefont
  {Marino}}]{Tonielli}%
  \BibitemOpen
  \bibfield  {author} {\bibinfo {author} {\bibfnamefont {F.}~\bibnamefont
  {Tonielli}}, \bibinfo {author} {\bibfnamefont {R.}~\bibnamefont {Fazio}},
  \bibinfo {author} {\bibfnamefont {S.}~\bibnamefont {Diehl}},\ and\ \bibinfo
  {author} {\bibfnamefont {J.}~\bibnamefont {Marino}},\ }\bibfield  {title}
  {\bibinfo {title} {Orthogonality catastrophe in dissipative quantum many-body
  systems},\ }\href {https://doi.org/10.1103/PhysRevLett.122.040604} {\bibfield
   {journal} {\bibinfo  {journal} {Phys. Rev. Lett.}\ }\textbf {\bibinfo
  {volume} {122}},\ \bibinfo {pages} {040604} (\bibinfo {year}
  {2019})}\BibitemShut {NoStop}%
\bibitem [{\citenamefont {Lupo}\ and\ \citenamefont
  {Schir\'o}(2016)}]{PRB_LupoSchiro}%
  \BibitemOpen
  \bibfield  {author} {\bibinfo {author} {\bibfnamefont {C.}~\bibnamefont
  {Lupo}}\ and\ \bibinfo {author} {\bibfnamefont {M.}~\bibnamefont
  {Schir\'o}},\ }\bibfield  {title} {\bibinfo {title} {Transient {Loschmidt}
  echo in quenched {Ising} chains},\ }\href
  {https://doi.org/10.1103/PhysRevB.94.014310} {\bibfield  {journal} {\bibinfo
  {journal} {Phys. Rev. B}\ }\textbf {\bibinfo {volume} {94}},\ \bibinfo
  {pages} {014310} (\bibinfo {year} {2016})}\BibitemShut {NoStop}%
\bibitem [{\citenamefont {Schir\'o}\ and\ \citenamefont
  {Mitra}(2014)}]{PRL_SchiroMitra}%
  \BibitemOpen
  \bibfield  {author} {\bibinfo {author} {\bibfnamefont {M.}~\bibnamefont
  {Schir\'o}}\ and\ \bibinfo {author} {\bibfnamefont {A.}~\bibnamefont
  {Mitra}},\ }\bibfield  {title} {\bibinfo {title} {Transient orthogonality
  catastrophe in a time-dependent nonequilibrium environment},\ }\href
  {https://doi.org/10.1103/PhysRevLett.112.246401} {\bibfield  {journal}
  {\bibinfo  {journal} {Phys. Rev. Lett.}\ }\textbf {\bibinfo {volume} {112}},\
  \bibinfo {pages} {246401} (\bibinfo {year} {2014})}\BibitemShut {NoStop}%
\bibitem [{\citenamefont {Berdanier}\ \emph {et~al.}(2019)\citenamefont
  {Berdanier}, \citenamefont {Marino},\ and\ \citenamefont
  {Altman}}]{berdanier2019universal}%
  \BibitemOpen
  \bibfield  {author} {\bibinfo {author} {\bibfnamefont {W.}~\bibnamefont
  {Berdanier}}, \bibinfo {author} {\bibfnamefont {J.}~\bibnamefont {Marino}},\
  and\ \bibinfo {author} {\bibfnamefont {E.}~\bibnamefont {Altman}},\
  }\bibfield  {title} {\bibinfo {title} {Universal dynamics of stochastically
  driven quantum impurities},\ }\href
  {https://doi.org/10.1103/PhysRevLett.123.230604} {\bibfield  {journal}
  {\bibinfo  {journal} {Phys. Rev. Lett.}\ }\textbf {\bibinfo {volume} {123}},\
  \bibinfo {pages} {230604} (\bibinfo {year} {2019})}\BibitemShut {NoStop}%
\bibitem [{\citenamefont {Jung}\ \emph {et~al.}(1999)\citenamefont {Jung},
  \citenamefont {M\"uller},\ and\ \citenamefont {Rotter}}]{PhysRevE.60.114}%
  \BibitemOpen
  \bibfield  {author} {\bibinfo {author} {\bibfnamefont {C.}~\bibnamefont
  {Jung}}, \bibinfo {author} {\bibfnamefont {M.}~\bibnamefont {M\"uller}},\
  and\ \bibinfo {author} {\bibfnamefont {I.}~\bibnamefont {Rotter}},\
  }\bibfield  {title} {\bibinfo {title} {Phase transitions in open quantum
  systems},\ }\href {https://doi.org/10.1103/PhysRevE.60.114} {\bibfield
  {journal} {\bibinfo  {journal} {Phys. Rev. E}\ }\textbf {\bibinfo {volume}
  {60}},\ \bibinfo {pages} {114} (\bibinfo {year} {1999})}\BibitemShut
  {NoStop}%
\bibitem [{\citenamefont {Burke}\ \emph {et~al.}(2020)\citenamefont {Burke},
  \citenamefont {Wiersig},\ and\ \citenamefont {Haque}}]{Haque}%
  \BibitemOpen
  \bibfield  {author} {\bibinfo {author} {\bibfnamefont {P.~C.}\ \bibnamefont
  {Burke}}, \bibinfo {author} {\bibfnamefont {J.}~\bibnamefont {Wiersig}},\
  and\ \bibinfo {author} {\bibfnamefont {M.}~\bibnamefont {Haque}},\ }\bibfield
   {title} {\bibinfo {title} {Non-{Hermitian} scattering on a tight-binding
  lattice},\ }\href {https://doi.org/10.1103/PhysRevA.102.012212} {\bibfield
  {journal} {\bibinfo  {journal} {Phys. Rev. A}\ }\textbf {\bibinfo {volume}
  {102}},\ \bibinfo {pages} {012212} (\bibinfo {year} {2020})}\BibitemShut
  {NoStop}%
\bibitem [{\citenamefont {Rosso}\ \emph {et~al.}(2020)\citenamefont {Rosso},
  \citenamefont {Iemini}, \citenamefont {Schir{\`o}},\ and\ \citenamefont
  {Mazza}}]{SciPost_Schiro}%
  \BibitemOpen
  \bibfield  {author} {\bibinfo {author} {\bibfnamefont {L.}~\bibnamefont
  {Rosso}}, \bibinfo {author} {\bibfnamefont {F.}~\bibnamefont {Iemini}},
  \bibinfo {author} {\bibfnamefont {M.}~\bibnamefont {Schir{\`o}}},\ and\
  \bibinfo {author} {\bibfnamefont {L.}~\bibnamefont {Mazza}},\ }\bibfield
  {title} {\bibinfo {title} {{Dissipative flow equations}},\ }\href
  {https://doi.org/10.21468/SciPostPhys.9.6.091} {\bibfield  {journal}
  {\bibinfo  {journal} {SciPost Phys.}\ }\textbf {\bibinfo {volume} {9}},\
  \bibinfo {pages} {091} (\bibinfo {year} {2020})}\BibitemShut {NoStop}%
\bibitem [{SM(2024)}]{SM}%
  \BibitemOpen
  \href@noop {} {\bibinfo {title} {{See the Supplemental Material at [URL will
  be inserted by publisher] for technical details.}}} (\bibinfo {year}
  {2024})\BibitemShut {NoStop}%
\bibitem [{\citenamefont {Wiseman}\ and\ \citenamefont
  {Milburn}(2009)}]{WisemanMilburn}%
  \BibitemOpen
  \bibfield  {author} {\bibinfo {author} {\bibfnamefont {H.~M.}\ \bibnamefont
  {Wiseman}}\ and\ \bibinfo {author} {\bibfnamefont {G.~J.}\ \bibnamefont
  {Milburn}},\ }\href@noop {} {\emph {\bibinfo {title} {Quantum Measurement and
  Control}}}\ (\bibinfo  {publisher} {Cambridge University Press},\ \bibinfo
  {address} {Cambridge},\ \bibinfo {year} {2009})\BibitemShut {NoStop}%
\bibitem [{\citenamefont {Nielsen}\ and\ \citenamefont
  {Chuang}(2010)}]{NielsenChuang}%
  \BibitemOpen
  \bibfield  {author} {\bibinfo {author} {\bibfnamefont {M.~A.}\ \bibnamefont
  {Nielsen}}\ and\ \bibinfo {author} {\bibfnamefont {I.~L.}\ \bibnamefont
  {Chuang}},\ }\href {https://doi.org/10.1017/CBO9780511976667} {\emph
  {\bibinfo {title} {Quantum Computation and Quantum Information: 10th
  Anniversary Edition}}}\ (\bibinfo  {publisher} {Cambridge University Press},\
  \bibinfo {address} {Cambridge},\ \bibinfo {year} {2010})\BibitemShut
  {NoStop}%
\bibitem [{\citenamefont {Turkeshi}\ and\ \citenamefont
  {Schir\'o}(2023)}]{SchiroNonHermPRB}%
  \BibitemOpen
  \bibfield  {author} {\bibinfo {author} {\bibfnamefont {X.}~\bibnamefont
  {Turkeshi}}\ and\ \bibinfo {author} {\bibfnamefont {M.}~\bibnamefont
  {Schir\'o}},\ }\bibfield  {title} {\bibinfo {title} {Entanglement and
  correlation spreading in non-{Hermitian} spin chains},\ }\href
  {https://doi.org/10.1103/PhysRevB.107.L020403} {\bibfield  {journal}
  {\bibinfo  {journal} {Phys. Rev. B}\ }\textbf {\bibinfo {volume} {107}},\
  \bibinfo {pages} {L020403} (\bibinfo {year} {2023})}\BibitemShut {NoStop}%
\bibitem [{\citenamefont {{Le Gal}}\ \emph {et~al.}(2023)\citenamefont {{Le
  Gal}}, \citenamefont {Turkeshi},\ and\ \citenamefont
  {Schir{\`o}}}]{SchiroNonHermSciPost}%
  \BibitemOpen
  \bibfield  {author} {\bibinfo {author} {\bibfnamefont {Y.}~\bibnamefont {{Le
  Gal}}}, \bibinfo {author} {\bibfnamefont {X.}~\bibnamefont {Turkeshi}},\ and\
  \bibinfo {author} {\bibfnamefont {M.}~\bibnamefont {Schir{\`o}}},\ }\bibfield
   {title} {\bibinfo {title} {{Volume-to-area law entanglement transition in a
  non-{Hermitian} free fermionic chain}},\ }\href
  {https://doi.org/10.21468/SciPostPhys.14.5.138} {\bibfield  {journal}
  {\bibinfo  {journal} {SciPost Phys.}\ }\textbf {\bibinfo {volume} {14}},\
  \bibinfo {pages} {138} (\bibinfo {year} {2023})}\BibitemShut {NoStop}%
\bibitem [{\citenamefont {Turkeshi}\ \emph {et~al.}(2021)\citenamefont
  {Turkeshi}, \citenamefont {Biella}, \citenamefont {Fazio}, \citenamefont
  {Dalmonte},\ and\ \citenamefont {Schir\'o}}]{TurkeshiPhysRevB.103.224210}%
  \BibitemOpen
  \bibfield  {author} {\bibinfo {author} {\bibfnamefont {X.}~\bibnamefont
  {Turkeshi}}, \bibinfo {author} {\bibfnamefont {A.}~\bibnamefont {Biella}},
  \bibinfo {author} {\bibfnamefont {R.}~\bibnamefont {Fazio}}, \bibinfo
  {author} {\bibfnamefont {M.}~\bibnamefont {Dalmonte}},\ and\ \bibinfo
  {author} {\bibfnamefont {M.}~\bibnamefont {Schir\'o}},\ }\bibfield  {title}
  {\bibinfo {title} {Measurement-induced entanglement transitions in the
  quantum {Ising} chain: From infinite to zero clicks},\ }\href
  {https://doi.org/10.1103/PhysRevB.103.224210} {\bibfield  {journal} {\bibinfo
   {journal} {Phys. Rev. B}\ }\textbf {\bibinfo {volume} {103}},\ \bibinfo
  {pages} {224210} (\bibinfo {year} {2021})}\BibitemShut {NoStop}%
\bibitem [{\citenamefont {Daley}(2014)}]{DaleyQuantumTrajectories}%
  \BibitemOpen
  \bibfield  {author} {\bibinfo {author} {\bibfnamefont {A.~J.}\ \bibnamefont
  {Daley}},\ }\bibfield  {title} {\bibinfo {title} {Quantum trajectories and
  open many-body quantum systems},\ }\href
  {https://doi.org/10.1080/00018732.2014.933502} {\bibfield  {journal}
  {\bibinfo  {journal} {Advances in Physics}\ }\textbf {\bibinfo {volume}
  {63}},\ \bibinfo {pages} {77} (\bibinfo {year} {2014})}\BibitemShut {NoStop}%
\bibitem [{Note1()}]{Note1}%
  \BibitemOpen
  \bibinfo {note} {{Up to a constant shift $K-\setbox \z@ \hbox {\mathsurround
  \z@ $\textstyle H$}\mathaccent "0365{H}\propto \protect \openone
  $.}}\BibitemShut {Stop}%
\bibitem [{\citenamefont {Foss-Feig}\ \emph {et~al.}(2012)\citenamefont
  {Foss-Feig}, \citenamefont {Daley}, \citenamefont {Thompson},\ and\
  \citenamefont {Rey}}]{Rey_Hot_reactive_fermions}%
  \BibitemOpen
  \bibfield  {author} {\bibinfo {author} {\bibfnamefont {M.}~\bibnamefont
  {Foss-Feig}}, \bibinfo {author} {\bibfnamefont {A.~J.}\ \bibnamefont
  {Daley}}, \bibinfo {author} {\bibfnamefont {J.~K.}\ \bibnamefont
  {Thompson}},\ and\ \bibinfo {author} {\bibfnamefont {A.~M.}\ \bibnamefont
  {Rey}},\ }\bibfield  {title} {\bibinfo {title} {Steady-state many-body
  entanglement of hot reactive fermions},\ }\href
  {https://doi.org/10.1103/PhysRevLett.109.230501} {\bibfield  {journal}
  {\bibinfo  {journal} {Phys. Rev. Lett.}\ }\textbf {\bibinfo {volume} {109}},\
  \bibinfo {pages} {230501} (\bibinfo {year} {2012})}\BibitemShut {NoStop}%
\bibitem [{\citenamefont {Naghiloo}\ \emph {et~al.}(2019)\citenamefont
  {Naghiloo}, \citenamefont {Abbasi}, \citenamefont {Joglekar},\ and\
  \citenamefont {Murch}}]{QubitExceptionalPoint}%
  \BibitemOpen
  \bibfield  {author} {\bibinfo {author} {\bibfnamefont {M.}~\bibnamefont
  {Naghiloo}}, \bibinfo {author} {\bibfnamefont {M.}~\bibnamefont {Abbasi}},
  \bibinfo {author} {\bibfnamefont {Y.~N.}\ \bibnamefont {Joglekar}},\ and\
  \bibinfo {author} {\bibfnamefont {K.~W.}\ \bibnamefont {Murch}},\ }\bibfield
  {title} {\bibinfo {title} {Quantum state tomography across the exceptional
  point in a single dissipative qubit},\ }\href
  {https://doi.org/10.1038/s41567-019-0652-z} {\bibfield  {journal} {\bibinfo
  {journal} {Nature Physics}\ }\textbf {\bibinfo {volume} {15}},\ \bibinfo
  {pages} {1232} (\bibinfo {year} {2019})}\BibitemShut {NoStop}%
\bibitem [{\citenamefont {Fr\"oml}(2020)}]{ThesisFroml}%
  \BibitemOpen
  \bibfield  {author} {\bibinfo {author} {\bibfnamefont {H.~F.}\ \bibnamefont
  {Fr\"oml}},\ }\emph {\bibinfo {title} {Localized dissipation in fermionic
  quantum wires}},\ \href@noop {} {Ph.D. thesis},\ \bibinfo  {school}
  {Universit\"at zu K\"oln} (\bibinfo {year} {2020})\BibitemShut {NoStop}%
\bibitem [{\citenamefont {Fr\"oml}\ \emph {et~al.}(2019)\citenamefont
  {Fr\"oml}, \citenamefont {Chiocchetta}, \citenamefont {Kollath},\ and\
  \citenamefont {Diehl}}]{FromlPRL}%
  \BibitemOpen
  \bibfield  {author} {\bibinfo {author} {\bibfnamefont {H.}~\bibnamefont
  {Fr\"oml}}, \bibinfo {author} {\bibfnamefont {A.}~\bibnamefont
  {Chiocchetta}}, \bibinfo {author} {\bibfnamefont {C.}~\bibnamefont
  {Kollath}},\ and\ \bibinfo {author} {\bibfnamefont {S.}~\bibnamefont
  {Diehl}},\ }\bibfield  {title} {\bibinfo {title} {Fluctuation-induced quantum
  {Zeno} effect},\ }\href {https://doi.org/10.1103/PhysRevLett.122.040402}
  {\bibfield  {journal} {\bibinfo  {journal} {Phys. Rev. Lett.}\ }\textbf
  {\bibinfo {volume} {122}},\ \bibinfo {pages} {040402} (\bibinfo {year}
  {2019})}\BibitemShut {NoStop}%
\bibitem [{\citenamefont {M\"uller}\ \emph {et~al.}(2021)\citenamefont
  {M\"uller}, \citenamefont {Gievers}, \citenamefont {Fr\"oml}, \citenamefont
  {Diehl},\ and\ \citenamefont {Chiocchetta}}]{FromlPRB2}%
  \BibitemOpen
  \bibfield  {author} {\bibinfo {author} {\bibfnamefont {T.}~\bibnamefont
  {M\"uller}}, \bibinfo {author} {\bibfnamefont {M.}~\bibnamefont {Gievers}},
  \bibinfo {author} {\bibfnamefont {H.}~\bibnamefont {Fr\"oml}}, \bibinfo
  {author} {\bibfnamefont {S.}~\bibnamefont {Diehl}},\ and\ \bibinfo {author}
  {\bibfnamefont {A.}~\bibnamefont {Chiocchetta}},\ }\bibfield  {title}
  {\bibinfo {title} {Shape effects of localized losses in quantum wires:
  Dissipative resonances and nonequilibrium universality},\ }\href
  {https://doi.org/10.1103/PhysRevB.104.155431} {\bibfield  {journal} {\bibinfo
   {journal} {Phys. Rev. B}\ }\textbf {\bibinfo {volume} {104}},\ \bibinfo
  {pages} {155431} (\bibinfo {year} {2021})}\BibitemShut {NoStop}%
\bibitem [{\citenamefont {Krapivsky}\ \emph {et~al.}(2019)\citenamefont
  {Krapivsky}, \citenamefont {Mallick},\ and\ \citenamefont
  {Sels}}]{LocalizedFermionSource}%
  \BibitemOpen
  \bibfield  {author} {\bibinfo {author} {\bibfnamefont {P.~L.}\ \bibnamefont
  {Krapivsky}}, \bibinfo {author} {\bibfnamefont {K.}~\bibnamefont {Mallick}},\
  and\ \bibinfo {author} {\bibfnamefont {D.}~\bibnamefont {Sels}},\ }\bibfield
  {title} {\bibinfo {title} {Free fermions with a localized source},\ }\href
  {https://doi.org/https://doi.org/10.1088/1742-5468/ab4e8e} {\bibfield
  {journal} {\bibinfo  {journal} {J. Stat. Mech.}\ ,\ \bibinfo {pages}
  {113108}} (\bibinfo {year} {2019})}\BibitemShut {NoStop}%
\bibitem [{\citenamefont {Dolgirev}\ \emph {et~al.}(2020)\citenamefont
  {Dolgirev}, \citenamefont {Marino}, \citenamefont {Sels},\ and\ \citenamefont
  {Demler}}]{Dolgirev}%
  \BibitemOpen
  \bibfield  {author} {\bibinfo {author} {\bibfnamefont {P.~E.}\ \bibnamefont
  {Dolgirev}}, \bibinfo {author} {\bibfnamefont {J.}~\bibnamefont {Marino}},
  \bibinfo {author} {\bibfnamefont {D.}~\bibnamefont {Sels}},\ and\ \bibinfo
  {author} {\bibfnamefont {E.}~\bibnamefont {Demler}},\ }\bibfield  {title}
  {\bibinfo {title} {Non-{Gaussian} correlations imprinted by local dephasing
  in fermionic wires},\ }\href {https://doi.org/10.1103/PhysRevB.102.100301}
  {\bibfield  {journal} {\bibinfo  {journal} {Phys. Rev. B}\ }\textbf {\bibinfo
  {volume} {102}},\ \bibinfo {pages} {100301(R)} (\bibinfo {year}
  {2020})}\BibitemShut {NoStop}%
\bibitem [{\citenamefont {Takasu}\ \emph {et~al.}(2020)\citenamefont {Takasu},
  \citenamefont {Yagami}, \citenamefont {Ashida}, \citenamefont {Hamazaki},
  \citenamefont {Kuno},\ and\ \citenamefont {Takahashi}}]{response_PTEP}%
  \BibitemOpen
  \bibfield  {author} {\bibinfo {author} {\bibfnamefont {Y.}~\bibnamefont
  {Takasu}}, \bibinfo {author} {\bibfnamefont {T.}~\bibnamefont {Yagami}},
  \bibinfo {author} {\bibfnamefont {Y.}~\bibnamefont {Ashida}}, \bibinfo
  {author} {\bibfnamefont {R.}~\bibnamefont {Hamazaki}}, \bibinfo {author}
  {\bibfnamefont {Y.}~\bibnamefont {Kuno}},\ and\ \bibinfo {author}
  {\bibfnamefont {Y.}~\bibnamefont {Takahashi}},\ }\bibfield  {title} {\bibinfo
  {title} {{PT-symmetric non-Hermitian quantum many-body system using ultracold
  atoms in an optical lattice with controlled dissipation}},\ }\href
  {https://doi.org/10.1093/ptep/ptaa094} {\bibfield  {journal} {\bibinfo
  {journal} {Progress of Theoretical and Experimental Physics}\ }\textbf
  {\bibinfo {volume} {2020}},\ \bibinfo {pages} {12A110} (\bibinfo {year}
  {2020})}\BibitemShut {NoStop}%
\bibitem [{\citenamefont {Krinner}\ \emph {et~al.}(2017)\citenamefont
  {Krinner}, \citenamefont {Esslinger},\ and\ \citenamefont
  {Brantut}}]{Esslinger_JPhys}%
  \BibitemOpen
  \bibfield  {author} {\bibinfo {author} {\bibfnamefont {S.}~\bibnamefont
  {Krinner}}, \bibinfo {author} {\bibfnamefont {T.}~\bibnamefont {Esslinger}},\
  and\ \bibinfo {author} {\bibfnamefont {J.-P.}\ \bibnamefont {Brantut}},\
  }\bibfield  {title} {\bibinfo {title} {Two-terminal transport measurements
  with cold atoms},\ }\href {https://doi.org/10.1088/1361-648X/aa74a1}
  {\bibfield  {journal} {\bibinfo  {journal} {Journal of Physics: Condensed
  Matter}\ }\textbf {\bibinfo {volume} {29}},\ \bibinfo {pages} {343003}
  (\bibinfo {year} {2017})}\BibitemShut {NoStop}%
\bibitem [{\citenamefont {Lebrat}\ \emph {et~al.}(2019)\citenamefont {Lebrat},
  \citenamefont {H\"ausler}, \citenamefont {Fabritius}, \citenamefont
  {Husmann}, \citenamefont {Corman},\ and\ \citenamefont
  {Esslinger}}]{Esslinger_PRL}%
  \BibitemOpen
  \bibfield  {author} {\bibinfo {author} {\bibfnamefont {M.}~\bibnamefont
  {Lebrat}}, \bibinfo {author} {\bibfnamefont {S.}~\bibnamefont {H\"ausler}},
  \bibinfo {author} {\bibfnamefont {P.}~\bibnamefont {Fabritius}}, \bibinfo
  {author} {\bibfnamefont {D.}~\bibnamefont {Husmann}}, \bibinfo {author}
  {\bibfnamefont {L.}~\bibnamefont {Corman}},\ and\ \bibinfo {author}
  {\bibfnamefont {T.}~\bibnamefont {Esslinger}},\ }\bibfield  {title} {\bibinfo
  {title} {Quantized conductance through a spin-selective atomic point
  contact},\ }\href {https://doi.org/10.1103/PhysRevLett.123.193605} {\bibfield
   {journal} {\bibinfo  {journal} {Phys. Rev. Lett.}\ }\textbf {\bibinfo
  {volume} {123}},\ \bibinfo {pages} {193605} (\bibinfo {year}
  {2019})}\BibitemShut {NoStop}%
\bibitem [{\citenamefont {Huang}\ \emph {et~al.}(2023)\citenamefont {Huang},
  \citenamefont {Mohan}, \citenamefont {Visuri}, \citenamefont {Fabritius},
  \citenamefont {Talebi}, \citenamefont {Wili}, \citenamefont {Uchino},
  \citenamefont {Giamarchi},\ and\ \citenamefont
  {Esslinger}}]{Esslinger_PRL_23}%
  \BibitemOpen
  \bibfield  {author} {\bibinfo {author} {\bibfnamefont {M.-Z.}\ \bibnamefont
  {Huang}}, \bibinfo {author} {\bibfnamefont {J.}~\bibnamefont {Mohan}},
  \bibinfo {author} {\bibfnamefont {A.-M.}\ \bibnamefont {Visuri}}, \bibinfo
  {author} {\bibfnamefont {P.}~\bibnamefont {Fabritius}}, \bibinfo {author}
  {\bibfnamefont {M.}~\bibnamefont {Talebi}}, \bibinfo {author} {\bibfnamefont
  {S.}~\bibnamefont {Wili}}, \bibinfo {author} {\bibfnamefont {S.}~\bibnamefont
  {Uchino}}, \bibinfo {author} {\bibfnamefont {T.}~\bibnamefont {Giamarchi}},\
  and\ \bibinfo {author} {\bibfnamefont {T.}~\bibnamefont {Esslinger}},\
  }\bibfield  {title} {\bibinfo {title} {Superfluid signatures in a dissipative
  quantum point contact},\ }\href
  {https://doi.org/10.1103/PhysRevLett.130.200404} {\bibfield  {journal}
  {\bibinfo  {journal} {Phys. Rev. Lett.}\ }\textbf {\bibinfo {volume} {130}},\
  \bibinfo {pages} {200404} (\bibinfo {year} {2023})}\BibitemShut {NoStop}%
\bibitem [{\citenamefont {Lebrat}\ \emph {et~al.}(2018)\citenamefont {Lebrat},
  \citenamefont {Gri\ifmmode~\check{s}\else \v{s}\fi{}ins}, \citenamefont
  {Husmann}, \citenamefont {H\"ausler}, \citenamefont {Corman}, \citenamefont
  {Giamarchi}, \citenamefont {Brantut},\ and\ \citenamefont
  {Esslinger}}]{Esslinger_PRX}%
  \BibitemOpen
  \bibfield  {author} {\bibinfo {author} {\bibfnamefont {M.}~\bibnamefont
  {Lebrat}}, \bibinfo {author} {\bibfnamefont {P.}~\bibnamefont
  {Gri\ifmmode~\check{s}\else \v{s}\fi{}ins}}, \bibinfo {author} {\bibfnamefont
  {D.}~\bibnamefont {Husmann}}, \bibinfo {author} {\bibfnamefont
  {S.}~\bibnamefont {H\"ausler}}, \bibinfo {author} {\bibfnamefont
  {L.}~\bibnamefont {Corman}}, \bibinfo {author} {\bibfnamefont
  {T.}~\bibnamefont {Giamarchi}}, \bibinfo {author} {\bibfnamefont {J.-P.}\
  \bibnamefont {Brantut}},\ and\ \bibinfo {author} {\bibfnamefont
  {T.}~\bibnamefont {Esslinger}},\ }\bibfield  {title} {\bibinfo {title} {Band
  and correlated insulators of cold fermions in a mesoscopic lattice},\ }\href
  {https://doi.org/10.1103/PhysRevX.8.011053} {\bibfield  {journal} {\bibinfo
  {journal} {Phys. Rev. X}\ }\textbf {\bibinfo {volume} {8}},\ \bibinfo {pages}
  {011053} (\bibinfo {year} {2018})}\BibitemShut {NoStop}%
\bibitem [{\citenamefont {Corman}\ \emph {et~al.}(2019)\citenamefont {Corman},
  \citenamefont {Fabritius}, \citenamefont {H\"ausler}, \citenamefont {Mohan},
  \citenamefont {Dogra}, \citenamefont {Husmann}, \citenamefont {Lebrat},\ and\
  \citenamefont {Esslinger}}]{Esslinger_PRA_theory}%
  \BibitemOpen
  \bibfield  {author} {\bibinfo {author} {\bibfnamefont {L.}~\bibnamefont
  {Corman}}, \bibinfo {author} {\bibfnamefont {P.}~\bibnamefont {Fabritius}},
  \bibinfo {author} {\bibfnamefont {S.}~\bibnamefont {H\"ausler}}, \bibinfo
  {author} {\bibfnamefont {J.}~\bibnamefont {Mohan}}, \bibinfo {author}
  {\bibfnamefont {L.~H.}\ \bibnamefont {Dogra}}, \bibinfo {author}
  {\bibfnamefont {D.}~\bibnamefont {Husmann}}, \bibinfo {author} {\bibfnamefont
  {M.}~\bibnamefont {Lebrat}},\ and\ \bibinfo {author} {\bibfnamefont
  {T.}~\bibnamefont {Esslinger}},\ }\bibfield  {title} {\bibinfo {title}
  {Quantized conductance through a dissipative atomic point contact},\ }\href
  {https://doi.org/10.1103/PhysRevA.100.053605} {\bibfield  {journal} {\bibinfo
   {journal} {Phys. Rev. A}\ }\textbf {\bibinfo {volume} {100}},\ \bibinfo
  {pages} {053605} (\bibinfo {year} {2019})}\BibitemShut {NoStop}%
\bibitem [{\citenamefont {Yan}\ \emph {et~al.}(2013)\citenamefont {Yan},
  \citenamefont {Moses}, \citenamefont {Gadway}, \citenamefont {Covey},
  \citenamefont {Hazzard}, \citenamefont {Rey}, \citenamefont {Jin},\ and\
  \citenamefont {Ye}}]{DipolarMolecules2BLoss}%
  \BibitemOpen
  \bibfield  {author} {\bibinfo {author} {\bibfnamefont {B.}~\bibnamefont
  {Yan}}, \bibinfo {author} {\bibfnamefont {S.~A.}\ \bibnamefont {Moses}},
  \bibinfo {author} {\bibfnamefont {B.}~\bibnamefont {Gadway}}, \bibinfo
  {author} {\bibfnamefont {J.~P.}\ \bibnamefont {Covey}}, \bibinfo {author}
  {\bibfnamefont {K.~R.~A.}\ \bibnamefont {Hazzard}}, \bibinfo {author}
  {\bibfnamefont {A.~M.}\ \bibnamefont {Rey}}, \bibinfo {author} {\bibfnamefont
  {D.~S.}\ \bibnamefont {Jin}},\ and\ \bibinfo {author} {\bibfnamefont
  {J.}~\bibnamefont {Ye}},\ }\bibfield  {title} {\bibinfo {title} {Observation
  of dipolar spin-exchange interactions with lattice-confined polar
  molecules},\ }\href {https://doi.org/10.1038/nature12483} {\bibfield
  {journal} {\bibinfo  {journal} {Nature}\ }\textbf {\bibinfo {volume} {501}},\
  \bibinfo {pages} {521} (\bibinfo {year} {2013})}\BibitemShut {NoStop}%
\bibitem [{\citenamefont {Zhu}\ \emph {et~al.}(2014)\citenamefont {Zhu},
  \citenamefont {Gadway}, \citenamefont {Foss-Feig}, \citenamefont
  {Schachenmayer}, \citenamefont {Wall}, \citenamefont {Hazzard}, \citenamefont
  {Yan}, \citenamefont {Moses}, \citenamefont {Covey}, \citenamefont {Jin},
  \citenamefont {Ye}, \citenamefont {Holland},\ and\ \citenamefont
  {Rey}}]{PRL_2BLoss_Zeno}%
  \BibitemOpen
  \bibfield  {author} {\bibinfo {author} {\bibfnamefont {B.}~\bibnamefont
  {Zhu}}, \bibinfo {author} {\bibfnamefont {B.}~\bibnamefont {Gadway}},
  \bibinfo {author} {\bibfnamefont {M.}~\bibnamefont {Foss-Feig}}, \bibinfo
  {author} {\bibfnamefont {J.}~\bibnamefont {Schachenmayer}}, \bibinfo {author}
  {\bibfnamefont {M.~L.}\ \bibnamefont {Wall}}, \bibinfo {author}
  {\bibfnamefont {K.~R.~A.}\ \bibnamefont {Hazzard}}, \bibinfo {author}
  {\bibfnamefont {B.}~\bibnamefont {Yan}}, \bibinfo {author} {\bibfnamefont
  {S.~A.}\ \bibnamefont {Moses}}, \bibinfo {author} {\bibfnamefont {J.~P.}\
  \bibnamefont {Covey}}, \bibinfo {author} {\bibfnamefont {D.~S.}\ \bibnamefont
  {Jin}}, \bibinfo {author} {\bibfnamefont {J.}~\bibnamefont {Ye}}, \bibinfo
  {author} {\bibfnamefont {M.}~\bibnamefont {Holland}},\ and\ \bibinfo {author}
  {\bibfnamefont {A.~M.}\ \bibnamefont {Rey}},\ }\bibfield  {title} {\bibinfo
  {title} {Suppressing the loss of ultracold molecules via the continuous
  quantum {Zeno} effect},\ }\href
  {https://doi.org/10.1103/PhysRevLett.112.070404} {\bibfield  {journal}
  {\bibinfo  {journal} {Phys. Rev. Lett.}\ }\textbf {\bibinfo {volume} {112}},\
  \bibinfo {pages} {070404} (\bibinfo {year} {2014})}\BibitemShut {NoStop}%
\bibitem [{\citenamefont {Sponselee}\ \emph {et~al.}(2018)\citenamefont
  {Sponselee}, \citenamefont {Freystatzky}, \citenamefont {Abeln},
  \citenamefont {Diem}, \citenamefont {Hundt}, \citenamefont {Kochanke},
  \citenamefont {Ponath}, \citenamefont {Santra}, \citenamefont {Mathey},
  \citenamefont {Sengstock},\ and\ \citenamefont
  {Becker}}]{DissipativeFermiHubbard}%
  \BibitemOpen
  \bibfield  {author} {\bibinfo {author} {\bibfnamefont {K.}~\bibnamefont
  {Sponselee}}, \bibinfo {author} {\bibfnamefont {L.}~\bibnamefont
  {Freystatzky}}, \bibinfo {author} {\bibfnamefont {B.}~\bibnamefont {Abeln}},
  \bibinfo {author} {\bibfnamefont {M.}~\bibnamefont {Diem}}, \bibinfo {author}
  {\bibfnamefont {B.}~\bibnamefont {Hundt}}, \bibinfo {author} {\bibfnamefont
  {A.}~\bibnamefont {Kochanke}}, \bibinfo {author} {\bibfnamefont
  {T.}~\bibnamefont {Ponath}}, \bibinfo {author} {\bibfnamefont
  {B.}~\bibnamefont {Santra}}, \bibinfo {author} {\bibfnamefont
  {L.}~\bibnamefont {Mathey}}, \bibinfo {author} {\bibfnamefont
  {K.}~\bibnamefont {Sengstock}},\ and\ \bibinfo {author} {\bibfnamefont
  {C.}~\bibnamefont {Becker}},\ }\bibfield  {title} {\bibinfo {title} {Dynamics
  of ultracold quantum gases in the dissipative {Fermi}--{Hubbard} model},\
  }\href {https://doi.org/10.1088/2058-9565/aadccd} {\bibfield  {journal}
  {\bibinfo  {journal} {Quantum Sci. Technol.}\ }\textbf {\bibinfo {volume}
  {4}},\ \bibinfo {pages} {014002} (\bibinfo {year} {2018})}\BibitemShut
  {NoStop}%
\bibitem [{\citenamefont {Honda}\ \emph {et~al.}(2023)\citenamefont {Honda},
  \citenamefont {Taie}, \citenamefont {Takasu}, \citenamefont {Nishizawa},
  \citenamefont {Nakagawa},\ and\ \citenamefont
  {Takahashi}}]{response_sign_reversal}%
  \BibitemOpen
  \bibfield  {author} {\bibinfo {author} {\bibfnamefont {K.}~\bibnamefont
  {Honda}}, \bibinfo {author} {\bibfnamefont {S.}~\bibnamefont {Taie}},
  \bibinfo {author} {\bibfnamefont {Y.}~\bibnamefont {Takasu}}, \bibinfo
  {author} {\bibfnamefont {N.}~\bibnamefont {Nishizawa}}, \bibinfo {author}
  {\bibfnamefont {M.}~\bibnamefont {Nakagawa}},\ and\ \bibinfo {author}
  {\bibfnamefont {Y.}~\bibnamefont {Takahashi}},\ }\bibfield  {title} {\bibinfo
  {title} {Observation of the sign reversal of the magnetic correlation in a
  driven-dissipative fermi gas in double wells},\ }\href
  {https://doi.org/10.1103/PhysRevLett.130.063001} {\bibfield  {journal}
  {\bibinfo  {journal} {Phys. Rev. Lett.}\ }\textbf {\bibinfo {volume} {130}},\
  \bibinfo {pages} {063001} (\bibinfo {year} {2023})}\BibitemShut {NoStop}%
\bibitem [{\citenamefont {Tomita}\ \emph {et~al.}(2019)\citenamefont {Tomita},
  \citenamefont {Nakajima}, \citenamefont {Takasu},\ and\ \citenamefont
  {Takahashi}}]{Bose_Hubb_2b_loss}%
  \BibitemOpen
  \bibfield  {author} {\bibinfo {author} {\bibfnamefont {T.}~\bibnamefont
  {Tomita}}, \bibinfo {author} {\bibfnamefont {S.}~\bibnamefont {Nakajima}},
  \bibinfo {author} {\bibfnamefont {Y.}~\bibnamefont {Takasu}},\ and\ \bibinfo
  {author} {\bibfnamefont {Y.}~\bibnamefont {Takahashi}},\ }\bibfield  {title}
  {\bibinfo {title} {{Dissipative Bose-Hubbard system with intrinsic two-body
  loss}},\ }\href {https://doi.org/10.1103/PhysRevA.99.031601} {\bibfield
  {journal} {\bibinfo  {journal} {Phys. Rev. A}\ }\textbf {\bibinfo {volume}
  {99}},\ \bibinfo {pages} {031601} (\bibinfo {year} {2019})}\BibitemShut
  {NoStop}%
\bibitem [{\citenamefont {Syassen}\ \emph {et~al.}(2008)\citenamefont
  {Syassen}, \citenamefont {Bauer}, \citenamefont {Lettner}, \citenamefont
  {Volz}, \citenamefont {Dietze}, \citenamefont {Garc{\'\i}a-Ripoll},
  \citenamefont {Cirac}, \citenamefont {Rempe},\ and\ \citenamefont
  {D{\"u}rr}}]{Science_2b_loss}%
  \BibitemOpen
  \bibfield  {author} {\bibinfo {author} {\bibfnamefont {N.}~\bibnamefont
  {Syassen}}, \bibinfo {author} {\bibfnamefont {D.~M.}\ \bibnamefont {Bauer}},
  \bibinfo {author} {\bibfnamefont {M.}~\bibnamefont {Lettner}}, \bibinfo
  {author} {\bibfnamefont {T.}~\bibnamefont {Volz}}, \bibinfo {author}
  {\bibfnamefont {D.}~\bibnamefont {Dietze}}, \bibinfo {author} {\bibfnamefont
  {J.~J.}\ \bibnamefont {Garc{\'\i}a-Ripoll}}, \bibinfo {author} {\bibfnamefont
  {J.~I.}\ \bibnamefont {Cirac}}, \bibinfo {author} {\bibfnamefont
  {G.}~\bibnamefont {Rempe}},\ and\ \bibinfo {author} {\bibfnamefont
  {S.}~\bibnamefont {D{\"u}rr}},\ }\bibfield  {title} {\bibinfo {title}
  {{Strong Dissipation Inhibits Losses and Induces Correlations in Cold
  Molecular Gases}},\ }\href {https://doi.org/10.1126/science.1155309}
  {\bibfield  {journal} {\bibinfo  {journal} {Science}\ }\textbf {\bibinfo
  {volume} {320}},\ \bibinfo {pages} {1329} (\bibinfo {year}
  {2008})}\BibitemShut {NoStop}%
\bibitem [{\citenamefont {Cetina}\ \emph {et~al.}(2016)\citenamefont {Cetina},
  \citenamefont {Jag}, \citenamefont {Lous}, \citenamefont {Fritsche},
  \citenamefont {Walraven}, \citenamefont {Grimm}, \citenamefont {Levinsen},
  \citenamefont {Parish}, \citenamefont {Schmidt}, \citenamefont {Knap},\ and\
  \citenamefont {Demler}}]{Science_Ramsey_interferometry}%
  \BibitemOpen
  \bibfield  {author} {\bibinfo {author} {\bibfnamefont {M.}~\bibnamefont
  {Cetina}}, \bibinfo {author} {\bibfnamefont {M.}~\bibnamefont {Jag}},
  \bibinfo {author} {\bibfnamefont {R.~S.}\ \bibnamefont {Lous}}, \bibinfo
  {author} {\bibfnamefont {I.}~\bibnamefont {Fritsche}}, \bibinfo {author}
  {\bibfnamefont {J.~T.~M.}\ \bibnamefont {Walraven}}, \bibinfo {author}
  {\bibfnamefont {R.}~\bibnamefont {Grimm}}, \bibinfo {author} {\bibfnamefont
  {J.}~\bibnamefont {Levinsen}}, \bibinfo {author} {\bibfnamefont {M.~M.}\
  \bibnamefont {Parish}}, \bibinfo {author} {\bibfnamefont {R.}~\bibnamefont
  {Schmidt}}, \bibinfo {author} {\bibfnamefont {M.}~\bibnamefont {Knap}},\ and\
  \bibinfo {author} {\bibfnamefont {E.}~\bibnamefont {Demler}},\ }\bibfield
  {title} {\bibinfo {title} {Ultrafast many-body interferometry of impurities
  coupled to a fermi sea},\ }\href {https://doi.org/10.1126/science.aaf5134}
  {\bibfield  {journal} {\bibinfo  {journal} {Science}\ }\textbf {\bibinfo
  {volume} {354}},\ \bibinfo {pages} {96} (\bibinfo {year} {2016})}\BibitemShut
  {NoStop}%
\bibitem [{\citenamefont {Cao}\ \emph {et~al.}(2019)\citenamefont {Cao},
  \citenamefont {Tilloy},\ and\ \citenamefont {Luca}}]{CaoTilloyDeLuca}%
  \BibitemOpen
  \bibfield  {author} {\bibinfo {author} {\bibfnamefont {X.}~\bibnamefont
  {Cao}}, \bibinfo {author} {\bibfnamefont {A.}~\bibnamefont {Tilloy}},\ and\
  \bibinfo {author} {\bibfnamefont {A.~D.}\ \bibnamefont {Luca}},\ }\bibfield
  {title} {\bibinfo {title} {Entanglement in a fermion chain under continuous
  monitoring},\ }\href {https://doi.org/10.21468/SciPostPhys.7.2.024}
  {\bibfield  {journal} {\bibinfo  {journal} {SciPost Phys.}\ }\textbf
  {\bibinfo {volume} {7}},\ \bibinfo {pages} {024} (\bibinfo {year}
  {2019})}\BibitemShut {NoStop}%
\bibitem [{Note2()}]{Note2}%
  \BibitemOpen
  \bibinfo {note} {The disagreement of the fitted exponent with respect to
  $\beta _b$ at larger $\gamma $ can be attributed to the local curvature of
  the dispersion around the Fermi energy, just as in the Hermitian case \cite
  {Giamarchi,GogolinNersesyanTsvelik}.}\BibitemShut {Stop}%
\bibitem [{Note3()}]{Note3}%
  \BibitemOpen
  \bibinfo {note} {Unless the density is too low or too high, so that the Fermi
  momentum crosses the “special momenta” $q^\ast $ reported in the
  following Section. There are also qualitative changes in the non-perturbative
  regime $\gamma >4J$ caused by the crossing of an exceptional point \cite
  {ThesisFroml,Haque}, not treated here.}\BibitemShut {Stop}%
\bibitem [{\citenamefont {Kane}\ \emph {et~al.}(1994)\citenamefont {Kane},
  \citenamefont {Matveev},\ and\ \citenamefont
  {Glazman}}]{FermiEdgeKaneMatveevGlazman}%
  \BibitemOpen
  \bibfield  {author} {\bibinfo {author} {\bibfnamefont {C.~L.}\ \bibnamefont
  {Kane}}, \bibinfo {author} {\bibfnamefont {K.~A.}\ \bibnamefont {Matveev}},\
  and\ \bibinfo {author} {\bibfnamefont {L.~I.}\ \bibnamefont {Glazman}},\
  }\bibfield  {title} {\bibinfo {title} {Fermi-edge singularities and
  backscattering in a weakly interacting one-dimensional electron gas},\ }\href
  {https://doi.org/10.1103/PhysRevB.49.2253} {\bibfield  {journal} {\bibinfo
  {journal} {Phys. Rev. B}\ }\textbf {\bibinfo {volume} {49}},\ \bibinfo
  {pages} {2253} (\bibinfo {year} {1994})}\BibitemShut {NoStop}%
\bibitem [{Note4()}]{Note4}%
  \BibitemOpen
  \bibinfo {note} {Indeed, it differs from a $\protect \mathcal {P}\protect
  \mathcal {T}$-symmetric \cite {ReviewNonHermitianUeda} Hamiltonian by an
  imaginary constant only, which is unobservable in the nonlinear dynamics
  Eq.~\protect \textup {\hbox {\mathsurround \z@ \protect \normalfont
  (\ignorespaces \ref {eq: nlse}\unskip \@@italiccorr )}}.}\BibitemShut {Stop}%
\bibitem [{\citenamefont {Zagoskin}\ and\ \citenamefont
  {Affleck}(1997)}]{Zagoskin_1997}%
  \BibitemOpen
  \bibfield  {author} {\bibinfo {author} {\bibfnamefont {A.~M.}\ \bibnamefont
  {Zagoskin}}\ and\ \bibinfo {author} {\bibfnamefont {I.}~\bibnamefont
  {Affleck}},\ }\bibfield  {title} {\bibinfo {title} {Fermi edge singularities:
  Bound states and finite-size effects},\ }\href
  {https://doi.org/10.1088/0305-4470/30/16/017} {\bibfield  {journal} {\bibinfo
   {journal} {J. Phys. A: Math. Gen.}\ }\textbf {\bibinfo {volume} {30}},\
  \bibinfo {pages} {5743} (\bibinfo {year} {1997})}\BibitemShut {NoStop}%
\bibitem [{\citenamefont {Xie}\ \emph {et~al.}(2024)\citenamefont {Xie},
  \citenamefont {Liang}, \citenamefont {Ma}, \citenamefont {Du}, \citenamefont
  {Peng}, \citenamefont {Li}, \citenamefont {Chen}, \citenamefont {Li},
  \citenamefont {Gao},\ and\ \citenamefont {Xue}}]{scaleFreeBoundStates}%
  \BibitemOpen
  \bibfield  {author} {\bibinfo {author} {\bibfnamefont {X.}~\bibnamefont
  {Xie}}, \bibinfo {author} {\bibfnamefont {G.}~\bibnamefont {Liang}}, \bibinfo
  {author} {\bibfnamefont {F.}~\bibnamefont {Ma}}, \bibinfo {author}
  {\bibfnamefont {Y.}~\bibnamefont {Du}}, \bibinfo {author} {\bibfnamefont
  {Y.}~\bibnamefont {Peng}}, \bibinfo {author} {\bibfnamefont {E.}~\bibnamefont
  {Li}}, \bibinfo {author} {\bibfnamefont {H.}~\bibnamefont {Chen}}, \bibinfo
  {author} {\bibfnamefont {L.}~\bibnamefont {Li}}, \bibinfo {author}
  {\bibfnamefont {F.}~\bibnamefont {Gao}},\ and\ \bibinfo {author}
  {\bibfnamefont {H.}~\bibnamefont {Xue}},\ }\href@noop {} {\bibinfo {title}
  {Observation of scale-free localized states induced by non-{Hermitian}
  defects}} (\bibinfo {year} {2024}),\ \Eprint
  {https://arxiv.org/abs/2402.04716} {arXiv:2402.04716 [cond-mat.mes-hall]}
  \BibitemShut {NoStop}%
\bibitem [{\citenamefont {Ashida}\ \emph {et~al.}(2020)\citenamefont {Ashida},
  \citenamefont {Gong},\ and\ \citenamefont {Ueda}}]{ReviewNonHermitianUeda}%
  \BibitemOpen
  \bibfield  {author} {\bibinfo {author} {\bibfnamefont {Y.}~\bibnamefont
  {Ashida}}, \bibinfo {author} {\bibfnamefont {Z.}~\bibnamefont {Gong}},\ and\
  \bibinfo {author} {\bibfnamefont {M.}~\bibnamefont {Ueda}},\ }\bibfield
  {title} {\bibinfo {title} {Non-{Hermitian} physics},\ }\href
  {https://doi.org/10.1080/00018732.2021.1876991} {\bibfield  {journal}
  {\bibinfo  {journal} {Advances in Physics}\ }\textbf {\bibinfo {volume}
  {69}},\ \bibinfo {pages} {249} (\bibinfo {year} {2020})}\BibitemShut
  {NoStop}%
\end{thebibliography}%


\begin{thebibliography}{41}%
\makeatletter
\providecommand \@ifxundefined [1]{%
 \@ifx{#1\undefined}
}%
\providecommand \@ifnum [1]{%
 \ifnum #1\expandafter \@firstoftwo
 \else \expandafter \@secondoftwo
 \fi
}%
\providecommand \@ifx [1]{%
 \ifx #1\expandafter \@firstoftwo
 \else \expandafter \@secondoftwo
 \fi
}%
\providecommand \natexlab [1]{#1}%
\providecommand \enquote  [1]{``#1''}%
\providecommand \bibnamefont  [1]{#1}%
\providecommand \bibfnamefont [1]{#1}%
\providecommand \citenamefont [1]{#1}%
\providecommand \href@noop [0]{\@secondoftwo}%
\providecommand \href [0]{\begingroup \@sanitize@url \@href}%
\providecommand \@href[1]{\@@startlink{#1}\@@href}%
\providecommand \@@href[1]{\endgroup#1\@@endlink}%
\providecommand \@sanitize@url [0]{\catcode `\\12\catcode `\$12\catcode
  `\&12\catcode `\#12\catcode `\^12\catcode `\_12\catcode `\%12\relax}%
\providecommand \@@startlink[1]{}%
\providecommand \@@endlink[0]{}%
\providecommand \url  [0]{\begingroup\@sanitize@url \@url }%
\providecommand \@url [1]{\endgroup\@href {#1}{\urlprefix }}%
\providecommand \urlprefix  [0]{URL }%
\providecommand \Eprint [0]{\href }%
\providecommand \doibase [0]{https://doi.org/}%
\providecommand \selectlanguage [0]{\@gobble}%
\providecommand \bibinfo  [0]{\@secondoftwo}%
\providecommand \bibfield  [0]{\@secondoftwo}%
\providecommand \translation [1]{[#1]}%
\providecommand \BibitemOpen [0]{}%
\providecommand \bibitemStop [0]{}%
\providecommand \bibitemNoStop [0]{.\EOS\space}%
\providecommand \EOS [0]{\spacefactor3000\relax}%
\providecommand \BibitemShut  [1]{\csname bibitem#1\endcsname}%
\let\auto@bib@innerbib\@empty
\bibitem [{\citenamefont {Wiseman}\ and\ \citenamefont
  {Milburn}(2009)}]{WisemanMilburn}%
  \BibitemOpen
  \bibfield  {author} {\bibinfo {author} {\bibfnamefont {H.~M.}\ \bibnamefont
  {Wiseman}}\ and\ \bibinfo {author} {\bibfnamefont {G.~J.}\ \bibnamefont
  {Milburn}},\ }\href@noop {} {\emph {\bibinfo {title} {Quantum Measurement and
  Control}}}\ (\bibinfo  {publisher} {Cambridge University Press},\ \bibinfo
  {address} {Cambridge},\ \bibinfo {year} {2009})\BibitemShut {NoStop}%
\bibitem [{\citenamefont {Garratt}\ \emph {et~al.}(2023)\citenamefont
  {Garratt}, \citenamefont {Weinstein},\ and\ \citenamefont
  {Altman}}]{AltmanMeasurementsTLL}%
  \BibitemOpen
  \bibfield  {author} {\bibinfo {author} {\bibfnamefont {S.~J.}\ \bibnamefont
  {Garratt}}, \bibinfo {author} {\bibfnamefont {Z.}~\bibnamefont {Weinstein}},\
  and\ \bibinfo {author} {\bibfnamefont {E.}~\bibnamefont {Altman}},\
  }\bibfield  {title} {\bibinfo {title} {Measurements conspire nonlocally to
  restructure critical quantum states},\ }\href
  {https://doi.org/10.1103/PhysRevX.13.021026} {\bibfield  {journal} {\bibinfo
  {journal} {Phys. Rev. X}\ }\textbf {\bibinfo {volume} {13}},\ \bibinfo
  {pages} {021026} (\bibinfo {year} {2023})}\BibitemShut {NoStop}%
\bibitem [{\citenamefont {Turkeshi}\ and\ \citenamefont
  {Schir\'o}(2023)}]{SchiroNonHermPRB}%
  \BibitemOpen
  \bibfield  {author} {\bibinfo {author} {\bibfnamefont {X.}~\bibnamefont
  {Turkeshi}}\ and\ \bibinfo {author} {\bibfnamefont {M.}~\bibnamefont
  {Schir\'o}},\ }\bibfield  {title} {\bibinfo {title} {Entanglement and
  correlation spreading in non-{Hermitian} spin chains},\ }\href
  {https://doi.org/10.1103/PhysRevB.107.L020403} {\bibfield  {journal}
  {\bibinfo  {journal} {Phys. Rev. B}\ }\textbf {\bibinfo {volume} {107}},\
  \bibinfo {pages} {L020403} (\bibinfo {year} {2023})}\BibitemShut {NoStop}%
\bibitem [{\citenamefont {{Le Gal}}\ \emph {et~al.}(2023)\citenamefont {{Le
  Gal}}, \citenamefont {Turkeshi},\ and\ \citenamefont
  {Schir{\`o}}}]{SchiroNonHermSciPost}%
  \BibitemOpen
  \bibfield  {author} {\bibinfo {author} {\bibfnamefont {Y.}~\bibnamefont {{Le
  Gal}}}, \bibinfo {author} {\bibfnamefont {X.}~\bibnamefont {Turkeshi}},\ and\
  \bibinfo {author} {\bibfnamefont {M.}~\bibnamefont {Schir{\`o}}},\ }\bibfield
   {title} {\bibinfo {title} {{Volume-to-area law entanglement transition in a
  non-{Hermitian} free fermionic chain}},\ }\href
  {https://doi.org/10.21468/SciPostPhys.14.5.138} {\bibfield  {journal}
  {\bibinfo  {journal} {SciPost Phys.}\ }\textbf {\bibinfo {volume} {14}},\
  \bibinfo {pages} {138} (\bibinfo {year} {2023})}\BibitemShut {NoStop}%
\bibitem [{\citenamefont {Turkeshi}\ \emph {et~al.}(2021)\citenamefont
  {Turkeshi}, \citenamefont {Biella}, \citenamefont {Fazio}, \citenamefont
  {Dalmonte},\ and\ \citenamefont {Schir\'o}}]{TurkeshiPhysRevB.103.224210}%
  \BibitemOpen
  \bibfield  {author} {\bibinfo {author} {\bibfnamefont {X.}~\bibnamefont
  {Turkeshi}}, \bibinfo {author} {\bibfnamefont {A.}~\bibnamefont {Biella}},
  \bibinfo {author} {\bibfnamefont {R.}~\bibnamefont {Fazio}}, \bibinfo
  {author} {\bibfnamefont {M.}~\bibnamefont {Dalmonte}},\ and\ \bibinfo
  {author} {\bibfnamefont {M.}~\bibnamefont {Schir\'o}},\ }\bibfield  {title}
  {\bibinfo {title} {Measurement-induced entanglement transitions in the
  quantum {Ising} chain: From infinite to zero clicks},\ }\href
  {https://doi.org/10.1103/PhysRevB.103.224210} {\bibfield  {journal} {\bibinfo
   {journal} {Phys. Rev. B}\ }\textbf {\bibinfo {volume} {103}},\ \bibinfo
  {pages} {224210} (\bibinfo {year} {2021})}\BibitemShut {NoStop}%
\bibitem [{\citenamefont {Ashida}\ \emph {et~al.}(2020)\citenamefont {Ashida},
  \citenamefont {Gong},\ and\ \citenamefont {Ueda}}]{ReviewNonHermitianUeda}%
  \BibitemOpen
  \bibfield  {author} {\bibinfo {author} {\bibfnamefont {Y.}~\bibnamefont
  {Ashida}}, \bibinfo {author} {\bibfnamefont {Z.}~\bibnamefont {Gong}},\ and\
  \bibinfo {author} {\bibfnamefont {M.}~\bibnamefont {Ueda}},\ }\bibfield
  {title} {\bibinfo {title} {Non-{Hermitian} physics},\ }\href
  {https://doi.org/10.1080/00018732.2021.1876991} {\bibfield  {journal}
  {\bibinfo  {journal} {Advances in Physics}\ }\textbf {\bibinfo {volume}
  {69}},\ \bibinfo {pages} {249} (\bibinfo {year} {2020})}\BibitemShut
  {NoStop}%
\bibitem [{\citenamefont {Fr\"oml}(2020)}]{ThesisFroml}%
  \BibitemOpen
  \bibfield  {author} {\bibinfo {author} {\bibfnamefont {H.~F.}\ \bibnamefont
  {Fr\"oml}},\ }\emph {\bibinfo {title} {Localized dissipation in fermionic
  quantum wires}},\ \href@noop {} {Ph.D. thesis},\ \bibinfo  {school}
  {Universit\"at zu K\"oln} (\bibinfo {year} {2020})\BibitemShut {NoStop}%
\bibitem [{\citenamefont {Fr\"oml}\ \emph {et~al.}(2019)\citenamefont
  {Fr\"oml}, \citenamefont {Chiocchetta}, \citenamefont {Kollath},\ and\
  \citenamefont {Diehl}}]{FromlPRL}%
  \BibitemOpen
  \bibfield  {author} {\bibinfo {author} {\bibfnamefont {H.}~\bibnamefont
  {Fr\"oml}}, \bibinfo {author} {\bibfnamefont {A.}~\bibnamefont
  {Chiocchetta}}, \bibinfo {author} {\bibfnamefont {C.}~\bibnamefont
  {Kollath}},\ and\ \bibinfo {author} {\bibfnamefont {S.}~\bibnamefont
  {Diehl}},\ }\bibfield  {title} {\bibinfo {title} {Fluctuation-induced quantum
  {Zeno} effect},\ }\href {https://doi.org/10.1103/PhysRevLett.122.040402}
  {\bibfield  {journal} {\bibinfo  {journal} {Phys. Rev. Lett.}\ }\textbf
  {\bibinfo {volume} {122}},\ \bibinfo {pages} {040402} (\bibinfo {year}
  {2019})}\BibitemShut {NoStop}%
\bibitem [{\citenamefont {Fr\"oml}\ \emph {et~al.}(2020)\citenamefont
  {Fr\"oml}, \citenamefont {Muckel}, \citenamefont {Kollath}, \citenamefont
  {Chiocchetta},\ and\ \citenamefont {Diehl}}]{FromlPRB1}%
  \BibitemOpen
  \bibfield  {author} {\bibinfo {author} {\bibfnamefont {H.}~\bibnamefont
  {Fr\"oml}}, \bibinfo {author} {\bibfnamefont {C.}~\bibnamefont {Muckel}},
  \bibinfo {author} {\bibfnamefont {C.}~\bibnamefont {Kollath}}, \bibinfo
  {author} {\bibfnamefont {A.}~\bibnamefont {Chiocchetta}},\ and\ \bibinfo
  {author} {\bibfnamefont {S.}~\bibnamefont {Diehl}},\ }\bibfield  {title}
  {\bibinfo {title} {Ultracold quantum wires with localized losses: Many-body
  quantum {Zeno} effect},\ }\href {https://doi.org/10.1103/PhysRevB.101.144301}
  {\bibfield  {journal} {\bibinfo  {journal} {Phys. Rev. B}\ }\textbf {\bibinfo
  {volume} {101}},\ \bibinfo {pages} {144301} (\bibinfo {year}
  {2020})}\BibitemShut {NoStop}%
\bibitem [{\citenamefont {M\"uller}\ \emph {et~al.}(2021)\citenamefont
  {M\"uller}, \citenamefont {Gievers}, \citenamefont {Fr\"oml}, \citenamefont
  {Diehl},\ and\ \citenamefont {Chiocchetta}}]{FromlPRB2}%
  \BibitemOpen
  \bibfield  {author} {\bibinfo {author} {\bibfnamefont {T.}~\bibnamefont
  {M\"uller}}, \bibinfo {author} {\bibfnamefont {M.}~\bibnamefont {Gievers}},
  \bibinfo {author} {\bibfnamefont {H.}~\bibnamefont {Fr\"oml}}, \bibinfo
  {author} {\bibfnamefont {S.}~\bibnamefont {Diehl}},\ and\ \bibinfo {author}
  {\bibfnamefont {A.}~\bibnamefont {Chiocchetta}},\ }\bibfield  {title}
  {\bibinfo {title} {Shape effects of localized losses in quantum wires:
  Dissipative resonances and nonequilibrium universality},\ }\href
  {https://doi.org/10.1103/PhysRevB.104.155431} {\bibfield  {journal} {\bibinfo
   {journal} {Phys. Rev. B}\ }\textbf {\bibinfo {volume} {104}},\ \bibinfo
  {pages} {155431} (\bibinfo {year} {2021})}\BibitemShut {NoStop}%
\bibitem [{\citenamefont {Krapivsky}\ \emph {et~al.}(2019)\citenamefont
  {Krapivsky}, \citenamefont {Mallick},\ and\ \citenamefont
  {Sels}}]{LocalizedFermionSource}%
  \BibitemOpen
  \bibfield  {author} {\bibinfo {author} {\bibfnamefont {P.~L.}\ \bibnamefont
  {Krapivsky}}, \bibinfo {author} {\bibfnamefont {K.}~\bibnamefont {Mallick}},\
  and\ \bibinfo {author} {\bibfnamefont {D.}~\bibnamefont {Sels}},\ }\bibfield
  {title} {\bibinfo {title} {Free fermions with a localized source},\ }\href
  {https://doi.org/https://doi.org/10.1088/1742-5468/ab4e8e} {\bibfield
  {journal} {\bibinfo  {journal} {J. Stat. Mech.}\ ,\ \bibinfo {pages}
  {113108}} (\bibinfo {year} {2019})}\BibitemShut {NoStop}%
\bibitem [{\citenamefont {Dolgirev}\ \emph {et~al.}(2020)\citenamefont
  {Dolgirev}, \citenamefont {Marino}, \citenamefont {Sels},\ and\ \citenamefont
  {Demler}}]{Dolgirev}%
  \BibitemOpen
  \bibfield  {author} {\bibinfo {author} {\bibfnamefont {P.~E.}\ \bibnamefont
  {Dolgirev}}, \bibinfo {author} {\bibfnamefont {J.}~\bibnamefont {Marino}},
  \bibinfo {author} {\bibfnamefont {D.}~\bibnamefont {Sels}},\ and\ \bibinfo
  {author} {\bibfnamefont {E.}~\bibnamefont {Demler}},\ }\bibfield  {title}
  {\bibinfo {title} {Non-{Gaussian} correlations imprinted by local dephasing
  in fermionic wires},\ }\href {https://doi.org/10.1103/PhysRevB.102.100301}
  {\bibfield  {journal} {\bibinfo  {journal} {Phys. Rev. B}\ }\textbf {\bibinfo
  {volume} {102}},\ \bibinfo {pages} {100301(R)} (\bibinfo {year}
  {2020})}\BibitemShut {NoStop}%
\bibitem [{\citenamefont {Tonielli}\ \emph {et~al.}(2019)\citenamefont
  {Tonielli}, \citenamefont {Fazio}, \citenamefont {Diehl},\ and\ \citenamefont
  {Marino}}]{Tonielli}%
  \BibitemOpen
  \bibfield  {author} {\bibinfo {author} {\bibfnamefont {F.}~\bibnamefont
  {Tonielli}}, \bibinfo {author} {\bibfnamefont {R.}~\bibnamefont {Fazio}},
  \bibinfo {author} {\bibfnamefont {S.}~\bibnamefont {Diehl}},\ and\ \bibinfo
  {author} {\bibfnamefont {J.}~\bibnamefont {Marino}},\ }\bibfield  {title}
  {\bibinfo {title} {Orthogonality catastrophe in dissipative quantum many-body
  systems},\ }\href {https://doi.org/10.1103/PhysRevLett.122.040604} {\bibfield
   {journal} {\bibinfo  {journal} {Phys. Rev. Lett.}\ }\textbf {\bibinfo
  {volume} {122}},\ \bibinfo {pages} {040604} (\bibinfo {year}
  {2019})}\BibitemShut {NoStop}%
\bibitem [{\citenamefont {Krinner}\ \emph {et~al.}(2017)\citenamefont
  {Krinner}, \citenamefont {Esslinger},\ and\ \citenamefont
  {Brantut}}]{Esslinger_JPhys}%
  \BibitemOpen
  \bibfield  {author} {\bibinfo {author} {\bibfnamefont {S.}~\bibnamefont
  {Krinner}}, \bibinfo {author} {\bibfnamefont {T.}~\bibnamefont {Esslinger}},\
  and\ \bibinfo {author} {\bibfnamefont {J.-P.}\ \bibnamefont {Brantut}},\
  }\bibfield  {title} {\bibinfo {title} {Two-terminal transport measurements
  with cold atoms},\ }\href {https://doi.org/10.1088/1361-648X/aa74a1}
  {\bibfield  {journal} {\bibinfo  {journal} {Journal of Physics: Condensed
  Matter}\ }\textbf {\bibinfo {volume} {29}},\ \bibinfo {pages} {343003}
  (\bibinfo {year} {2017})}\BibitemShut {NoStop}%
\bibitem [{\citenamefont {Lebrat}\ \emph {et~al.}(2019)\citenamefont {Lebrat},
  \citenamefont {H\"ausler}, \citenamefont {Fabritius}, \citenamefont
  {Husmann}, \citenamefont {Corman},\ and\ \citenamefont
  {Esslinger}}]{Esslinger_PRL}%
  \BibitemOpen
  \bibfield  {author} {\bibinfo {author} {\bibfnamefont {M.}~\bibnamefont
  {Lebrat}}, \bibinfo {author} {\bibfnamefont {S.}~\bibnamefont {H\"ausler}},
  \bibinfo {author} {\bibfnamefont {P.}~\bibnamefont {Fabritius}}, \bibinfo
  {author} {\bibfnamefont {D.}~\bibnamefont {Husmann}}, \bibinfo {author}
  {\bibfnamefont {L.}~\bibnamefont {Corman}},\ and\ \bibinfo {author}
  {\bibfnamefont {T.}~\bibnamefont {Esslinger}},\ }\bibfield  {title} {\bibinfo
  {title} {Quantized conductance through a spin-selective atomic point
  contact},\ }\href {https://doi.org/10.1103/PhysRevLett.123.193605} {\bibfield
   {journal} {\bibinfo  {journal} {Phys. Rev. Lett.}\ }\textbf {\bibinfo
  {volume} {123}},\ \bibinfo {pages} {193605} (\bibinfo {year}
  {2019})}\BibitemShut {NoStop}%
\bibitem [{\citenamefont {Huang}\ \emph {et~al.}(2023)\citenamefont {Huang},
  \citenamefont {Mohan}, \citenamefont {Visuri}, \citenamefont {Fabritius},
  \citenamefont {Talebi}, \citenamefont {Wili}, \citenamefont {Uchino},
  \citenamefont {Giamarchi},\ and\ \citenamefont
  {Esslinger}}]{Esslinger_PRL_23}%
  \BibitemOpen
  \bibfield  {author} {\bibinfo {author} {\bibfnamefont {M.-Z.}\ \bibnamefont
  {Huang}}, \bibinfo {author} {\bibfnamefont {J.}~\bibnamefont {Mohan}},
  \bibinfo {author} {\bibfnamefont {A.-M.}\ \bibnamefont {Visuri}}, \bibinfo
  {author} {\bibfnamefont {P.}~\bibnamefont {Fabritius}}, \bibinfo {author}
  {\bibfnamefont {M.}~\bibnamefont {Talebi}}, \bibinfo {author} {\bibfnamefont
  {S.}~\bibnamefont {Wili}}, \bibinfo {author} {\bibfnamefont {S.}~\bibnamefont
  {Uchino}}, \bibinfo {author} {\bibfnamefont {T.}~\bibnamefont {Giamarchi}},\
  and\ \bibinfo {author} {\bibfnamefont {T.}~\bibnamefont {Esslinger}},\
  }\bibfield  {title} {\bibinfo {title} {Superfluid signatures in a dissipative
  quantum point contact},\ }\href
  {https://doi.org/10.1103/PhysRevLett.130.200404} {\bibfield  {journal}
  {\bibinfo  {journal} {Phys. Rev. Lett.}\ }\textbf {\bibinfo {volume} {130}},\
  \bibinfo {pages} {200404} (\bibinfo {year} {2023})}\BibitemShut {NoStop}%
\bibitem [{\citenamefont {Lebrat}\ \emph {et~al.}(2018)\citenamefont {Lebrat},
  \citenamefont {Gri\ifmmode~\check{s}\else \v{s}\fi{}ins}, \citenamefont
  {Husmann}, \citenamefont {H\"ausler}, \citenamefont {Corman}, \citenamefont
  {Giamarchi}, \citenamefont {Brantut},\ and\ \citenamefont
  {Esslinger}}]{Esslinger_PRX}%
  \BibitemOpen
  \bibfield  {author} {\bibinfo {author} {\bibfnamefont {M.}~\bibnamefont
  {Lebrat}}, \bibinfo {author} {\bibfnamefont {P.}~\bibnamefont
  {Gri\ifmmode~\check{s}\else \v{s}\fi{}ins}}, \bibinfo {author} {\bibfnamefont
  {D.}~\bibnamefont {Husmann}}, \bibinfo {author} {\bibfnamefont
  {S.}~\bibnamefont {H\"ausler}}, \bibinfo {author} {\bibfnamefont
  {L.}~\bibnamefont {Corman}}, \bibinfo {author} {\bibfnamefont
  {T.}~\bibnamefont {Giamarchi}}, \bibinfo {author} {\bibfnamefont {J.-P.}\
  \bibnamefont {Brantut}},\ and\ \bibinfo {author} {\bibfnamefont
  {T.}~\bibnamefont {Esslinger}},\ }\bibfield  {title} {\bibinfo {title} {Band
  and correlated insulators of cold fermions in a mesoscopic lattice},\ }\href
  {https://doi.org/10.1103/PhysRevX.8.011053} {\bibfield  {journal} {\bibinfo
  {journal} {Phys. Rev. X}\ }\textbf {\bibinfo {volume} {8}},\ \bibinfo {pages}
  {011053} (\bibinfo {year} {2018})}\BibitemShut {NoStop}%
\bibitem [{\citenamefont {Corman}\ \emph
  {et~al.}(2019{\natexlab{a}})\citenamefont {Corman}, \citenamefont
  {Fabritius}, \citenamefont {H\"ausler}, \citenamefont {Mohan}, \citenamefont
  {Dogra}, \citenamefont {Husmann}, \citenamefont {Lebrat},\ and\ \citenamefont
  {Esslinger}}]{Esslinger_PRA_theory}%
  \BibitemOpen
  \bibfield  {author} {\bibinfo {author} {\bibfnamefont {L.}~\bibnamefont
  {Corman}}, \bibinfo {author} {\bibfnamefont {P.}~\bibnamefont {Fabritius}},
  \bibinfo {author} {\bibfnamefont {S.}~\bibnamefont {H\"ausler}}, \bibinfo
  {author} {\bibfnamefont {J.}~\bibnamefont {Mohan}}, \bibinfo {author}
  {\bibfnamefont {L.~H.}\ \bibnamefont {Dogra}}, \bibinfo {author}
  {\bibfnamefont {D.}~\bibnamefont {Husmann}}, \bibinfo {author} {\bibfnamefont
  {M.}~\bibnamefont {Lebrat}},\ and\ \bibinfo {author} {\bibfnamefont
  {T.}~\bibnamefont {Esslinger}},\ }\bibfield  {title} {\bibinfo {title}
  {Quantized conductance through a dissipative atomic point contact},\ }\href
  {https://doi.org/10.1103/PhysRevA.100.053605} {\bibfield  {journal} {\bibinfo
   {journal} {Phys. Rev. A}\ }\textbf {\bibinfo {volume} {100}},\ \bibinfo
  {pages} {053605} (\bibinfo {year} {2019}{\natexlab{a}})}\BibitemShut
  {NoStop}%
\bibitem [{\citenamefont {Naghiloo}\ \emph {et~al.}(2019)\citenamefont
  {Naghiloo}, \citenamefont {Abbasi}, \citenamefont {Joglekar},\ and\
  \citenamefont {Murch}}]{QubitExceptionalPoint}%
  \BibitemOpen
  \bibfield  {author} {\bibinfo {author} {\bibfnamefont {M.}~\bibnamefont
  {Naghiloo}}, \bibinfo {author} {\bibfnamefont {M.}~\bibnamefont {Abbasi}},
  \bibinfo {author} {\bibfnamefont {Y.~N.}\ \bibnamefont {Joglekar}},\ and\
  \bibinfo {author} {\bibfnamefont {K.~W.}\ \bibnamefont {Murch}},\ }\bibfield
  {title} {\bibinfo {title} {Quantum state tomography across the exceptional
  point in a single dissipative qubit},\ }\href
  {https://doi.org/10.1038/s41567-019-0652-z} {\bibfield  {journal} {\bibinfo
  {journal} {Nature Physics}\ }\textbf {\bibinfo {volume} {15}},\ \bibinfo
  {pages} {1232} (\bibinfo {year} {2019})}\BibitemShut {NoStop}%
\bibitem [{\citenamefont {Takasu}\ \emph {et~al.}(2020)\citenamefont {Takasu},
  \citenamefont {Yagami}, \citenamefont {Ashida}, \citenamefont {Hamazaki},
  \citenamefont {Kuno},\ and\ \citenamefont {Takahashi}}]{response_PTEP}%
  \BibitemOpen
  \bibfield  {author} {\bibinfo {author} {\bibfnamefont {Y.}~\bibnamefont
  {Takasu}}, \bibinfo {author} {\bibfnamefont {T.}~\bibnamefont {Yagami}},
  \bibinfo {author} {\bibfnamefont {Y.}~\bibnamefont {Ashida}}, \bibinfo
  {author} {\bibfnamefont {R.}~\bibnamefont {Hamazaki}}, \bibinfo {author}
  {\bibfnamefont {Y.}~\bibnamefont {Kuno}},\ and\ \bibinfo {author}
  {\bibfnamefont {Y.}~\bibnamefont {Takahashi}},\ }\bibfield  {title} {\bibinfo
  {title} {{PT-symmetric non-Hermitian quantum many-body system using ultracold
  atoms in an optical lattice with controlled dissipation}},\ }\href
  {https://doi.org/10.1093/ptep/ptaa094} {\bibfield  {journal} {\bibinfo
  {journal} {Progress of Theoretical and Experimental Physics}\ }\textbf
  {\bibinfo {volume} {2020}},\ \bibinfo {pages} {12A110} (\bibinfo {year}
  {2020})}\BibitemShut {NoStop}%
\bibitem [{\citenamefont {Mahan}(1993)}]{Mahan}%
  \BibitemOpen
  \bibfield  {author} {\bibinfo {author} {\bibfnamefont {G.~D.}\ \bibnamefont
  {Mahan}},\ }\href@noop {} {\emph {\bibinfo {title} {Many-Particle
  Physics}}},\ \bibinfo {edition} {2nd}\ ed.\ (\bibinfo  {publisher} {Plenum
  press},\ \bibinfo {address} {New York},\ \bibinfo {year} {1993})\BibitemShut
  {NoStop}%
\bibitem [{\citenamefont {Giamarchi}(2003)}]{Giamarchi}%
  \BibitemOpen
  \bibfield  {author} {\bibinfo {author} {\bibfnamefont {T.}~\bibnamefont
  {Giamarchi}},\ }\href@noop {} {\emph {\bibinfo {title} {Quantum Physics in
  One Dimension}}},\ \bibinfo {edition} {1st}\ ed.\ (\bibinfo  {publisher}
  {Clarendon Press},\ \bibinfo {address} {Oxford},\ \bibinfo {year}
  {2003})\BibitemShut {NoStop}%
\bibitem [{\citenamefont {Gogolin}\ \emph {et~al.}(1998)\citenamefont
  {Gogolin}, \citenamefont {Nersesyan},\ and\ \citenamefont
  {Tsvelik}}]{GogolinNersesyanTsvelik}%
  \BibitemOpen
  \bibfield  {author} {\bibinfo {author} {\bibfnamefont {A.~O.}\ \bibnamefont
  {Gogolin}}, \bibinfo {author} {\bibfnamefont {A.~A.}\ \bibnamefont
  {Nersesyan}},\ and\ \bibinfo {author} {\bibfnamefont {A.~M.}\ \bibnamefont
  {Tsvelik}},\ }\href@noop {} {\emph {\bibinfo {title} {Bosonization and
  Strongly Correlated Systems}}}\ (\bibinfo  {publisher} {Cambridge University
  Press},\ \bibinfo {address} {Cambridge},\ \bibinfo {year} {1998})\BibitemShut
  {NoStop}%
\bibitem [{\citenamefont {Knap}\ \emph {et~al.}(2012)\citenamefont {Knap},
  \citenamefont {Shashi}, \citenamefont {Nishida}, \citenamefont {Imambekov},
  \citenamefont {Abanin},\ and\ \citenamefont {Demler}}]{PhysRevX.2.041020}%
  \BibitemOpen
  \bibfield  {author} {\bibinfo {author} {\bibfnamefont {M.}~\bibnamefont
  {Knap}}, \bibinfo {author} {\bibfnamefont {A.}~\bibnamefont {Shashi}},
  \bibinfo {author} {\bibfnamefont {Y.}~\bibnamefont {Nishida}}, \bibinfo
  {author} {\bibfnamefont {A.}~\bibnamefont {Imambekov}}, \bibinfo {author}
  {\bibfnamefont {D.~A.}\ \bibnamefont {Abanin}},\ and\ \bibinfo {author}
  {\bibfnamefont {E.}~\bibnamefont {Demler}},\ }\bibfield  {title} {\bibinfo
  {title} {Time-dependent impurity in ultracold fermions: Orthogonality
  catastrophe and beyond},\ }\href {https://doi.org/10.1103/PhysRevX.2.041020}
  {\bibfield  {journal} {\bibinfo  {journal} {Phys. Rev. X}\ }\textbf {\bibinfo
  {volume} {2}},\ \bibinfo {pages} {041020} (\bibinfo {year}
  {2012})}\BibitemShut {NoStop}%
\bibitem [{\citenamefont {Catani}\ \emph {et~al.}(2012)\citenamefont {Catani},
  \citenamefont {Lamporesi}, \citenamefont {Naik}, \citenamefont {Gring},
  \citenamefont {Inguscio}, \citenamefont {Minardi}, \citenamefont {Kantian},\
  and\ \citenamefont {Giamarchi}}]{PhysRevA.85.023623}%
  \BibitemOpen
  \bibfield  {author} {\bibinfo {author} {\bibfnamefont {J.}~\bibnamefont
  {Catani}}, \bibinfo {author} {\bibfnamefont {G.}~\bibnamefont {Lamporesi}},
  \bibinfo {author} {\bibfnamefont {D.}~\bibnamefont {Naik}}, \bibinfo {author}
  {\bibfnamefont {M.}~\bibnamefont {Gring}}, \bibinfo {author} {\bibfnamefont
  {M.}~\bibnamefont {Inguscio}}, \bibinfo {author} {\bibfnamefont
  {F.}~\bibnamefont {Minardi}}, \bibinfo {author} {\bibfnamefont
  {A.}~\bibnamefont {Kantian}},\ and\ \bibinfo {author} {\bibfnamefont
  {T.}~\bibnamefont {Giamarchi}},\ }\bibfield  {title} {\bibinfo {title}
  {Quantum dynamics of impurities in a one-dimensional bose gas},\ }\href
  {https://doi.org/10.1103/PhysRevA.85.023623} {\bibfield  {journal} {\bibinfo
  {journal} {Phys. Rev. A}\ }\textbf {\bibinfo {volume} {85}},\ \bibinfo
  {pages} {023623} (\bibinfo {year} {2012})}\BibitemShut {NoStop}%
\bibitem [{\citenamefont {Cetina}\ \emph {et~al.}(2016)\citenamefont {Cetina},
  \citenamefont {Jag}, \citenamefont {Lous}, \citenamefont {Fritsche},
  \citenamefont {Walraven}, \citenamefont {Grimm}, \citenamefont {Levinsen},
  \citenamefont {Parish}, \citenamefont {Schmidt}, \citenamefont {Knap},\ and\
  \citenamefont {Demler}}]{Cetina-Science}%
  \BibitemOpen
  \bibfield  {author} {\bibinfo {author} {\bibfnamefont {M.}~\bibnamefont
  {Cetina}}, \bibinfo {author} {\bibfnamefont {M.}~\bibnamefont {Jag}},
  \bibinfo {author} {\bibfnamefont {R.~S.}\ \bibnamefont {Lous}}, \bibinfo
  {author} {\bibfnamefont {I.}~\bibnamefont {Fritsche}}, \bibinfo {author}
  {\bibfnamefont {J.~T.~M.}\ \bibnamefont {Walraven}}, \bibinfo {author}
  {\bibfnamefont {R.}~\bibnamefont {Grimm}}, \bibinfo {author} {\bibfnamefont
  {J.}~\bibnamefont {Levinsen}}, \bibinfo {author} {\bibfnamefont {M.~M.}\
  \bibnamefont {Parish}}, \bibinfo {author} {\bibfnamefont {R.}~\bibnamefont
  {Schmidt}}, \bibinfo {author} {\bibfnamefont {M.}~\bibnamefont {Knap}},\ and\
  \bibinfo {author} {\bibfnamefont {E.}~\bibnamefont {Demler}},\ }\bibfield
  {title} {\bibinfo {title} {Ultrafast many-body interferometry of impurities
  coupled to a {Fermi} sea},\ }\href {https://doi.org/10.1126/science.aaf5134}
  {\bibfield  {journal} {\bibinfo  {journal} {Science}\ }\textbf {\bibinfo
  {volume} {354}},\ \bibinfo {pages} {96} (\bibinfo {year} {2016})},\ \Eprint
  {https://arxiv.org/abs/https://www.science.org/doi/pdf/10.1126/science.aaf5134}
  {https://www.science.org/doi/pdf/10.1126/science.aaf5134} \BibitemShut
  {NoStop}%
\bibitem [{\citenamefont {Corman}\ \emph
  {et~al.}(2019{\natexlab{b}})\citenamefont {Corman}, \citenamefont
  {Fabritius}, \citenamefont {H\"ausler}, \citenamefont {Mohan}, \citenamefont
  {Dogra}, \citenamefont {Husmann}, \citenamefont {Lebrat},\ and\ \citenamefont
  {Esslinger}}]{PhysRevA.100.053605}%
  \BibitemOpen
  \bibfield  {author} {\bibinfo {author} {\bibfnamefont {L.}~\bibnamefont
  {Corman}}, \bibinfo {author} {\bibfnamefont {P.}~\bibnamefont {Fabritius}},
  \bibinfo {author} {\bibfnamefont {S.}~\bibnamefont {H\"ausler}}, \bibinfo
  {author} {\bibfnamefont {J.}~\bibnamefont {Mohan}}, \bibinfo {author}
  {\bibfnamefont {L.~H.}\ \bibnamefont {Dogra}}, \bibinfo {author}
  {\bibfnamefont {D.}~\bibnamefont {Husmann}}, \bibinfo {author} {\bibfnamefont
  {M.}~\bibnamefont {Lebrat}},\ and\ \bibinfo {author} {\bibfnamefont
  {T.}~\bibnamefont {Esslinger}},\ }\bibfield  {title} {\bibinfo {title}
  {Quantized conductance through a dissipative atomic point contact},\ }\href
  {https://doi.org/10.1103/PhysRevA.100.053605} {\bibfield  {journal} {\bibinfo
   {journal} {Phys. Rev. A}\ }\textbf {\bibinfo {volume} {100}},\ \bibinfo
  {pages} {053605} (\bibinfo {year} {2019}{\natexlab{b}})}\BibitemShut
  {NoStop}%
\bibitem [{\citenamefont {Yan}\ \emph {et~al.}(2013)\citenamefont {Yan},
  \citenamefont {Moses}, \citenamefont {Gadway}, \citenamefont {Covey},
  \citenamefont {Hazzard}, \citenamefont {Rey}, \citenamefont {Jin},\ and\
  \citenamefont {Ye}}]{DipolarMolecules2BLoss}%
  \BibitemOpen
  \bibfield  {author} {\bibinfo {author} {\bibfnamefont {B.}~\bibnamefont
  {Yan}}, \bibinfo {author} {\bibfnamefont {S.~A.}\ \bibnamefont {Moses}},
  \bibinfo {author} {\bibfnamefont {B.}~\bibnamefont {Gadway}}, \bibinfo
  {author} {\bibfnamefont {J.~P.}\ \bibnamefont {Covey}}, \bibinfo {author}
  {\bibfnamefont {K.~R.~A.}\ \bibnamefont {Hazzard}}, \bibinfo {author}
  {\bibfnamefont {A.~M.}\ \bibnamefont {Rey}}, \bibinfo {author} {\bibfnamefont
  {D.~S.}\ \bibnamefont {Jin}},\ and\ \bibinfo {author} {\bibfnamefont
  {J.}~\bibnamefont {Ye}},\ }\bibfield  {title} {\bibinfo {title} {Observation
  of dipolar spin-exchange interactions with lattice-confined polar
  molecules},\ }\href {https://doi.org/10.1038/nature12483} {\bibfield
  {journal} {\bibinfo  {journal} {Nature}\ }\textbf {\bibinfo {volume} {501}},\
  \bibinfo {pages} {521} (\bibinfo {year} {2013})}\BibitemShut {NoStop}%
\bibitem [{\citenamefont {Zhu}\ \emph {et~al.}(2014)\citenamefont {Zhu},
  \citenamefont {Gadway}, \citenamefont {Foss-Feig}, \citenamefont
  {Schachenmayer}, \citenamefont {Wall}, \citenamefont {Hazzard}, \citenamefont
  {Yan}, \citenamefont {Moses}, \citenamefont {Covey}, \citenamefont {Jin},
  \citenamefont {Ye}, \citenamefont {Holland},\ and\ \citenamefont
  {Rey}}]{PRL_2BLoss_Zeno}%
  \BibitemOpen
  \bibfield  {author} {\bibinfo {author} {\bibfnamefont {B.}~\bibnamefont
  {Zhu}}, \bibinfo {author} {\bibfnamefont {B.}~\bibnamefont {Gadway}},
  \bibinfo {author} {\bibfnamefont {M.}~\bibnamefont {Foss-Feig}}, \bibinfo
  {author} {\bibfnamefont {J.}~\bibnamefont {Schachenmayer}}, \bibinfo {author}
  {\bibfnamefont {M.~L.}\ \bibnamefont {Wall}}, \bibinfo {author}
  {\bibfnamefont {K.~R.~A.}\ \bibnamefont {Hazzard}}, \bibinfo {author}
  {\bibfnamefont {B.}~\bibnamefont {Yan}}, \bibinfo {author} {\bibfnamefont
  {S.~A.}\ \bibnamefont {Moses}}, \bibinfo {author} {\bibfnamefont {J.~P.}\
  \bibnamefont {Covey}}, \bibinfo {author} {\bibfnamefont {D.~S.}\ \bibnamefont
  {Jin}}, \bibinfo {author} {\bibfnamefont {J.}~\bibnamefont {Ye}}, \bibinfo
  {author} {\bibfnamefont {M.}~\bibnamefont {Holland}},\ and\ \bibinfo {author}
  {\bibfnamefont {A.~M.}\ \bibnamefont {Rey}},\ }\bibfield  {title} {\bibinfo
  {title} {Suppressing the loss of ultracold molecules via the continuous
  quantum {Zeno} effect},\ }\href
  {https://doi.org/10.1103/PhysRevLett.112.070404} {\bibfield  {journal}
  {\bibinfo  {journal} {Phys. Rev. Lett.}\ }\textbf {\bibinfo {volume} {112}},\
  \bibinfo {pages} {070404} (\bibinfo {year} {2014})}\BibitemShut {NoStop}%
\bibitem [{\citenamefont {Sponselee}\ \emph {et~al.}(2018)\citenamefont
  {Sponselee}, \citenamefont {Freystatzky}, \citenamefont {Abeln},
  \citenamefont {Diem}, \citenamefont {Hundt}, \citenamefont {Kochanke},
  \citenamefont {Ponath}, \citenamefont {Santra}, \citenamefont {Mathey},
  \citenamefont {Sengstock},\ and\ \citenamefont
  {Becker}}]{DissipativeFermiHubbard}%
  \BibitemOpen
  \bibfield  {author} {\bibinfo {author} {\bibfnamefont {K.}~\bibnamefont
  {Sponselee}}, \bibinfo {author} {\bibfnamefont {L.}~\bibnamefont
  {Freystatzky}}, \bibinfo {author} {\bibfnamefont {B.}~\bibnamefont {Abeln}},
  \bibinfo {author} {\bibfnamefont {M.}~\bibnamefont {Diem}}, \bibinfo {author}
  {\bibfnamefont {B.}~\bibnamefont {Hundt}}, \bibinfo {author} {\bibfnamefont
  {A.}~\bibnamefont {Kochanke}}, \bibinfo {author} {\bibfnamefont
  {T.}~\bibnamefont {Ponath}}, \bibinfo {author} {\bibfnamefont
  {B.}~\bibnamefont {Santra}}, \bibinfo {author} {\bibfnamefont
  {L.}~\bibnamefont {Mathey}}, \bibinfo {author} {\bibfnamefont
  {K.}~\bibnamefont {Sengstock}},\ and\ \bibinfo {author} {\bibfnamefont
  {C.}~\bibnamefont {Becker}},\ }\bibfield  {title} {\bibinfo {title} {Dynamics
  of ultracold quantum gases in the dissipative {Fermi}--{Hubbard} model},\
  }\href {https://doi.org/10.1088/2058-9565/aadccd} {\bibfield  {journal}
  {\bibinfo  {journal} {Quantum Sci. Technol.}\ }\textbf {\bibinfo {volume}
  {4}},\ \bibinfo {pages} {014002} (\bibinfo {year} {2018})}\BibitemShut
  {NoStop}%
\bibitem [{\citenamefont {Xie}\ \emph {et~al.}(2024)\citenamefont {Xie},
  \citenamefont {Liang}, \citenamefont {Ma}, \citenamefont {Du}, \citenamefont
  {Peng}, \citenamefont {Li}, \citenamefont {Chen}, \citenamefont {Li},
  \citenamefont {Gao},\ and\ \citenamefont {Xue}}]{scaleFreeBoundStates}%
  \BibitemOpen
  \bibfield  {author} {\bibinfo {author} {\bibfnamefont {X.}~\bibnamefont
  {Xie}}, \bibinfo {author} {\bibfnamefont {G.}~\bibnamefont {Liang}}, \bibinfo
  {author} {\bibfnamefont {F.}~\bibnamefont {Ma}}, \bibinfo {author}
  {\bibfnamefont {Y.}~\bibnamefont {Du}}, \bibinfo {author} {\bibfnamefont
  {Y.}~\bibnamefont {Peng}}, \bibinfo {author} {\bibfnamefont {E.}~\bibnamefont
  {Li}}, \bibinfo {author} {\bibfnamefont {H.}~\bibnamefont {Chen}}, \bibinfo
  {author} {\bibfnamefont {L.}~\bibnamefont {Li}}, \bibinfo {author}
  {\bibfnamefont {F.}~\bibnamefont {Gao}},\ and\ \bibinfo {author}
  {\bibfnamefont {H.}~\bibnamefont {Xue}},\ }\href@noop {} {\bibinfo {title}
  {Observation of scale-free localized states induced by non-{Hermitian}
  defects}} (\bibinfo {year} {2024}),\ \Eprint
  {https://arxiv.org/abs/2402.04716} {arXiv:2402.04716 [cond-mat.mes-hall]}
  \BibitemShut {NoStop}%
\bibitem [{\citenamefont {Cao}\ \emph {et~al.}(2019)\citenamefont {Cao},
  \citenamefont {Tilloy},\ and\ \citenamefont {Luca}}]{CaoTilloyDeLuca}%
  \BibitemOpen
  \bibfield  {author} {\bibinfo {author} {\bibfnamefont {X.}~\bibnamefont
  {Cao}}, \bibinfo {author} {\bibfnamefont {A.}~\bibnamefont {Tilloy}},\ and\
  \bibinfo {author} {\bibfnamefont {A.~D.}\ \bibnamefont {Luca}},\ }\bibfield
  {title} {\bibinfo {title} {Entanglement in a fermion chain under continuous
  monitoring},\ }\href {https://doi.org/10.21468/SciPostPhys.7.2.024}
  {\bibfield  {journal} {\bibinfo  {journal} {SciPost Phys.}\ }\textbf
  {\bibinfo {volume} {7}},\ \bibinfo {pages} {024} (\bibinfo {year}
  {2019})}\BibitemShut {NoStop}%
\bibitem [{\citenamefont {Kane}\ \emph {et~al.}(1994)\citenamefont {Kane},
  \citenamefont {Matveev},\ and\ \citenamefont
  {Glazman}}]{FermiEdgeKaneMatveevGlazman}%
  \BibitemOpen
  \bibfield  {author} {\bibinfo {author} {\bibfnamefont {C.~L.}\ \bibnamefont
  {Kane}}, \bibinfo {author} {\bibfnamefont {K.~A.}\ \bibnamefont {Matveev}},\
  and\ \bibinfo {author} {\bibfnamefont {L.~I.}\ \bibnamefont {Glazman}},\
  }\bibfield  {title} {\bibinfo {title} {Fermi-edge singularities and
  backscattering in a weakly interacting one-dimensional electron gas},\ }\href
  {https://doi.org/10.1103/PhysRevB.49.2253} {\bibfield  {journal} {\bibinfo
  {journal} {Phys. Rev. B}\ }\textbf {\bibinfo {volume} {49}},\ \bibinfo
  {pages} {2253} (\bibinfo {year} {1994})}\BibitemShut {NoStop}%
\bibitem [{\citenamefont {Imambekov}\ \emph {et~al.}(2012)\citenamefont
  {Imambekov}, \citenamefont {Schmidt},\ and\ \citenamefont
  {Glazman}}]{ImambekovSchmidtGlazman}%
  \BibitemOpen
  \bibfield  {author} {\bibinfo {author} {\bibfnamefont {A.}~\bibnamefont
  {Imambekov}}, \bibinfo {author} {\bibfnamefont {T.~L.}\ \bibnamefont
  {Schmidt}},\ and\ \bibinfo {author} {\bibfnamefont {L.~I.}\ \bibnamefont
  {Glazman}},\ }\bibfield  {title} {\bibinfo {title} {One-dimensional quantum
  liquids: Beyond the {Luttinger} liquid paradigm},\ }\href
  {https://doi.org/10.1103/RevModPhys.84.1253} {\bibfield  {journal} {\bibinfo
  {journal} {Rev. Mod. Phys.}\ }\textbf {\bibinfo {volume} {84}},\ \bibinfo
  {pages} {1253} (\bibinfo {year} {2012})}\BibitemShut {NoStop}%
\bibitem [{\citenamefont {Blanes}\ \emph {et~al.}(2009)\citenamefont {Blanes},
  \citenamefont {Casas}, \citenamefont {Oteo},\ and\ \citenamefont
  {Ros}}]{MagnusExpansion}%
  \BibitemOpen
  \bibfield  {author} {\bibinfo {author} {\bibfnamefont {S.}~\bibnamefont
  {Blanes}}, \bibinfo {author} {\bibfnamefont {F.}~\bibnamefont {Casas}},
  \bibinfo {author} {\bibfnamefont {J.}~\bibnamefont {Oteo}},\ and\ \bibinfo
  {author} {\bibfnamefont {J.}~\bibnamefont {Ros}},\ }\bibfield  {title}
  {\bibinfo {title} {The {Magnus} expansion and some of its applications},\
  }\href {https://doi.org/https://doi.org/10.1016/j.physrep.2008.11.001}
  {\bibfield  {journal} {\bibinfo  {journal} {Physics Reports}\ }\textbf
  {\bibinfo {volume} {470}},\ \bibinfo {pages} {151} (\bibinfo {year}
  {2009})}\BibitemShut {NoStop}%
\bibitem [{\citenamefont {Jung}\ \emph {et~al.}(1999)\citenamefont {Jung},
  \citenamefont {M\"uller},\ and\ \citenamefont {Rotter}}]{PhysRevE.60.114}%
  \BibitemOpen
  \bibfield  {author} {\bibinfo {author} {\bibfnamefont {C.}~\bibnamefont
  {Jung}}, \bibinfo {author} {\bibfnamefont {M.}~\bibnamefont {M\"uller}},\
  and\ \bibinfo {author} {\bibfnamefont {I.}~\bibnamefont {Rotter}},\
  }\bibfield  {title} {\bibinfo {title} {Phase transitions in open quantum
  systems},\ }\href {https://doi.org/10.1103/PhysRevE.60.114} {\bibfield
  {journal} {\bibinfo  {journal} {Phys. Rev. E}\ }\textbf {\bibinfo {volume}
  {60}},\ \bibinfo {pages} {114} (\bibinfo {year} {1999})}\BibitemShut
  {NoStop}%
\bibitem [{\citenamefont {Burke}\ \emph {et~al.}(2020)\citenamefont {Burke},
  \citenamefont {Wiersig},\ and\ \citenamefont {Haque}}]{Haque}%
  \BibitemOpen
  \bibfield  {author} {\bibinfo {author} {\bibfnamefont {P.~C.}\ \bibnamefont
  {Burke}}, \bibinfo {author} {\bibfnamefont {J.}~\bibnamefont {Wiersig}},\
  and\ \bibinfo {author} {\bibfnamefont {M.}~\bibnamefont {Haque}},\ }\bibfield
   {title} {\bibinfo {title} {Non-{Hermitian} scattering on a tight-binding
  lattice},\ }\href {https://doi.org/10.1103/PhysRevA.102.012212} {\bibfield
  {journal} {\bibinfo  {journal} {Phys. Rev. A}\ }\textbf {\bibinfo {volume}
  {102}},\ \bibinfo {pages} {012212} (\bibinfo {year} {2020})}\BibitemShut
  {NoStop}%
\bibitem [{\citenamefont {Rosso}\ \emph {et~al.}(2020)\citenamefont {Rosso},
  \citenamefont {Iemini}, \citenamefont {Schir{\`o}},\ and\ \citenamefont
  {Mazza}}]{SciPost_Schiro}%
  \BibitemOpen
  \bibfield  {author} {\bibinfo {author} {\bibfnamefont {L.}~\bibnamefont
  {Rosso}}, \bibinfo {author} {\bibfnamefont {F.}~\bibnamefont {Iemini}},
  \bibinfo {author} {\bibfnamefont {M.}~\bibnamefont {Schir{\`o}}},\ and\
  \bibinfo {author} {\bibfnamefont {L.}~\bibnamefont {Mazza}},\ }\bibfield
  {title} {\bibinfo {title} {{Dissipative flow equations}},\ }\href
  {https://doi.org/10.21468/SciPostPhys.9.6.091} {\bibfield  {journal}
  {\bibinfo  {journal} {SciPost Phys.}\ }\textbf {\bibinfo {volume} {9}},\
  \bibinfo {pages} {091} (\bibinfo {year} {2020})}\BibitemShut {NoStop}%
\bibitem [{\citenamefont {Combescot}\ and\ \citenamefont
  {Nozi\`eres}(1971)}]{CombescotNozieres}%
  \BibitemOpen
  \bibfield  {author} {\bibinfo {author} {\bibfnamefont {M.}~\bibnamefont
  {Combescot}}\ and\ \bibinfo {author} {\bibfnamefont {P.}~\bibnamefont
  {Nozi\`eres}},\ }\bibfield  {title} {\bibinfo {title} {Infrared catastrophy
  and excitons in the {X-ray} spectra of metals},\ }\href@noop {} {\bibfield
  {journal} {\bibinfo  {journal} {J. Phys. (France)}\ }\textbf {\bibinfo
  {volume} {32}},\ \bibinfo {pages} {913} (\bibinfo {year} {1971})}\BibitemShut
  {NoStop}%
\bibitem [{\citenamefont {Nozi\`eres}\ and\ \citenamefont
  {De~Dominicis}(1969)}]{NozieresDeDominicis}%
  \BibitemOpen
  \bibfield  {author} {\bibinfo {author} {\bibfnamefont {P.}~\bibnamefont
  {Nozi\`eres}}\ and\ \bibinfo {author} {\bibfnamefont {C.~T.}\ \bibnamefont
  {De~Dominicis}},\ }\bibfield  {title} {\bibinfo {title} {Singularities in the
  {X-ray} absorption and emission of metals. iii. one-body theory exact
  solution},\ }\href {https://doi.org/10.1103/PhysRev.178.1097} {\bibfield
  {journal} {\bibinfo  {journal} {Phys. Rev.}\ }\textbf {\bibinfo {volume}
  {178}},\ \bibinfo {pages} {1097} (\bibinfo {year} {1969})}\BibitemShut
  {NoStop}%
\bibitem [{\citenamefont {Ng}(1995)}]{PRB_Ng}%
  \BibitemOpen
  \bibfield  {author} {\bibinfo {author} {\bibfnamefont {T.-K.}\ \bibnamefont
  {Ng}},\ }\bibfield  {title} {\bibinfo {title} {{X-ray}-edge singularity in
  nonequilibrium systems},\ }\href {https://doi.org/10.1103/PhysRevB.51.2009}
  {\bibfield  {journal} {\bibinfo  {journal} {Phys. Rev. B}\ }\textbf {\bibinfo
  {volume} {51}},\ \bibinfo {pages} {2009} (\bibinfo {year}
  {1995})}\BibitemShut {NoStop}%
\end{thebibliography}%
\end{document}


\title{Supplemental material -- Orthogonality catastrophe beyond bosonization from post-selection}
\author{Martino Stefanini}
\affiliation{Institut f\"ur Physik, Johannes Gutenberg-Universit\"at Mainz, D-55099 Mainz, Germany}
\author{Jamir Marino}
\affiliation{Institut f\"ur Physik, Johannes Gutenberg-Universit\"at Mainz, D-55099 Mainz, Germany}

\date{\today}
\maketitle
\section{Realization of the dynamics}
As mentioned in the main text, there are two main approaches for reproducing the nonlinear Schr\"odinger equation
\begin{align}\label{eq: nlse}
    \ii \dv{}{t}\ket{\psi(t)}=\Big[H-\ii\frac{\gamma}{2}n_0+\ii\frac{\gamma}{2}\ev{n_0}{\psi(t)}\Big]\ket{\psi(t)}
\end{align}
that governs the dynamics of the non-Hermitian orthogonality catastrophe (OC). The first approach requires to perform a series of weak measurements of the fermion number at site $j_0$, chosen such that the set of Kraus operators \cite{WisemanMilburn} that describe the measurements includes the desired one,
\begin{equation}\label{eq: M}
    M\propto \ee^{-\frac{\gamma\delta t}{2}n_0}~.
\end{equation}
The state of the system will evolve stochastically, as a function of the random measurement outcomes, and the desired dynamics \eqref{eq: nlse} will be obtained by post-selecting the state trajectories for which the outcomes correspond to \eqref{eq: M}. This strategy is known as "no-click" scenario \cite{AltmanMeasurementsTLL,SchiroNonHermPRB,SchiroNonHermSciPost,TurkeshiPhysRevB.103.224210}. While this approach is conceptually straightforward, it is experimentally demanding as it requires to measure a local observable on a fine grid of times and to keep track of all outcomes of the measurements.
A possible choice of a Kraus set yielding the desired non-Hermitian dynamics is given by the set
\begin{equation}\label{eq: dephasing M}
    \begin{cases}
    M_0(\delta t)=\ee^{-\frac{\gamma\delta t}{2}n_0}\\
    M_1(\delta t)=(1-\ee^{-\gamma\delta t})^{1/2}n_0
    \end{cases}~,
\end{equation}
with outcomes labeled as $0$ and $1$, which correspond to finding no fermions or one fermion at the measured site, respectively. This set is a common unraveling of the Lindblad equation with local dephasing $\mathcal{J}=\gamma^{1/2}n_0$ \cite{WisemanMilburn}. However, \eq~\eqref{eq: dephasing M} is slightly less general than the model considered in the main text as it requires $\gamma>0$. An example of Kraus set that requires $\gamma<0$ is
\begin{equation}
        \begin{cases}
        M_0(\delta t)&=\ee^{\frac{\gamma\delta t}{2}}\ee^{-\frac{\gamma\delta t}{2}n_0}\\
        M_1(\delta t)&=(1-\ee^{\gamma\delta t})^{1/2}c_0^\dag
        \end{cases}~,
\end{equation}
which is an unraveling of a Lindblad dynamics with localized gain $\mathcal{J}=\abs{\gamma}^{1/2}c_{j_0}^\dag$. Other examples can be found that allow for both signs of $\gamma$. Since we are not interested in any specific realization of the measurements but rather only on the action of $M_0$ [\eq~\eqref{eq: M}], for generality we keep the sign of $\gamma$ unspecified.
\par The second approach uses a suitable dissipative (Lindbladian) dynamics to “reconstruct” \eq~\eqref{eq: nlse}. Choosing for simplicity a situation with just one jump operator,the dissipative dynamics reads
\begin{equation}\label{eq: lindblad}
    \dv{}{t}\rho(t)=-\ii\comm{H}{\rho}+\mathcal{J}\rho \mathcal{J}^\dag -\frac{1}{2}\{\mathcal{J}^\dag \mathcal{J},\rho\}~.
\end{equation}
The following procedure is simply an adaptation of the standard projection technique \cite{ReviewNonHermitianUeda} to the dissipative case. We assume the existence of an observable $Q$ (which is in general extensive) that is conserved by the Hamiltonian and by the combination $\mathcal{J}^\dag \mathcal{J}$ (i.e. $\comm{Q}{\mathcal{J}^\dag \mathcal{J}}=0$), but not by the jump $\mathcal{J}$ itself: $\comm{\mathcal{J}}{Q}\neq0$. A prototypical situation is that of a localized loss, $\mathcal{J}\propto c_{j_0}$, for which $\mathcal{J}^\dag \mathcal{J}\propto n_{j_0}$ and the observable $Q$ is simply the total number $Q=\sum_j c_j^\dag c_j$. We can use the eigenstates of $Q$ to partition the Hilbert space in orthogonal subspaces, according to the (generally degenerate) eigenvalues $q$ of $Q$. We are going to denote as $\mathcal{P}_q$ the projectors on these subspaces. We are interested in characterizing the dynamics that occurs within a target subspace defined by the initial value of $Q$, $q_0$ (the method makes sense only if the initial state belongs to an eigenspace of $Q$). The target subspace will have a projector $\mathcal{P}_0\equiv \mathcal{P}_{q_0}$, and we define its complementary projector $\mathcal{P}_1\equiv\sum_{q\neq q_0}\mathcal{P}_q$. Introducing the notation $\mathcal{O}_{ab}\equiv \mathcal{P}_a\mathcal{O}\mathcal{P}_b$ with $a,\,b\in\{0,\,1\}$, the conservation of $Q$ implies that $H=H_{00}+H_{11}$, and analogously for $\mathcal{J}^\dag \mathcal{J}$, while we shall assume that the jumps are purely off-diagonal $\mathcal{J}_{aa}=0$. If we project the Lindblad equation \eqref{eq: lindblad} on the two subspaces defined by $\mathcal{P}_{0}$ and $\mathcal{P}_1$, we obtain
\begin{equation}
    \begin{aligned}
        \dv{}{t}\rho_{00}&=-\ii \widetilde{H}_{00}\rho_{00}+\ii\rho_{00}\widetilde{H}_{00}+\mathcal{J}_{01}\rho_{11}(\mathcal{J}^\dag)_{10}~,\\
        \dv{}{t}\rho_{11}&=-\ii \widetilde{H}_{11}\rho_{11}+\ii\rho_{11}\widetilde{H}_{11}+\mathcal{J}_{10}\rho_{00}(\mathcal{J}^\dag)_{01}~,\\
    \end{aligned}
\end{equation}
where $\widetilde{H}\equiv H-\ii\mathcal{J}^\dag\mathcal{J}/2$. We can solve the second equation for $\rho_{11}$ to obtain a closed equation for $\rho_{00}$:
\begin{equation}
    \dv{}{t}\rho_{00}=-\ii \widetilde{H}_{00}\rho_{00}+\ii\rho_{00}\widetilde{H}_{00}+\int_0^t\dd{s}\mathcal{J}_{01}\ee^{-\ii\widetilde{H}_{11}(t-s)}\mathcal{J}_{10}\rho_{00}(s)(\mathcal{J}^\dag)_{01}\ee^{\ii\widetilde{H}_{11}^\dag(t-s)}(\mathcal{J}^\dag)_{10}~,
\end{equation}
where we used the assumption that $\rho(0)$ belongs to the target subspace and so $\rho_{11}(0)=0$. The first two terms on the right-hand side of the \eq~above describe a non-Hermitian evolution, while the third contains a memory effect caused by processes that go to from the target subspace to its complement and back. The last assumption that we need is that the jump obeys the condition $\mathcal{J}_{qq^\prime}\neq0\Rightarrow \mathcal{J}_{q^\prime q}=0$ (using the generalized notation $\mathcal{O}_{qq^\prime}\equiv \mathcal{P}_q\mathcal{O}\mathcal{P}_{q^\prime}$). Then, the third term of the \eq~above vanishes, since $\mathcal{J}_{01}\ee^{-\ii\widetilde{H}_{11}(t-s)}\mathcal{J}_{10}=\sum_{q\neq q_0}\mathcal{J}_{q_0 q}\ee^{-\ii\widetilde{H}_{qq}(t-s)}\mathcal{J}_{q q_0}=0$ (and analogously for $(\mathcal{J}^\dag)_{01}\ee^{\ii\widetilde{H}_{11}^\dag(t-s)}(\mathcal{J}^\dag)_{10}$), and we are left with $\rho_{00}(t)$ evolving according to the non-Hermitian Hamiltonian $\widetilde{H}$. The condition $\mathcal{J}_{qq^\prime}\neq0\Rightarrow \mathcal{J}_{q^\prime q}=0$ could be satisfied by a jump operator $\mathcal{J}$ that is a triangular matrix with respect to $q$---as in the case of the single-body loss. Vice versa, for a Hermitian jump we have $\mathcal{J}_{qq^\prime}=\mathcal{J}_{q^\prime q}^\ast$ and the whole projection procedure can not be implemented. Experimentally, the system is initialized in the $Q=q_0$ subspace, then it is left to evolve according to \eq~\eqref{eq: lindblad} for a time $t$, at which $Q$ is measured again to yield a value $q$. If the new value of $Q$ is the same as the beginning the experimental run is kept, otherwise it is discarded---a form of post-selection. In this way, at each time the post-selected system is described by the projected density matrix $\rho_{00}(t)=\ee^{-\ii\widetilde{H} t}\rho(0)\ee^{\ii\widetilde{H} t}$ (recalling that $\rho(0)\equiv\rho_{00}(0)$). If the initial state is pure $\rho(0)=\ketbra{\psi(0)}$, then it remains pure during the dynamics: $\rho_{00}(t)=\ketbra*{\tilde{\psi}(t)}$, where $\ket*{\tilde{\psi}(t)}=\ee^{-\ii \widetilde{H} t}\ket{\psi(0)}$. The probability of observing this post-selected state is given by $\Tr\rho_{00}(t)$, which in the case of a pure state is the squared norm $\norm*{\sket{\tilde{\psi}(t)}}^2=\snorm{\tilde{\psi}(t)}$. Measuring this probability allows to reconstruct the desired state $\ket{\psi(t)}=\sket{\tilde{\psi}(t)}/\norm*{\sket{\tilde{\psi}(t)}}$. 
\par Suitable setups for this approach are the case of a localized single-particle loss $\mathcal{J}=\gamma^{1/2}c_{j_0}$ \cite{ThesisFroml,FromlPRL,FromlPRB1,FromlPRB2} mentioned above, and the case of a localized single-particle gain $\mathcal{J}=\abs{\gamma}^{1/2}c_{j_0}^\dag$ \cite{LocalizedFermionSource}, both of which yield the proper $\widetilde{H}=K$ (with $\gamma>0$ in the former case, and $\gamma<0$ in the latter\footnote{Indeed, $\widetilde{H}=H-\ii\abs{\gamma}/2\,c_{j_0} c_{j_0}^\dag=H-\ii\abs{\gamma}/2\,(\one-n_0)=H+\ii\abs{
\gamma}/2\, n_0-\ii\abs{\gamma}/2$, with the imaginary constant term playing no role in the dynamics \eqref{eq: nlse}. However, the constant term guarantees that all eigenvalues of $\widetilde{H}$ are negative, and so that $\widetilde{H}$ generates a stable dynamics.}) and for which the global observable $Q$ that provides the post-selection is simply the total number of fermions.
\par The case of localized dephasing $\mathcal{J}=\gamma^{1/2}n_0$ \cite{Dolgirev, Tonielli} generates the correct non-Hermitian Hamiltonian $K$, but is not suitable for the procedure outlined above, since there is no $Q$ with the right properties---in other words, there is no way of reconstructing \textit{a posteriori} whether a dissipative event (a jump) has occurred. The only way to proceed would be to fall back to the continuous monitoring of \eq~\eqref{eq: dephasing M}, which is indeed a stochastic unraveling of a Lindblad equation with local dephasing. 
\par We notice that the dissipative strategy outlined above does not simply reduce to the one based on measurements: at a given time $t$ one generally has to measure a global observable (in our case, the total number of fermions) \emph{once}, and then perform a post-selection based on the result, whereas the first strategy requires the recording a local observable (the outcome of the density measurement) for each instant before $t$. At a conceptual level, the main difference is that the local monitoring requires coupling the system to one or more ancillary systems, and then measuring appropriate observables on them, while in the dissipative scenario the measurement is performed on the system itself, while the environment providing the dissipation is never looked at.
\par The dissipative strategy explained above for a system with a localized loss is probably the most promising experimental scenario for the realization of the results reported in the main text. Fermionic systems with tunable single-body losses in have been already realized \cite{Esslinger_JPhys,Esslinger_PRL,Esslinger_PRL_23,Esslinger_PRX}, and the observed dynamics has been interpreted in terms of a non-Hermitian Hamiltonian even in the absence of post-selection \cite{Esslinger_PRA_theory}. The post-selection proposed above is in the spirit of \rref~\cite{QubitExceptionalPoint}, and performing it on the measured number of atoms should be within reach of near-term experiments, as in \cite{response_PTEP} or in the aforementioned works \cite{Esslinger_JPhys,Esslinger_PRL,Esslinger_PRL_23,Esslinger_PRX}.
\par While the previous two strategies are fairly standard, we propose here a novel approach which employs dissipative dynamics to reconstruct \eq~\eqref{eq: nlse}, \emph{without explicit post-selection}. The idea is to introduce an “impurity” in the system, which triggers the dynamics and serves as a probe of the rest of the system, in analogy with the usual OC case \cite{Mahan,Giamarchi,GogolinNersesyanTsvelik,PhysRevX.2.041020}. In more detail, we assume that at the measured site $j_0$ it is possible to introduce an immobile\footnote{Immobile impurities can be easily employed with ultracold atoms: for example, see \cite{PhysRevX.2.041020,PhysRevA.85.023623,Cetina-Science}, and \cite{PhysRevA.100.053605} for an experimental setup with a localized loss.} impurity $d$ that interacts with a “bath” fermion $c_{j_0}$ at the site so that both fermions are then lost from the system. This type of two-body loss can be realized with ultra-cold atoms \cite{DipolarMolecules2BLoss, PRL_2BLoss_Zeno, DissipativeFermiHubbard}. The Lindblad equation for the combined impurity-bath system is then \eq~\eqref{eq: lindblad} with $\mathcal{J}=\gamma^{1/2}d\, c_{j_0}$. The non-Hermitian generator of the Lindbladian dynamics is $\widetilde{H}=H-\ii\gamma/2\, d^\dag d\, n_0$, from which we see that the presence of the impurity ($d^\dag d=1$) induces the desired non-Hermitian dynamics \eqref{eq: nlse} with $\widetilde{H}=H-\ii\gamma/2\, n_0\equiv K$. Not only the impurity degree of freedom serves as a switch of the imaginary potential, but it can also be used as a probe to extract the full return amplitude of the non-Hermitian dynamics of the \emph{bath}. This can be obtained by choosing the initial impurity state in a superposition of occupied ($\ket{1}$) and empty ($\ket{0}$).
\par We can decompose the density matrix as $\rho(t)=\sum_{\alpha,\beta\in\{0,1\}}R_{\alpha\beta}(t)\ketbra{\alpha}{\beta}$, where $\{\ket{0},\,\ket{1}\}$ form the basis of the impurity states and $R_{\alpha\beta}(t)$ are operators acting on the $c$ fermions' Hilbert space. Since $\rho(t)$ is Hermitian, we have $R_{\alpha\alpha}^\dag=R_{\alpha\alpha}$ and $R_{10}^\dag=R_{01}$. Plugging this decomposition in the Lindblad equation we find that the $R$ operators satisfy 
\begin{equation}
    \begin{aligned}
        &\dv{}{t}R_{00}(t)=-\ii\comm{H}{R_{00}(t)}+\gamma c_{j_0} R_{11}(t) c_{j_0}^\dag~,\\
        &\dv{}{t}R_{10}(t)=-\ii K R_{10}(t)+\ii R_{10}(t)H~,\\
        &\dv{}{t}R_{11}(t)=-\ii K R_{11}(t)+\ii R_{11}(t)K^\dag~,
    \end{aligned}
\end{equation}
that are solved by
\begin{equation}\label{eqs: Rs}
    \begin{aligned}
        R_{00}(t)&=\ee^{-\ii H t}R_{00}(0)\ee^{\ii H t}+\gamma \int_0^t \dd{s}\ee^{-\ii H (t-s)}c_{j_0} R_{11}(s) c_{j_0}^\dag \ee^{\ii H (t-s)}~,\\
        R_{10}(t)&=\ee^{-\ii K t}R_{10}(0)\ee^{\ii H t}~,\\
        R_{11}(t)&=\ee^{-\ii K t}R_{11}(0)\ee^{\ii K^\dag t}~.
    \end{aligned}
\end{equation}
Let us assume that the impurity and bath are initially prepared in uncorrelated pure states $\rho(0)=\ket{FS}_c\bra{FS}\otimes \ketbra{\varphi}$, where $\ket{FS}_c$ is the ground state of the bath and $\ket{\varphi}=\sum_{\alpha=0}^1{\varphi_\alpha \ket{\alpha}}$ is a generic state of the impurity. Then $R_{\alpha\beta}(0)=\varphi_\alpha\varphi_\beta^\ast \ket{FS}_c\bra{FS}$ and \eqs~\eqref{eqs: Rs} yield
\begin{equation}
    \begin{aligned}
        R_{00}(t)&=\abs{\varphi_0}^2\ket{FS}_c\bra{FS}+\abs{\varphi_1}^2\gamma \int_0^t \dd{s}\ee^{-\ii H (t-s)}c_{j_0} \sket{\tilde{\psi}(s)}_c\sbra{\tilde{\psi}(s)}c_{j_0}^\dag \ee^{\ii H (t-s)}~,\\
        R_{10}(t)&=\varphi_1\varphi_0^\ast \sket{\tilde{\psi}(t)}_c\bra{FS}\ee^{\ii E_{FS} t}~,\\
        R_{11}(t)&=\abs{\varphi_1}^2\sket{\tilde{\psi}(t)}_c\sbra{\tilde{\psi}(t)}~,
    \end{aligned}
\end{equation}
in which we introduced the ground-state energy of the bath $E_{FS}\equiv \ev{H}{FS}$ and the un-normalized state $\sket{\tilde{\psi}_C(t)}\equiv \ee^{-\ii K t}\ket{FS}_c$. Already at this level, we observe that the desired non-Hermitian dynamics is contained in the full state of the system. Actually, in order to extract the return amplitude $\mathcal{L}(t)\equiv _c\sbraket{FS}{\tilde{\psi}(t)}_c/_c\snorm{\tilde{\psi}(t)}_c^{1/2}$ only the density matrix $\rho^d$ of the impurity is needed. Indeed, from $\rho^d_{\alpha\beta}(t)\equiv\Tr_c R_{\alpha\beta}(t)$ we obtain:
\begin{equation}
    \rho^d(t)=\mqty(1-\abs{\varphi_1}^2\, _c\snorm{\tilde{\psi}(t)}_c & \varphi_0\varphi_1^\ast \,_c\tilde{\mathcal{L}}^\ast(t)\ee^{-\ii E_{FS} t}\\ \varphi_1\varphi_0^\ast \tilde{\mathcal{L}}(t)\ee^{\ii E_{FS} t} & \abs{\varphi_1}^2\, _c\snorm{\tilde{\psi}(t)}_c)~,
\end{equation}
where $\tilde{\mathcal{L}}(t)\equiv\,_c\sbraket{FS}{\tilde{\psi}(t)}_c$, and the $\rho_{00}^d$ element has been simplified using the property
\begin{equation*}
    _c\snorm{\tilde{\psi}(t)}_c=1-\gamma\int_0^t \dd{s}\, _c\sbra{\tilde{\psi}(t)}n_0\sket{\tilde{\psi}(t)}_c~,
\end{equation*} 
that can be proven from $\ii\dv*{}{t}\sket{\tilde{\psi}(t)}=K\sket{\tilde{\psi}(t)}$. Thus, a measurement of the impurity population gives access to the norm $_c\snorm{\tilde{\psi}(t)}_c$, while the off-diagonal elements of $\rho_d$ give the overlap $_c\sbraket{FS}{\tilde{\psi}(t)}_c$. One can also employ a Ramsey interferometry scheme \cite{PhysRevX.2.041020} to measure $_c\sbraket{FS}{\tilde{\psi}(t)}_c$ from the impurity population after the application of a $\pi/2$-pulse. This scheme has been already implemented in experiments with ultracold atoms \cite{Cetina-Science}.
\par We would like to mention that the single-particle physics of our non-Hermitian Hamiltonian $K=H-\ii\gamma n_0/2$ has been recently realized \cite{scaleFreeBoundStates} with electrical circuits, confirming the existence of the quasi-bound modes that invalidate bosonization in the many body system. 
\par Concluding this Section, we remark that all the approaches presented incur in some form of post-selection problem, since in all cases a fraction of the experimental runs has to be discarded, and in general this fraction increases exponentially on a timescale $(\bar{n}\gamma)^{-1}$. However, this does not necessarily preclude the possibility to observe the phenomena reported in the main text, because it may still possible to find a regime of density and measurement rate in which they are visible before the post-selection overhead becomes too large. The exact determination of the extent of this regime requires will not be pursued further.
\section{Particle-hole transformation}
Let us introduce the basis of the fermionic Fock space for spinless fermions moving on a 1D lattice of $L$ sites as $\ket{\bm{n}}$, where $\bm{n}$ is a string of $0$s and $1$s enumerating the occupation of the lattice sites $j=0\dots L-1$. The length of $\bm{n}$ is $L$. In other words, $\ket{\bm{n}}=\ket{n_0 \dots n_{L-1}}\equiv(c_0^\dag)^{n_0}\dots (c_{L-1}^\dag)^{n_{L-1}}\ket{0}$, where $\ket{0}=\ket{\bm{0}}=\ket{0\dots0}$ is the vacuum. We choose to order the application of the fermion creation operators as the order of the sites, but the choice of this convention does not change the results. Then, the particle-hole (ph) transformation is implemented by the operator $\mathcal{C}=\mathcal{S}\mathcal{P}$, where
\begin{subequations}
    \begin{align}
        \mathcal{S}&\equiv \ee^{\ii \pi \sum_{j}c_j^\dag c_j}~,\\
        \mathcal{P}&\equiv\sum_{\bm{n}}\ketbra{\mathtt{NOT}(\bm{n})}{\bm{n}}~,
    \end{align}
\end{subequations}
in which $\mathtt{NOT}$ is the logical $\mathtt{NOT}$ operation that swaps the values of $0$ and $1$ in the strings. Both the above operators are self-adjoint and square to the identity. The two operators commute if the number of sites is even, and anticommute otherwise: $\mathcal{S}\mathcal{P}=(-1)^L \mathcal{P}\mathcal{S}$. The operator $\mathcal{C}^\prime=\mathcal{P}\mathcal{S}$ would be equally good as a ph transformation. 
\par The action of $\mathcal{P}$ is evident: it changes every occupied site (particle) to an empty site (hole), and vice-versa. It induces the following transformation on the fermionic operators:
\begin{equation}
    \begin{aligned}
        \mathcal{P}c_j\mathcal{P}^\dag&=\sum_{\bm{n},\bm{m}}\ketbra{\mathtt{NOT}(\bm{m})}{\bm{m}}c_j\ketbra{\bm{n}}{\mathtt{NOT}(\bm{n})}=\sum_{\bm{n},\bm{m}\vert n_j=1}\ket{\mathtt{NOT}(\bm{m})}\delta_{\bm{m},\bm{n}-(1)_j}(-1)^{\sum_{k<j}n_k}\bra{\mathtt{NOT}(\bm{n})}=\\ 
    &=\sum_{\bm{n}\vert n_j=0}(-1)^{\sum_{k<j}(1-n_k)}\ketbra{\bm{n}+(1)_j}{\bm{n}}=c_j^\dag \sum_{\bm{n}}(-1)^{\sum_{k<j}1}\ketbra{\bm{n}}{\bm{n}}=(-1)^{j-1}c_j^\dag~,
    \end{aligned}
\end{equation}
where the notation $\ket{\bm{n}\pm(-1)_j}$ stands for $\ket{n_0\dots n_j\pm1\dots n_{L-1}}$, and we have used
\begin{equation}
    \begin{aligned}
        c_j\ket{\bm{n}}&=\delta_{n_j,1}(-1)^{\sum_{k<j}n_k}\ket{\bm{n}-(1)_j}\\
        c_j^\dag\ket{\bm{n}}&=\delta_{n_j,0}(-1)^{\sum_{k<j}n_k}\ket{\bm{n}+(1)_j}
    \end{aligned}
\end{equation}
The action of $\mathcal{S}$ adds minus sign $\mathcal{S}c_j\mathcal{S}^\dag=-c_j$, so that the full particle-hole transformation of the fermion operators has the usual form: $\mathcal{C}c_j\mathcal{C}^\dag=(-1)^{j}c_j^\dag$. In momentum space, this transformation reads 
\begin{equation}\label{eq: phs p}
    \mathcal{C}c_p\mathcal{C}^\dag=c_{-p-\pi}^\dag=
    \begin{cases}
        &c^\dag_{\pi-p}\quad\text{for }p\in[0,\pi]\\
        &c^\dag_{3\pi-p}\quad\text{for }p\in]\pi,2\pi[
    \end{cases}~.
\end{equation}
We can now see how this transformation acts on the nonlinear Schr\"odinger equation \eqref{eq: nlse} that governs the post-selected dynamics. Its solution can be obtained as $\ket{\psi(t)}=\sket{\tilde{\psi}(t)}/\snorm{\tilde{\psi}(t)}^{1/2}$, where $\sket{\tilde{\psi}(t)}=\ee^{-\ii K t}\ket{\psi(0)}$. Suppose that we start from the state $\ket{\psi^{\mathcal{C}}(0)}=\mathcal{C}\ket{\psi(0)}$. If $\ket{\psi(0)}$ has a fixed number of fermions $N_f$, then the new state has the “dual” number of fermions $N_f^\mathcal{C}=L-N_f$. Then 
\begin{equation*}
    \sket{\tilde{\psi}^{\mathcal{C}}(t)}=\mathcal{C}\ee^{-\ii\mathcal{C} K \mathcal{C}t}\ket{\psi(0)}=\ee^{-\gamma t/2}\mathcal{C}\ee^{-\ii K^\dag t}\ket{\psi(0)}~,
\end{equation*}
since $\mathcal{C}(H-\ii\gamma/2\, n_0)\mathcal{C}=H+\ii\gamma/2 n_0-\ii\gamma/2=K^\dag-\ii\gamma/2$. The normalized state is therefore
\begin{equation*}
    \ket{\psi^{\mathcal{C}}(t)}=\frac{\mathcal{C}\ee^{-\ii K^\dag t}\ket{\psi(0)}}{\ev{\ee^{\ii K t}\ee^{-\ii K^\dag t}}{\psi(0)}^{1/2}}=\mathcal{C}\ket{\psi_{-\gamma}(t)}~,
\end{equation*}
which implies that the $\ket{\psi^{\mathcal{C}}(t)}$ is the ph-transformed solution $\ket{\psi_{-\gamma}(t)}$ of the nonlinear Schr\"odinger equation with $\gamma\to-\gamma$:
\begin{equation*}
    \ii\dv{}{t}\ket{\psi_{-\gamma}(t)}=\Big[K^\dag-\ii\frac{\gamma}{2}\ev{n_0}{\psi_{-\gamma}(t)}\Big]\ket{\psi_{-\gamma}(t)}
\end{equation*}
and initial condition $\ket{\psi_{-\gamma}(0)}=\ket{\psi(0)}$. This completes the proof of our claim that the post-selected dynamics is not symmetric under a ph transformation, and that such a ph transformation is equivalent to a change of sign of $\gamma$. This would be also valid for a Hermitian impurity, $H_V=H+V n_0$. Indeed, the $V<0$ scenario features a bound state \cite{Mahan} that is absent from the $V>0$ situation (in which there is actually an anti-bound state). As a consequence of the results above, we have that
\begin{align*}
    \mathcal{L}^{\mathcal{C}}(t)&\equiv \braket{\psi^{\mathcal{C}}(0)}{\psi^{\mathcal{C}}(t)}=\braket{\psi_{-\gamma}(0)}{\psi_{-\gamma}(t)}\\
    \intertext{and}
    E^{\mathcal{C}}(t)&\equiv \ev{H}{\psi^\mathcal{C}(t)}=\ev{H}{\psi_{-\gamma}(t)},
\end{align*}
showing the absence of ph symmetry in these quantities and the property that their behavior starting from the ph conjugate state $\ket{\psi^{\mathcal{C}}(0)}=\mathcal{C}\ket{\psi(0)}$ is equivalent to the behavior of the solution with the original initial state but with the opposite sign of $\gamma$. 
\begin{figure}
    \centering
    \includegraphics[width=0.7\linewidth]{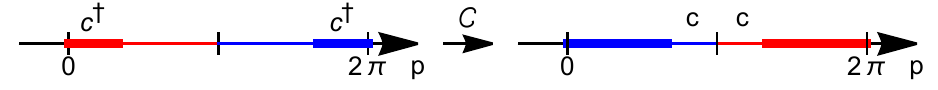}
    \caption{Action of the particle-hole transformation on the Fermi sea. Thick and thin lines correspond to occupied and empty states, respectively. Notice that the order of the states is also reversed by $\mathcal{C}$.}
    \label{fig: phs FS}
\end{figure}
\par In all our simulations, the initial state is the Fermi sea $\ket{FS}$. In the main text we claimed that the observables that we measure for a certain $\gamma$ and filling $\bar{n}=N_f/L$ are the same as those for $-\gamma$ and filling $\bar{n}^{\mathcal{C}}=1-\bar{n}$. This can be seen by proving that $\mathcal{C}\ket{FS}$ is the Fermi sea with the complementary filling $1-\bar{n}$. As shown graphically in \fig~\ref{fig: phs FS} using \eq~\eqref{eq: phs p}, this is exactly what happens.
\section{Details of numerical implementation}
\subsection{Numerical algorithm}
We repeat here the numerical method that we employed to compute the dynamics, taken from \rref~\cite{CaoTilloyDeLuca}. A general Gaussian pure state can be written as 
\begin{equation}
    \ket{\psi}=\prod_{i=1}^{N_f} \sum_{j=0}^{L-1}U_{ji}c_j^\dag \ket{0}~,
\end{equation}
where $\ket{0}$ is the fermionic vacuum and $U_{ji}$ is simply the collection of the single-particle wave functions. The time evolution generated by a quadratic Hamiltonian [such as our nonlinear Schr\"odinger equation  \eqref{eq: nlse}] will conserve the above form of the state, since it will transform the $c_j$s linearly according the single-particle Hamiltonian: $\ee^{-\ii \sum_{ii^\prime}h_{ii^\prime} c_{i}^\dag c_{i^\prime}t} c_j^\dag \ee^{\ii \sum_{ii^\prime}h_{ii^\prime} c_{i}^\dag c_{i^\prime} t}=\sum_{j^\prime}(\ee^{-\ii h t})_{jj^\prime} c_{j^\prime}^\dag$. The result is that we only need to compute the single-particle time evolution operator $\ee^{-\ii h t}$ to evolve $U$: $\tilde{U}(t+\delta t)=\ee^{-\ii h \delta t} U(t)$. The normalization of the state requires that $U$ is an isometry, $U^\dag (t) U(t)=\one_{N_f}$. This requirement can be imposed at each time step by a QR decomposition, $\tilde{U}(t+\delta t)=QR$, and setting the new $U$ matrix to $U(t+\delta t)=Q$. A useful by-product of this method is that the correlation function of the fermions is easily computed from $U$: $\langle c_i^\dag c_j\rangle=(U U^\dag)_{ij}$.
\par In our case, the $U$ matrix reads $U_{ji}(t)=\phi_{p_i}(j,t)$, where the $p_i$s run over the initial occupied momentum states. The wave functions obey the initial condition $\phi_{p_i}(j,t=0)=\ee^{\ii p_i j}/L^{1/2}$, and the time evolution is conveniently Trotterized as $\ee^{-\ii \kappa \delta t}\approx \ee^{-\ii h \delta t/2} \ee^{-\gamma/2\, n \delta t} \ee^{-\ii h \delta t/2}$, where $\kappa$, $h$ and $n$ are the single-particle versions of $K$, $H$ and $n_0$, respectively. The unitary part, $\ee^{-\ii h \delta t/2}$, can be easily computed in the momentum basis where it is diagonal, while $\ee^{-\gamma/2\, n \delta t}$ is diagonal in real space. Different time steps $\delta t$ have been taken in the interval $\delta t=0.01$ to $0.04$ (in the units where $2J=1$) to verify convergence. In particular, the data for $L=1000$ are for $\delta t=0.04$. We simulate only systems with an odd number of fermions $N_f$, because an even $N_f$ would cause the ground state to be doubly degenerate in pbc. Alternatively, one could employ anti-periodic boundary conditions with even number of fermions. We have verified that the choice of boundary conditions and fermion numbers do not affect the physics qualitatively.
\par The code that we used and the original data can be provided by the corresponding author upon reasonable request.
\subsection{Numerical fits of return amplitude}
\begin{figure}
    \centering
    \subfloat[\label{fig:fit beta}]{\includegraphics[width=0.45\linewidth]{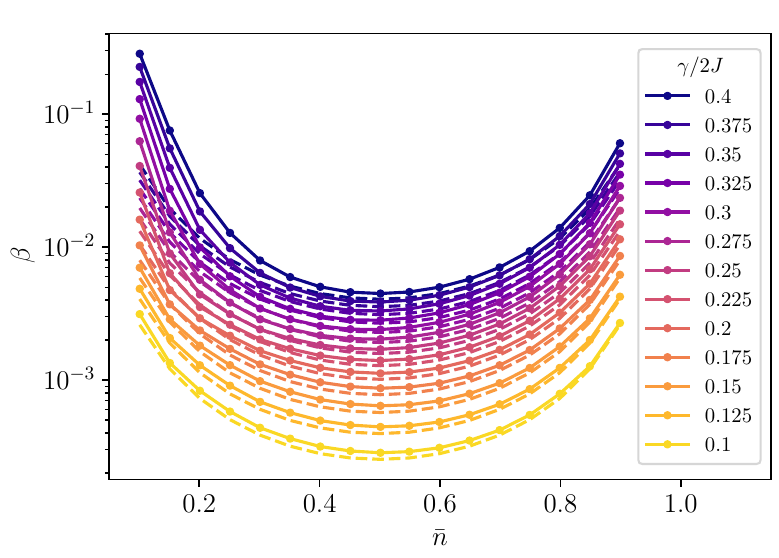}}
    \subfloat[\label{fig:fit Gamma}]{\includegraphics[width=0.45\linewidth]{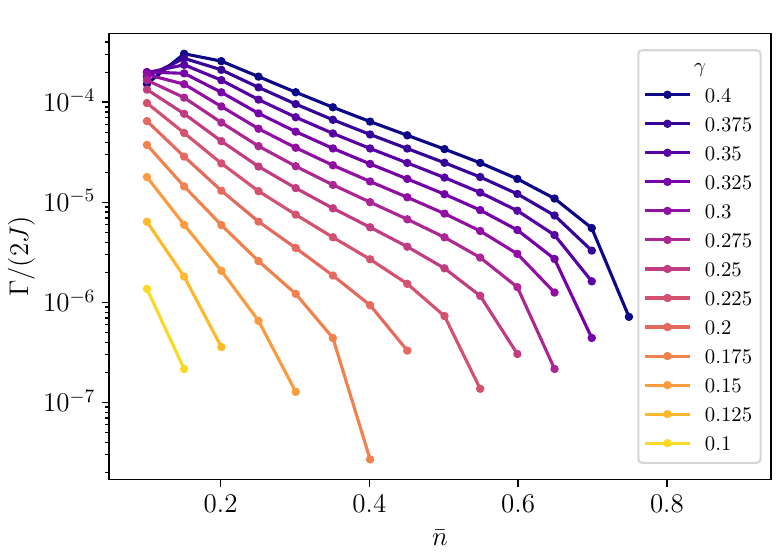}}\\
    \subfloat[\label{fig: Gamma nonpert}]{\includegraphics[width=0.45\linewidth]{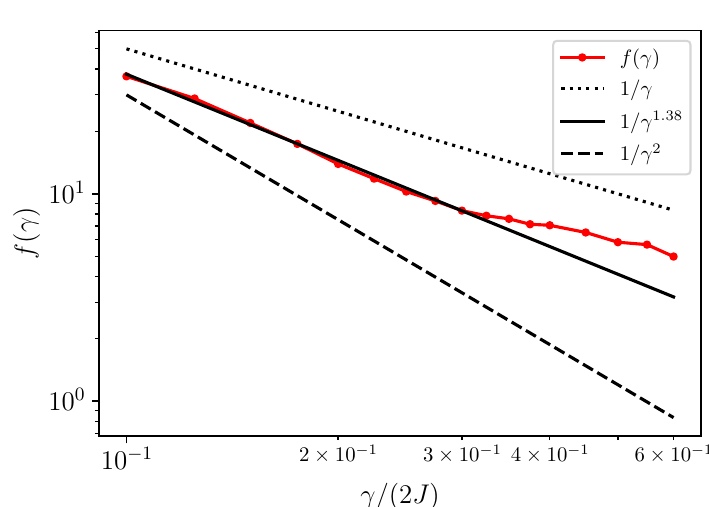}}
    \caption{Parameters fitted from the return amplitude. (a) The OC parameter $\beta$, compared with the bosonization prediction (dashed lines). (b) The exponential decay rate $\Gamma$. (c) The fitted dependence of $\Gamma$ on the density, $\ln\Gamma =a(\gamma)-f(\gamma)\bar{n}$, showing the probable power-law increase of $f(\gamma)$ at small $\gamma$ and hence the non-perturbative nature of $\Gamma$.}
    \label{fig:fits}
\end{figure}
As stated in the main text, we assume that the return amplitude has the long-time profile $\abs{\mathcal{L}(t)}=A t^{-\beta}\ee^{-\Gamma t}$. To find the values of $A$, $\beta$ and $\Gamma$ we have performed a numerical fit of $\ln\abs{\mathcal{L}(t)}$ to $\ln A -\beta \ln t-\Gamma t$ using a simple least-squares minimization. The results are shown in \figs~\ref{fig:fit beta} and \ref{fig:fit Gamma}. The uncertainty on the fitted parameters is computed as $\delta a_j=[\chi^2 (Q^{-1})_{jj}/(N-3)]^{1/2}$, where $\bm{a}=(A,\beta,\Gamma)$, $\chi^2=\sum_i(\ln\abs{\mathcal{L}(t_i)}-\ln A +\beta \ln t_i+\Gamma t_i)^2$ is the sum of the squared residuals, $N$ is the number of data, and $Q=M^T M$, in which $M$ is defined by $M_{i1}=1$, $M_{i2}=\ln t_i$ and $M_{i3}=t_i$. The uncertainties computed in this way are always smaller than the plotted points themselves.
\par The meaning of the behavior of $\beta$ and $\Gamma$ as a function of the density has been discussed at length in the main text. Here we just notice that if the density is high enough, the fitted value of $\Gamma$ becomes negative. In the absence of a quantitative prediction for $\Gamma$, we have not been able to give a meaning to such non-physical prediction---it could be a $\gamma$-dependent finite-size effect, a numerical instability caused by a value of $\Gamma$ which is too small, or even a hint that the functional form of the decay is not exponential. Indeed, we find that for large $\gamma$, when the momentum of the special modes $q^\ast(\gamma/(2J))$ becomes comparable to the Fermi momentum $k_F$ the decay is most likely a stretched exponential, albeit the analysis is obscured by the presence of oscillations. 
\par Notwithstanding these caveats, it is relevant to understand the relation of $\Gamma$ to the measurement rate, $\gamma$. Noticing the exponential decrease of $\Gamma$ with the density, we assume $\ln\Gamma=a(\gamma)-f(\gamma)\bar{n}$ and we extract the coefficients with a linear fit against the density. The term $a(\gamma)$ is found to be an approximately linear function, whereas $f(\gamma)$ has a more interesting behavior that is shown in \fig~\ref{fig: Gamma nonpert}. We find that $f(\gamma)$ grows at small $\gamma$, suggesting that the extra exponential decay may have a non-perturbative origin. Indeed, the double logarithmic plot of \fig~\ref{fig: Gamma nonpert} shows a plausible power-law behavior, which we extract through a linear fit of the first $10$ points as $f(\gamma)\approx 0.453/\gamma^{1.38}$. An accurate determination of $\Gamma$ at small $\gamma$ requires further work.   
\section{Bosonization}\label{sect: bosonization}
In this Section we bosonize the non-Hermitian Hamiltonian $K=H-\ii \gamma/2 n_0$ along the lines of \cite{GogolinNersesyanTsvelik} and \cite{FermiEdgeKaneMatveevGlazman}. We assume that the measured site is at $j_0=0$, and that the number of sites $L$ is odd\footnote{For an even number of sites one needs to add another symmetric mode, $c_{\pi s}=c_{\pi}$ (assuming a Brillouin zone $p\in ]-\pi,\pi]$). The presence of this mode does not affect the continuum results.}, and we define the symmetric $c_{ks}$ and antisymmetric $c_{ka}$ fermion operators according to
\begin{equation}\label{eq: symmetrization}
c_{\pm k}=
    \begin{cases}
    \frac{1}{\sqrt{2}}(c_{ks}\pm c_{ka})~,&\text{ if } k>0~,\\
    c_{0s}~,&\text{ if } k=0~,
    \end{cases}
\end{equation}
which correspond to decomposing the real-space operators in operators which are symmetric or antisymmetric with respect to the origin:
\begin{equation}
    c_j=\frac{1}{\sqrt{L}}\sum_k\ee^{\ii k j}c_k=\frac{1}{\sqrt{L}}c_{0s}+\sum_{k>0}\sqrt{\tfrac{2}{L}}\cos(k j)c_{ks}+\sum_{k>0}\ii\sqrt{\tfrac{2}{L}}\sin(kj)c_{ka}\equiv c_{js}+c_{ja}~.
\end{equation}
Correspondingly, as long as the band dispersion is even in momentum, $\ce_{-k}=\ce_k$, the free Hamiltonian separates into two symmetry sectors $H=H_s+H_a\equiv\sum_{k\geq0} \ce_k c_{ks}^\dag c_{ks}+\sum_{k>0} \ce_k c_{ka}^\dag c_{ka}$, and the ground state factorizes into two Fermi seas $\ket{FS}=\ket{FS}_s \ket{FS}_a$, both filled from momentum $k=0$ to the Fermi momentum $k=k_F$. The imaginary potential scattering is proportional to $n_0=c_{j=0,s}^\dag c_{j=0,s}$, which involves the symmetric states only, hence the asymmetric states do not participate in the dynamics. As usual in bosonization, we need to adopt a continuum description in terms of a fermion field $\psi(x=ja)\equiv a^{-1/2}c_j$, with vanishing lattice spacing $a\to0$. This field can be decomposed as above, $\psi(x)=\psi_s(x)+\psi_a(x)$. Then, we assume that all the dynamics occurs in the vicinity of the \emph{single} Fermi point of the symmetric modes, so that we can restrict the degrees of freedom to those around $k=k_F$
\begin{equation}
    \psi_s(x)=\frac{1}{\sqrt{L}}c_{0s}+\sum_{k>0}\frac{\ee^{\ii k x}+\ee^{-\ii k x}}{\sqrt{2L}}c_{ks}\approx\frac{1}{\sqrt{2}}\big[\ee^{\ii k_F x}R(x)+\ee^{-\ii k_F x}L(x)\big]~,
\end{equation}
where the right-movers $R(x)$ and left-movers $L(x)$ are
\begin{subequations}
    \begin{align}
    &R(x)=\sum_{k:\,\abs{k-k_F}<\alpha^{-1}}\frac{\ee^{\ii (k-k_F) x}}{\sqrt{L}}c_{ks}~,\\
    &L(x)=\sum_{k:\,\abs{k-k_F}<\alpha^{-1}}\frac{\ee^{-\ii (k-k_F) x}}{\sqrt{L}}c_{ks}~.
\end{align}
\end{subequations}
with a suitable cutoff $\alpha$. The left-movers are not independent of the right-movers: $L(x)=R(-x)$, so that there is actually only one chiral species of fermions. This happens because the symmetric sector has only one Fermi point. Notice also that the symmetrized fields have to be considered for $x>0$ (or, equivalently, $x<0$) only, to avoid doubling the degrees of freedom. Then, we can use $R(-x)=L(x)$ to define $R(x)$ on the whole axis. Finally, we linearize the dispersion around the Fermi point and express it in terms of $R(x)$ alone:
\begin{equation}
    H_s-E_\mathrm{gs}\approx\sum_{\abs{k-k_F}<1/\alpha}v_F(k-k_F)c^\dag_{ks}c_{ks}=\int_{-La/2}^{La/2}\dd{x}R^\dag(x)(-\ii v_F\partial)R(x)~,
\end{equation}
where $v_F\equiv \partial_k\ce_k\eval_{k=k_F}$ is the Fermi velocity, equal to $2J\sin(\pi\bar{n})$ for lattice fermions. The one above is the standard Hamiltonian for chiral fermions, which can be bosonized in terms of a single bosonic field $\chi(x)$ with the usual formulae \cite{Giamarchi}:
\begin{subequations}
    \begin{align}
        &R(x)=\frac{1}{\sqrt{2\pi\alpha}}F\ee^{-\ii\chi(x)}~,\\
        &\chi(x)=\phi(x)-\theta(x)=\frac{2\ii \pi}{\sqrt{L}}\sum_{q>0}\frac{V_q}{q}(b_q\ee^{i q x}-b_q^\dag \ee^{-\ii q x})~.
    \end{align}
\end{subequations}
In the above, $F$ is the Klein operator, while $\phi$ and $\theta$ are the usual TLL fields in the notation of \cite{Giamarchi}.
\par The description of the system in terms of a single chiral field has an important technical consequence: the local imaginary potential is proportional to $\psi^\dag(0)\psi(0)=\psi_s^\dag(0)\psi_s(0)=2 R^\dag(0)R(0)$, which does not contain any backscattering terms like $R^\dag(0) L(0)+L^\dag(0) R(0)$, which would give rise to nonlinear terms $\propto\cos(2\chi(0))$. Thus, using the standard relation \cite{Giamarchi,GogolinNersesyanTsvelik} $R^\dag(x)R(x)=\frac{1}{2}\bar{\rho}-\frac{1}{2\pi}\partial\chi(x)$ [with $\bar{\rho}\equiv N_f/(La)$ the density of fermions], we find that the bosonized non-Hermitian Hamiltonian (for the symmetric modes) is quadratic in $\chi(x)$:
\begin{align}
    K_{TLL}&=\frac{v_F}{2\pi}\int\dd{x}[\partial\chi(x)]^2-\ii\gamma_c\Big[\frac{1}{2}\bar{\rho}-\frac{1}{2\pi}\partial\chi(0)\Big]=\sum_{q>0}\Big[\ce_q b_q^\dag b_q-\ii\frac{\gamma_c}{\sqrt{L}}V_q(b_q+b_q^\dag)\Big]-\ii\gamma_c\frac{1}{2}\bar{\rho}~,
\end{align}
where $\gamma_c=\gamma a$, $\ce_q\equiv v_F q$ and $V_q\equiv q^{1/2} \ee^{-\alpha q}/(2\pi)$. We are dropping the finite-size contributions containing the Klein factors and we put the TLL subscript to emphasize that this Hamiltonian describes the system within the usual framework of Tomonaga-Luttinger liquids (TLLs). Indeed, we could generate nonlinear terms in $K_{TLL}$ by considering the curvature of the dispersion relation \cite{ImambekovSchmidtGlazman}, but this would bring us beyond the scope of TLLs.
\par To sum up, we have obtained a quadratic non-Hermitian Hamiltonian, and therefore we can obtain the full dynamics of the system. Using the Magnus expansion and the Baker-Campbell-Hausdorff formula \cite{MagnusExpansion}, we can write
\begin{align}
    \ee^{-\ii K_{TLL}t}&=\ee^{\ii\Im F(t)-\gamma_c\frac{1}{2}\bar{\rho}t}\,\ee^{-\ii H_{TLL}t}\,\ee^{\sum_q(\alpha_q^\ast(t) b_q^\dag +\alpha_q(t)b_q)}=\notag\\
    &=\ee^{F(t)-\gamma_c\frac{1}{2}\bar{\rho}t}\,\ee^{-\ii H_{TLL}t}\,\ee^{\sum_q\alpha_q^\ast(t) b_q^\dag}\,\ee^{\sum_q\alpha_q(t) b_q}~,
\end{align}
where $H_{TLL}\equiv \sum_{q>0}\ce_q b_q^\dag b_q$ and we introduced 
\begin{equation}\label{eq: F}
    F(t)\equiv\frac{\gamma_c^2}{L}\sum_{q>0}V_q^2\frac{1-\ii\ce_q t-\ee^{-\ii \ce_q t}}{\ce_q^2}=\frac{\gamma_c^2}{(2\pi v_F)^2}\Big(\ln\frac{\alpha+\ii v_F t}{\alpha}-\ii \frac{v_F t}{\alpha}\Big)
\end{equation}
(where the second equality refers to the continuum limit) and
\begin{equation}
    \alpha_q(t)\equiv\ii\frac{\gamma_c}{\sqrt{L}}V_q \frac{1-\ee^{-\ii\ce_q t}}{\ce_q}~.
\end{equation} 
In the bosonic language, the initial state $\ket{FS}_s$ is simply the boson vacuum $\ket{\omega}$, hence the unnormalized state is
\begin{equation}
    \sket{\tilde{\psi}(t)}\equiv \ee^{-\ii K_{TLL} t}\ket{\omega}=\ee^{2\Re F(t)-\gamma_c\frac{1}{2}\bar{\rho}t}\ee^{\ii\Im F(t)}\ket{\mathrm{coh}[\alpha_q^\ast(t)\ee^{-\ii\ce_q}]}~,
\end{equation}
where $\ket{\mathrm{coh}[z_q]}\equiv \ee^{\sum_q(z_q b_q^\dag-z^\ast_q b_q)}\ket{\omega}$ is a bosonic coherent state, and we used 
\begin{align*}
    &\ee^{\sum_q \alpha_q^\ast b_q^\dag}\ket{\omega}=\ee^{\frac{1}{2}\sum_q\abs{\alpha_q}^2}\ket{\mathrm{coh}[\alpha_q^\ast]}~,\\
    &\ee^{-\ii\sum_q\theta_q b_q^\dag b_q}\ket{\mathrm{coh}[z_q]}=\ket{\mathrm{coh}[z_q\ee^{-\ii\theta_q}]}
\end{align*}
and $\frac{1}{2}\sum_q\abs{\alpha_q(t)}^2=\Re F(t)$. Finally, we obtain the normalized state
\begin{equation}
    \ket{\psi(t)}=\frac{\sket{\tilde{\psi}(t)}}{\snorm{\tilde{\psi}(t)}^{1/2}}=\ee^{\ii\Im F(t)}\ket{\mathrm{coh}[\alpha_q^\ast(t)\ee^{-\ii\ce_q t}]}~.
\end{equation}
We notice that this result is almost identical to the one that we would obtain in the Hermitian case (i.e. for $\gamma\to-\ii V$).
\par It is easy to compute the return amplitude using the property of the coherent states that $\braket{\omega}{\mathrm{coh}[z_q]}=\ee^{-\tfrac{1}{2}\sum_q\abs{z_q}^2}$:
\begin{equation}
    \mathcal{L}(t)=\braket{\omega}{\psi(t)}=\ee^{-F(t)}~,
\end{equation}
and from \eq~\eqref{eq: F} we can read off that for times $v_F t \gg \alpha$ we have $\abs{\mathcal{L}(t)}\sim [\alpha/(v_F t)]^{\beta_b}$, where
\begin{equation}
    \beta_b=\frac{\gamma_c^2}{(2\pi v_F)^2}
\end{equation}
is the bosonization prediction quoted in the main text. The above result is the same as in the Hermitian case of a scattering potential with strength $\gamma_c/2$.
\par The energy injected by the measurement can be computed as follows:
\begin{equation}
    \begin{aligned}
        E(t)&=\ev{H_{TLL}}{\psi(t)}
        =\sum_{q>0}\ce_q \abs{\alpha_q(t)}^2=\frac{\gamma_c^2}{L}\sum_{q>0}\ce_q V_q^2\,2\frac{1-\cos(\ce_q t)}{\ce_q^2}=\\
        &=2\frac{\gamma_c^2}{(2\pi)^2 v_F}\int_0^{+\infty}\dd{q}\ee^{-\alpha q}[1-\cos(q v_F t)]=\frac{\gamma_c^2}{2\pi^2 v_F}\Big[\frac{1}{\alpha}-\Re\frac{1}{\alpha+\ii v_F t}\Big]\xrightarrow[t\to+\infty]{}\frac{\gamma_c^2}{2\pi^2 v_F \alpha}~. 
    \end{aligned}
\end{equation}
Although the exact value of the energy is cutoff-dependent (as expected for an effective low-energy description such as the TLL), it is predicted to saturate.
\begin{figure}
    \centering
    \includegraphics[width=0.45\linewidth]{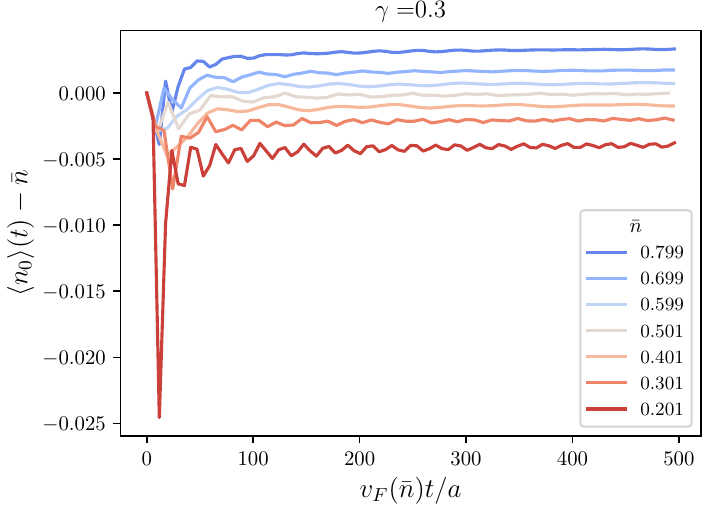}
    \caption{Time evolution of the fermion density at the impurity site.}
    \label{fig: n0}
\end{figure}
\par The time-dependence of the density at the measured site can be computed similarly:
\begin{equation}\label{eq: bsnz n0}
    \begin{aligned}
        \expval{\psi^\dag(0)\psi(0)}_t-\bar{\rho}&=2\expval{R^\dag(0)R(0)}_t-\bar{\rho}=-\frac{1}{\pi}\expval{\partial\chi(0)}_t=\frac{2}{L^{1/2}}\sum_{q>0}V_q\expval{b_q+b_q^\dag}_t=\\
        &=\Im \frac{4\gamma_c}{L}\sum_{q>0}V_q^2\frac{1-\ee^{-\ii\ce_q t}}{\ce_q}=\Im \frac{\gamma_c}{\pi^2 v_F}\int_0^{+\infty}\dd{q}\ee^{-\alpha q}(1-\ee^{-\ii q v_F t})=-\frac{\gamma_c}{\pi^2 v_F}\frac{v_F t}{\alpha^2+(v_F t)^2}~.
    \end{aligned}
\end{equation}
Thus, the prediction of the bosonized model is that the density at the measured site should decrease almost linearly up to a cutoff-dependent time $\alpha/v_F$, after which it would start increasing again until returning asymptotically to the initial value. This is not what the numerics shows. In \fig~\ref{fig: n0} we report the typical time evolution of the density at the measured site $\expval{n_0}(t)$ for different initial densities. Also in this observable, we find a monotonic behavior as a function of the density, with $\expval{n_0}(t)$ saturating to a value which grows with $\bar{n}$, and that is smaller (larger) than $\bar{n}$ approximately below (above) half-filling. So, the bosonization prediction \eqref{eq: bsnz n0}, in which $\expval{n_0}(t)$ should return back to the original value after a transient, is valid only in the vicinity of half-filling (including the sharp decrease at small times, which is however not universal). We notice that the interpretation of the dynamics of the system in real space is not intuitive, as the time evolution induced by non-Hermitian Hamiltonians is nonlocal due to the lack of a Lieb-Robinson bound \cite{ReviewNonHermitianUeda}.  
\par We can have an intuition about why the bosonized model does not agree with the numerical results by noticing that it has a larger symmetry than the discrete one. Indeed, if we drop the constant shift $-\ii\gamma_c\bar{\rho}/2$, the non-hermitian hamiltonian $K_{TLL}$ is $\mathcal{P}\mathcal{T}$ symmetric \cite{ReviewNonHermitianUeda}, where $\mathcal{T}$ is the anti-unitary time-reversal operator, while the parity operator $\mathcal{P}$ changes the sign of $\chi(x)$. As long as the symmetry is unbroken, it implies that the eigenvalues of $K_{TLL}$ are all in the form $\mathcal{E}-\ii\gamma_c\bar{\rho}/2$, where $\mathcal{E}$ is real. In other words, all eigenstates of the $K_{TLL}$ decay with the same lifetime, $\gamma_c\bar{\rho}/2$. On the other hand, as we will show in the next section, the discrete non-Hermitian Hamiltonian does not have such a symmetry, since the imaginary part of its spectrum is not constant\footnote{An unbroken $\mathcal{P}\mathcal{T}$ symmetry is a sufficient condition \cite{ReviewNonHermitianUeda} for a \emph{real} spectrum, but here we are extending the notion to within an \emph{imaginary} shift of the Hamiltonian. Such a constant shift is unobservable in the nonlinear Schr\"odinger equation, just as a real energy shift does not have any observable consequences for Hermitian Hamiltonians.}. The existence of a variety of lifetimes implies that the nonlinear Schr\"odinger equation \eqref{eq: nlse} will bring the state to converge to the many-body eigenstate of $K$ that has the lowest imaginary part (or the highest, for $\gamma<0$). This convergence will happen on a timescale that is extensive in the system size (we have numerically verified that it is usually much larger than $L/v_F$), so that the dynamics that we are interested in occurs in a time regime (i.e. for times smaller than $L/(2v_F)$) in which the system state is still very far from the stationary one. Nonetheless, this equilibration dynamics is completely absent in the bosonized model, since a constant imaginary part of the eigenvalues of $K$ cancels out of the dynamics in \eq~\eqref{eq: nlse}. As a consequence, we expect $K_{TLL}$ to behave effectively as a Hermitian Hamiltonian, and this is what we observed in the explicit solution outlined above, which is analogous to the one in the usual OC. As a final remark, we will show that the discrete version of $K$ does have an approximate $\mathcal{P}\mathcal{T}$ symmetry, since it features a large number of eigenvalues that have essentially the same imaginary part. We believe that this is the reason why the bosonized mode is able to correctly predict the OC behavior [i.e. the algebraic decay of $\mathcal{L}(t)$] that we observe.  
\section{Spectral properties of non-Hermitian Hamiltonian}
The non-Hermitian Hamiltonian $K$ has been already considered in \cite{PhysRevE.60.114,Haque,SciPost_Schiro,FromlPRB1}, but from points of view that differ from the present work. The first one considered the single-particle scattering properties of $K$, but employed open boundary conditions. The second one focused on the “strong-coupling” regime $\gamma\geq4J$ and on the continuum limit, which is not relevant for the results presented in this work, since we are especially interested in the scattering properties in the “weak-coupling” regime $\gamma<4J$, which require to consider a finite-size system. Thus, although many of the results of this section are known, it is useful to gather them together for completeness.
\subsection{Analytic relations}
It is convenient to work in momentum space, in which $K=\sum_{p,q}c_p^\dag \kappa_{pq} c_q$, where the single-particle Hamiltonian $\kappa$ reads $\kappa_{pq}=\ce_q \delta_{pq}-\frac{\ii\gamma}{2L}$. The single-particle right eigenfunctions $\braket{k}{r_\alpha}$ satisfy $\sum_k \kappa_{pk}\braket{k}{r_\alpha}=\lambda_\alpha \braket{p}{r_\alpha}$, or
\begin{equation*}
     \ce_p \braket{p}{r_\alpha}-\ii\frac{\gamma}{2L}\sum_k \braket{k}{r_\alpha} = \lambda_\alpha \braket{p}{r_\alpha}~, 
\end{equation*}
from which we extract
\begin{equation}
     \braket{p}{r_\alpha}= \ii\frac{\gamma}{2L}\frac{S_\alpha}{\ce_p-\lambda_\alpha}~, 
\end{equation}
where $S_\alpha\equiv\sum_k \braket{k}{r_\alpha}$. Introducing the real-space eigenfunctions $\braket{j}{r_\alpha}=L^{-1/2}\sum_p \ee^{\ii pj}\braket{p}{r_\alpha}$ we recognize that\footnote{In this section, we set $j_0=0$.} $S_\alpha=\sqrt{L}\braket{j=0}{r_\alpha}$. Self-consistency requires either $S_\alpha=0$ or
\begin{equation}\label{eq: eigenvalues K}
    1=\ii\frac{\gamma}{2L}\sum_p\frac{1}{\ce_p-\lambda_\alpha}~,
\end{equation}
which determines the nontrivial eigenvalues. The solutions with $S_\alpha=0$ are just the antisymmetric modes, which are unaffected by the impurity. Taking the conjugate of \eq~\eqref{eq: eigenvalues K} we find that $\lambda_\alpha(-\gamma)=\lambda_\alpha(\gamma)^\ast$.
\par The left eigenstates satisfy $\sum_k \braket{l_\alpha}{k}\kappa_{kp}=\lambda_\alpha \braket{l_\alpha}{p}$, but since $\kappa=\kappa^T$ is symmetric, $\braket{l_\alpha}{k}$ is determined by the same equation as $\braket{p}{r_\alpha}$, so $\braket{l_\alpha}{p}=\braket{p}{r_\alpha}$. The $S_\alpha$ are fixed by normalizing the states according to $\braket{l_\alpha}{r_\beta}=\delta_{\alpha\beta}$:
\begin{equation*}
    \braket{l_\alpha}{r_\beta}=\bigg(\frac{\ii\gamma}{2L}\bigg)^2 S_\alpha S_\beta\sum_p\frac{1}{\ce_p-\lambda_\alpha}\frac{1}{\ce_p-\lambda_\beta}~.
\end{equation*}
For $\lambda_\alpha\neq\lambda_\beta$ the sum is
\begin{equation*}
    \ii\frac{\gamma}{2L}\sum_p\frac{1}{\ce_p-\lambda_\alpha}\frac{1}{\ce_p-\lambda_\beta}=\frac{1}{\lambda_\alpha-\lambda_\beta}\ii\frac{\gamma}{2L}\sum_p\Big[\frac{1}{\ce_p-\lambda_\alpha}-\frac{1}{\ce_p-\lambda_\beta}\Big]=0~,
\end{equation*}
where the last equality comes from \eq~\eqref{eq: eigenvalues K}. Numerically, the spectrum is nondegenerate, so 
\begin{equation}
    S_\alpha^{-1}=\ii\frac{\gamma}{2L}\bigg[\sum_p\frac{1}{(\ce_p-\lambda_\alpha)^2}\bigg]^{1/2}~.
\end{equation}
An important quantity is the overlap $\braket{r_\alpha}{r_\beta}$:
\begin{equation*}
    \braket{r_\alpha}{r_\beta}=\Big(\frac{\gamma}{2L}\Big)^2\sum_p\frac{1}{\ce_p-\lambda_\alpha^\ast}\frac{1}{\ce_p-\lambda_\beta}~.
\end{equation*}
The sum can be computed as above:
\begin{equation*}
    \ii\frac{\gamma}{2L}\sum_p\frac{1}{\ce_p-\lambda_\alpha^\ast}\frac{1}{\ce_p-\lambda_\beta}=\frac{1}{\lambda_\alpha^\ast-\lambda_\beta}\ii\frac{\gamma}{2L}\sum_p\Big[\frac{1}{\ce_p-\lambda_\alpha^\ast}-\frac{1}{\ce_p-\lambda_\beta}\Big]=-\frac{2}{\lambda_\alpha^\ast-\lambda_\beta}~,
\end{equation*}
and so we obtain
\begin{equation}
     \braket{r_\alpha}{r_\beta}=\ii\frac{\gamma}{L}\frac{S_\alpha^\ast S_\beta}{\lambda_\alpha^\ast-\lambda_\beta}~.
\end{equation}
As $\mathrm{sgn}(\Im\lambda_\alpha)=-\mathrm{sgn}(\gamma)$, $\lambda_\alpha^\ast\neq\lambda_\beta$ is always valid.
\par The normalization constants satisfy $\sum_\alpha S_\alpha^2=L$, since
\begin{equation*}
    \sum_\alpha S_\alpha^2=\sum_\alpha \sum_{p,k}\braket{p}{r_\alpha}\braket{k}{r_\alpha}=\sum_\alpha \sum_{p,k}\braket{p}{r_\alpha}\braket{l_\alpha}{k}=\sum_{p,k}\bra{p}\sum_\alpha\ketbra{r_\alpha}{l_\alpha}\,\ket{k}=\sum_{p,k}\braket{p}{k}=L~,
\end{equation*}
where we used the completeness relation \cite{ReviewNonHermitianUeda} $\sum_\alpha\ketbra{r_\alpha}{l_\alpha}=\one$. Another useful relation can be found by taking the derivative of \eq~\eqref{eq: eigenvalues K} with respect to $\gamma$: we find that
\begin{equation}\label{eq: derivative lambda}
    \pdv{\lambda_\alpha}{\gamma}=-\frac{\ii}{2L}S_\alpha^2(\{\lambda_\beta(\gamma)\},\gamma)=-\frac{\ii}{2}\braket{j=0}{r_\alpha}^2~,
\end{equation}
which connects the spectrum with the value of the right eigenfunctions at the measured site. 
\subsection{Spectrum of non-Hermitian Hamiltonian}
Equation \eqref{eq: eigenvalues K} needs to be solved numerically to determine the eigenvalues, or, equivalently, one can diagonalize $\kappa_{pq}$ numerically. Most of the qualitative results are largely known from \cite{Haque,FromlPRB1}, and we repeat them in some detail. 
\begin{figure}
    \centering
    \subfloat[]
    {\includegraphics[width=0.45\linewidth]{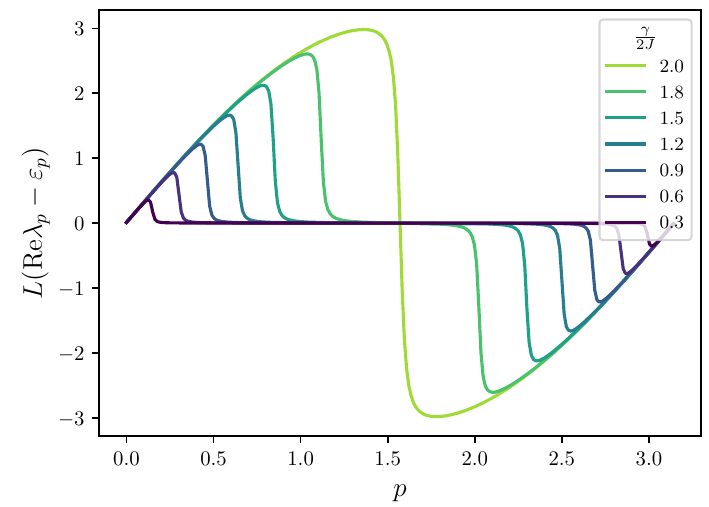}}
    \subfloat[]
    {\includegraphics[width=0.45\linewidth]{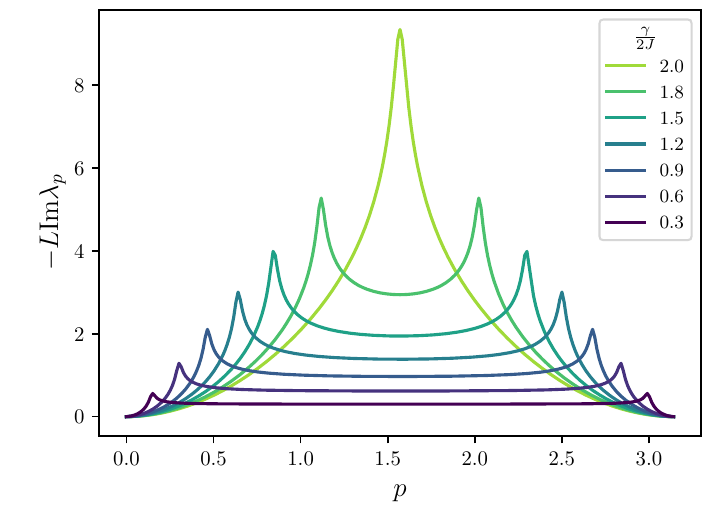}}
    \caption{Single-particle spectrum of $K$ up to the exceptional point, $\gamma\leq 4J$. (a) Real part (b) Imaginary part. Notice that the values have been multiplied by the system size $L$.}
    \label{fig: spectrum K}
\end{figure}
In the “weak-coupling” regime $\gamma\leq 4J$, the eigenstates are similar to the scattering states in the Hermitian case \cite{CombescotNozieres}: the eigenvalues $\lambda_\alpha$ occur close to the original eigenvalues $\ce_p$, hence we can label them with the corresponding momentum: 
\begin{equation}\label{eq: def delta}
    \lambda_p=\ce_p-v_p\frac{2\pi}{L}\frac{\delta_p}{\pi}~,
\end{equation}
where $v_p\equiv \partial_p\ce_p=2J\sin p$ is the local group velocity, and as customary we have parametrized the correction to the energies in terms of a scattering phase $\delta_p$, which in this non-Hermitian case is \emph{complex}. As discussed in the previous Section, only the symmetric modes interact with the impurity, so the momenta $p$ belong to the halved Brillouin zone $p\in[0,\pi]$ (in which $v_p\geq0$). 
\par We notice that the correction $\lambda_p-\ce_p$ is of order $L^{-1}$, implying that in the $\gamma\leq 4J$ regime one cannot directly take the continuum limit $L^{-1}\sum_p\to(2\pi)^{-1}\int\dd{p}$ in \eq~\eqref{eq: eigenvalues K}, since $\lambda_p$ depends on $L$ as well [and hence the sum in \eq~\eqref{eq: eigenvalues K} is not a Riemann sum]. The approximate method described in \cite{CombescotNozieres} may be used, instead. Another point that may need some clarification is the large size limit $L\to+\infty$. If we take this limit with \eq~\eqref{eq: def delta} we obtain $\lambda_p\to\ce_p$, which seems to imply that in the large size limit we would recover the physics of noninteracting fermions. This actually implies only that the large size limit is not suitable for a single particle. But we are interested in a many-particle system in the thermodynamic limit, in which not only the system size grows but also the number of particles $N_f$ does, such that the density $\bar{n}=N_f/L$ remains constant. In this limit, the \emph{many-body} eigenvalues of $K$ are well-defined, since they are sums over the occupied states $\bm{p}=(p_1,\dots,p_{N_f})$, $\Lambda_{\bm{p}}=\sum_{j=1}^{N_f}\lambda_{p_j}$, for which the correction is $\Lambda_{\bm{p}}-\sum_{j=1}^{N_f}\ce_{p_j}=2\overline{v\delta}\,N_f/L=2\overline{v\delta}\,\bar{n}$ ($\overline{v\delta}$ is the arithmetic average of $v_p \delta_p$ over the occupied states), which is finite in the thermodynamic limit. 
\par Figure \ref{fig: spectrum K} shows the spectrum of $\kappa$ for a system of $L=500$ sites for various values of $\gamma\leq 4J$. The left panel displays the real part of $\lambda_p-\ce_p$ (proportional to the real part of $\delta$). We observe that, except for $\gamma$ close to $4J$, there are corrections mainly at the edges of the band (i.e. for $p$ close to $0$ or $\pi$). 
\par The right panel of \fig~\ref{fig: spectrum K} shows the imaginary part of $\lambda_p$ (which is completely contained in $\delta_p$), and displays most of the physics that we are interested in. For all $\gamma<4J$, $-\Im\lambda_p$ has a common structure: it is almost vanishing at the very edges of the band, then it has a peak up to about $2\gamma/L$ at some special momenta $q^\ast$ and $\pi-q^\ast$, until it settles to an almost constant value close to\footnote{This value can be easily computed from first order degenerate perturbation theory.} $\gamma/L$ for all momenta in a region in the middle of the band. Thus, we see that the description provided by bosonization is accurate for the states not too far from mid-band, which have roughly the same imaginary part of $\lambda_p$ and are expected to behave similarly to a Hermitian system, as discussed at the end of the previous Section. We can also say that there is an emergent $\mathcal{P}\mathcal{T}$ symmetry. But this symmetry is explicitly broken close to the band edges, where $\Im\lambda_p$ varies abruptly. In all the numerical results that we presented in the main text, the Fermi surface lies far from the band edges. Therefore, the linearization of the spectrum close to $p=k_F$ cannot possibly account for these special modes, and we understand what is being missed by bosonization. Curiously, the feature that brings about the failure of bosonization is not in the real part of the spectrum, that is, the band dispersion, but rather in the imaginary part, hence the lifetime of the states.
\begin{figure}
    \centering
    \includegraphics[width=0.5\linewidth]{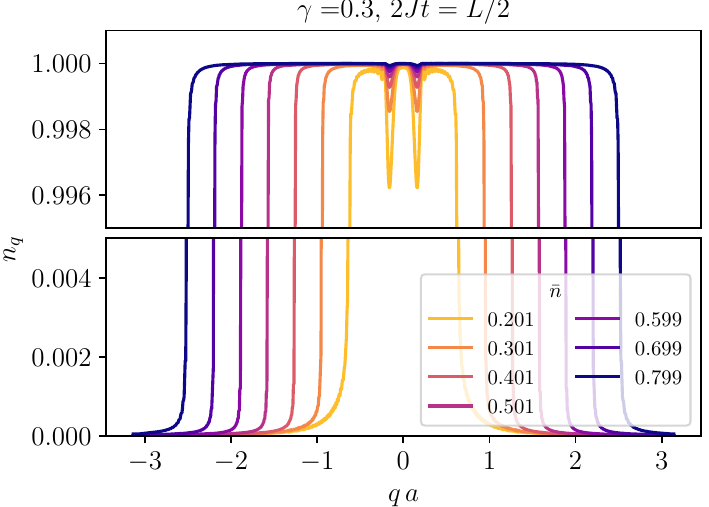}
    \caption{Momentum occupancy $n_q(t)=\ev{c_q^\dag c_q}{\psi(t)}$ at a fixed time, showing the extra depletion at the special momenta deep within the Fermi sea.}
    \label{fig: momentum occupancy}
\end{figure}
\par To show that this intuition is correct, and that the enhanced decay rate at the special momenta $q^\ast$ does affect the dynamics, we report in \fig~\ref{fig: momentum occupancy} the momentum occupancy $n_q\equiv\expval{c_q^\dag c_q}$ in the full Brillouin zone for $L=1000$ sites at a given time and at a fixed value of $\gamma/(2J)$, while the density is varied from one curve to the other. This occupancy can be computed from the Fourier transform of the real-space correlation function $\sexpval{c_i^\dag c_j}$, which is easily computed with the algorithm of \cite{CaoTilloyDeLuca}. From the \fig, we see that the Fermi surface is still well-defined, and only rounded by the potential scattering. This is exactly the kind of process that is described by bosonization. On the other hand, there is an extra depletion of population at two momenta close to $q=0$, and it turns out that these momenta are exactly corresponding to the special momenta $\pm q^\ast$ ($p=q^\ast$ in the half-Brillouin zone of the symmetric modes corresponds to $p=\pm q^\ast$ in the whole Brillouin zone). Unless the density is sufficiently high, the special states at the top of the band, $p=\pm(\pi-q^\ast)$ are not populated enough to show any nontrivial dynamics. The important observation on \fig~\ref{fig: momentum occupancy} is that the rate of depletion of the special modes is larger at smaller densities. This property can be taken as a clue that the special modes play a more relevant role at lower densities. Such conclusion fits precisely into our intuition that the special modes are the ones responsible for the breakdown of bosonization and it is agreement with the numerical results showing that the return amplitude and the energy absorption depart from bosonization more strongly at lower densities. Actually, the momentum occupation is directly related to the absorbed energy, $E(t)=\sum_q \ce_q n_q(t)$, and it is easy to see that $E(t)$ has to rise faster than what bosonization predicts, since the states removed from around $\abs{q}=q^\ast$, which have low energy, must be moved to unoccupied states at high energy $\ce_p>\ce_F$ because of particle number conservation and the exclusion principle. On the other hand, there is no direct relation between $n_q$ and the return amplitude $\mathcal{L}(t)$.
\par We have not been able to derive a quantitative relation between the density and the effect of the special modes. Indeed, the momentum of the depletions in $n_q(t)$ decreases in time towards $\pm q^\ast$, and the value of $q^\ast$ itself does not depend on $\bar{n}$ since it is a property of the single-particle states. The numerics and the approximate formula for $\lambda_p$ that we derive later suggest that $q^\ast$ is (at least approximately) determined by the condition that the group velocity $v_p$ of the fermions is equal to the strength of the potential $\gamma/2$\footnote{From \eq~\eqref{eq: def delta} and the observation $-\Im\lambda_{q^\ast}\approx2\gamma/L$ we deduce that $\delta_{q^\ast}\approx 2\ii$.}, i.e. 
\begin{equation}\label{eq: special q}
    q^\ast\approx\arcsin\frac{\gamma}{4J}~.
\end{equation} 
This relation has already emerged in \rref~\cite{Haque}, as the momentum at which the imaginary potential scatters the incoming single-particle wave packets most effectively.
\begin{figure}
    \centering
    \subfloat[]
    {\includegraphics[width=0.45\linewidth]{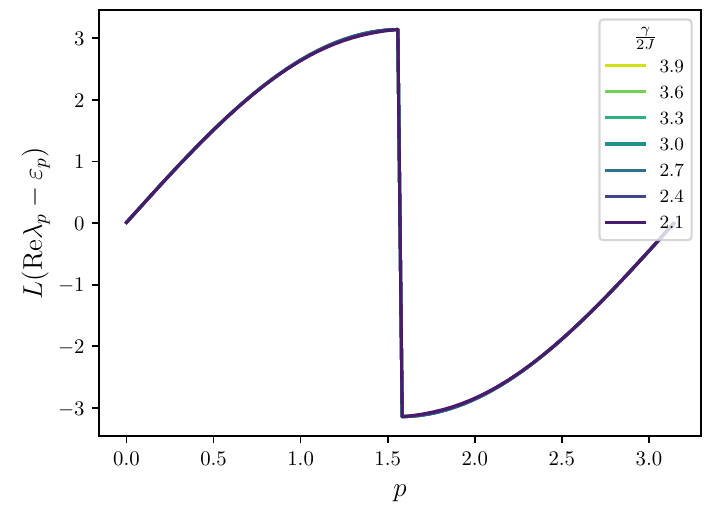}}
    \subfloat[]
    {\includegraphics[width=0.45\linewidth]{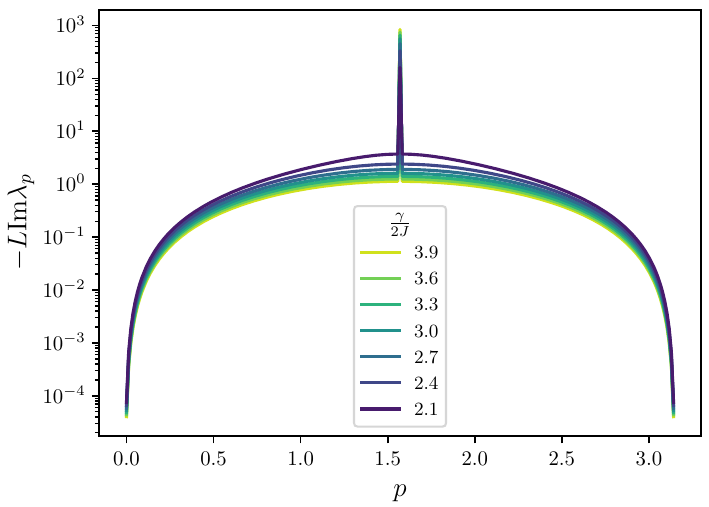}}
    \caption{Single-particle spectrum of $K$ above the critical point, $\gamma>4J$. (a) Real part, which is almost insensitive to the value of $\gamma$ (b) Imaginary part, showing the large peak at the bound state (notice the logarithmic scale).}
    \label{fig: spectrum K large gamma}
\end{figure}
\par Of course, the above  equation suggests that something else must be happening when $\gamma\geq4J$. Although this “strong-coupling” regime is beyond the numerical work presented in the main text, we report what it happens for completeness. Most of the properties of this regime have been already analyzed in \cite{FromlPRB1}, in the limit $L\to+\infty$. From \fig~\ref{fig: spectrum K} and \eq~\eqref{eq: special q} we observe that for $\gamma=4J$ the two peaks merge into one. This value of $\gamma$ marks an exceptional point \cite{ReviewNonHermitianUeda}, namely a point in the parameter space of $K$ in which the non-Hermitian Hamiltonian is not diagonalizable because the two special eigenstates coalesce to the same vector. For $\gamma$ above the critical value $4J$ the matrix becomes diagonalizable again and the spectrum reorganizes completely, as shown in \fig~\ref{fig: spectrum K large gamma}: almost all of the imaginary part (right panel in \fig~\ref{fig: spectrum K large gamma}) becomes concentrated\footnote{The sum of all imaginary parts is fixed by the trace of $K$: $\sum_p \Im\lambda_p=\Im\Tr K=-\gamma/2$.} on a single state in the middle of the band $p=\pi/2$, which acquires a decay rate of the order of $\gamma/2$, whereas the rest of the eigenvalues have a rather small imaginary part (notice the logarithmic scale of the Figure). The corrections to the real part of the eigenstates (left panel in \fig~\ref{fig: spectrum K large gamma}) remain small and scarcely dependent on $\gamma$. The analytic dependence on $\gamma$ of the imaginary part of the special eigenvalue can be found in \cite{FromlPRB1,ThesisFroml}, since it is the only one that survives also in the large size limit $L\to+\infty$, so that it can be obtained by transforming the sum in \eq~\eqref{eq: eigenvalues K} in an integral over momenta and solving for $\lambda$.
\begin{figure}
    \centering
    \includegraphics[width=0.5\linewidth]{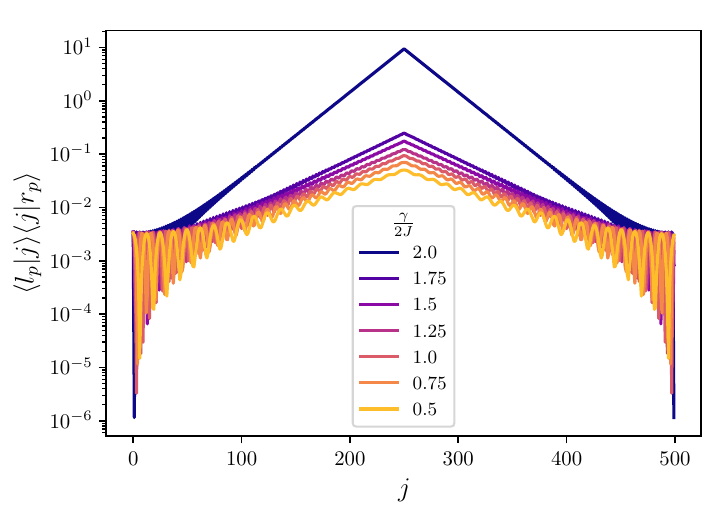}
    \caption{Special modes in real space. We are plotting $\braket{l_p}{j}\braket{j}{r_p}=\abs{\braket{j}{r_p}}^2$ as a function of $j$ for $p=q^\ast=\arcsin(\gamma/(4J))$.  We have taken $L=500$ sites, with the impurity sitting at $j_0=L/2$. We observe an exponentially decaying profile close to $j_0$, reminiscent of a bound state, although the decay length is proportional to $L$. A true bound state emerges only at the exceptional point, $\gamma=4J$, shown here as the upper blue curve.}
    \label{fig: special modes real space}
\end{figure}
\par The above discussion provides a reasonably clear picture of why bosonization fails in our model of non-Hermitian OC, but it leaves the physical nature of the special modes somewhat unspecified. A better understanding comes from considering their wavefunctions\footnote{Since $\kappa$ is symmetric, we can consider equally the right or the left eigenstates.} in real space, as shown in \fig~\ref{fig: special modes real space}. Close to $j=j_0=L/2$ we can observe the typical exponential envelope of a bound state. The special modes are quasi-bound states (or scale-free localized states \cite{scaleFreeBoundStates}), whose decay length is proportional to the length of the system. They are pinned at the impurity site, experiencing a stronger potential and hence an enhanced decay rate. At the exceptional point the low- and high-momentum states merge and a true bound state is created \cite{FromlPRB1}.
\par We can give an approximate expression for the scattering phase $\delta_p$, using the relation \eqref{eq: derivative lambda}. Following \cite{CombescotNozieres}, we can approximate $S_p$ as
\begin{equation}
    S_p\approx-\ii\frac{\sqrt{8}v_p}{\gamma}\sin\delta_p~,
\end{equation}
which is quite close to the numerical data. Substituting the equation in \eq~\eqref{eq: derivative lambda} and solving the resulting differential equation we obtain
\begin{equation}
    \cot\delta_p(\gamma)+\frac{2\ii v_p}{\gamma}\approx\theta_p~,
\end{equation}
in which $\theta_p$ is a function that does not depend on $\gamma$. Unfortunately, the limit $\gamma\to0$, in which $\delta_p\to0$, cannot be used to set the value of $\theta_p$, because both $\cot\delta_p$ and $1/\gamma$ diverge. If we set $\theta_p=0$ we obtain
\begin{equation}
\delta_p\approx
    \begin{cases}
        -\frac{\pi}{2}\mathrm{sgn}(\frac{\pi}{2}-p)+\ii\,\mathrm{arctanh}\frac{\gamma}{2v_p}\quad\text{for }\gamma<v_p~,\\
        \ii\,\mathrm{arctanh}\frac{2v_p}{\gamma}\quad\text{for }\gamma>v_p~,
    \end{cases}
\end{equation}
which is a rough approximation for $\Re\delta_p$ but a quite good one for $\Im\delta_p$ (except for $v_p$ close to $\gamma/2$, i.e. the special modes).
In particular, we find that $\delta_p\approx\ii\frac{\gamma}{2 v_p}$ to lowest order in $\gamma$ and in the middle of the band (i.e for $\gamma< v_p$), which yields the result of bosonization: $\beta=\frac{\abs{\delta_{k_F}^2}}{\pi^2}\approx\frac{\gamma^2}{(2\pi v_F)^2}$, using the exact relation between $\beta$ and $\delta_p$ \cite{NozieresDeDominicis,CombescotNozieres}.
\subsection{The role of band curvature}
\begin{figure}
    \centering
    \subfloat[width=0.45\linewidth][\label{fig: im spectrum tomo}]{\includegraphics[width=0.5\linewidth]{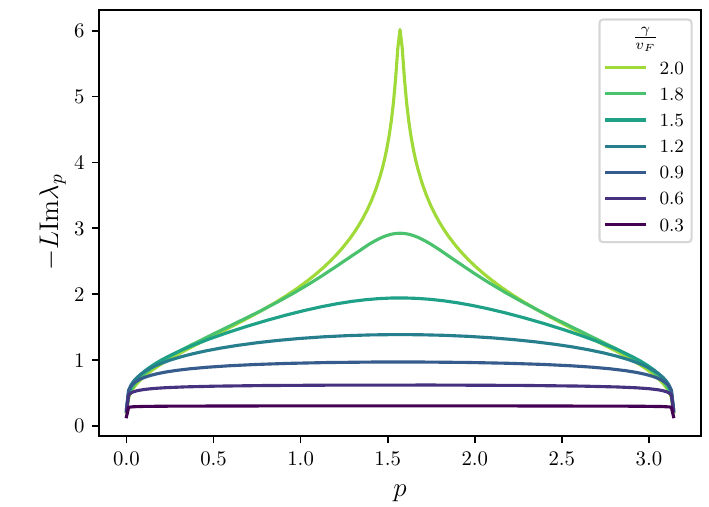}}
    \subfloat[width=0.45\linewidth][\label{fig: mom occ tomo}]{\includegraphics[width=0.5\linewidth]{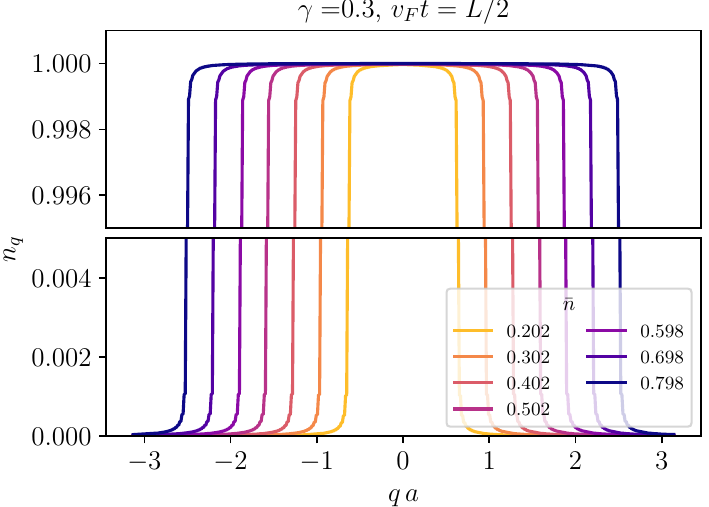}}
    \caption{Properties of the non-Hermitian Tomonaga Hamiltonian in momentum space. (a) Imaginary part of the spectrum of $K$ for $L=500$ and $v_F=1$, below the exceptional point $\gamma\leq 2v_F$. The imaginary part is essentially flat, except near the $\gamma=2v_F$. (b) Momentum occupancy at a fixed time for $L=500$, $v_F=1$, showing the absence of dynamics deep in the Fermi sea.}
    \label{figs: tomonaga 1}
\end{figure}
There is another aspect that is important to mention: the condition \eqref{eq: special q} implies that the presence of the special, quasi-bound modes is due to the variation of the group velocity of the fermions---in other words, it is a consequence of the band curvature. To confirm this picture, we have repeated the various numerical calculations with an artificial model that features linearly-dispersing fermions, i.e. we have set $\ce_q=v_F\abs{q}$ for $\abs{q}\leq\pi$ and periodically repeated for other momenta. Following \cite{Mahan}, we will refer to this setup as the Tomonaga model. In \fig~\ref{fig: im spectrum tomo} we show the imaginary part of the single-particle eigenvalues of $K$ for $L=500$ and increasing values of $\gamma/v_F$. The appearance of the spectrum is quite different from the previous case, with an imaginary part which is almost completely flat for small $\gamma$ (except for very close to the band edges). The decay rates acquire some curvature only in the vicinity of the exceptional point $\gamma/2=v_F$ [which is the same condition as \eq~\eqref{eq: special q}], at which point a bound state forms, analogously to the previous case. In agreement with the flatness of the single-particle decay rates, we observe no extra depletions in the momentum occupancies, as shown in \fig~\ref{fig: mom occ tomo}. 
\begin{figure}
    \centering
    \subfloat[]
    {\includegraphics[width=0.95\linewidth]{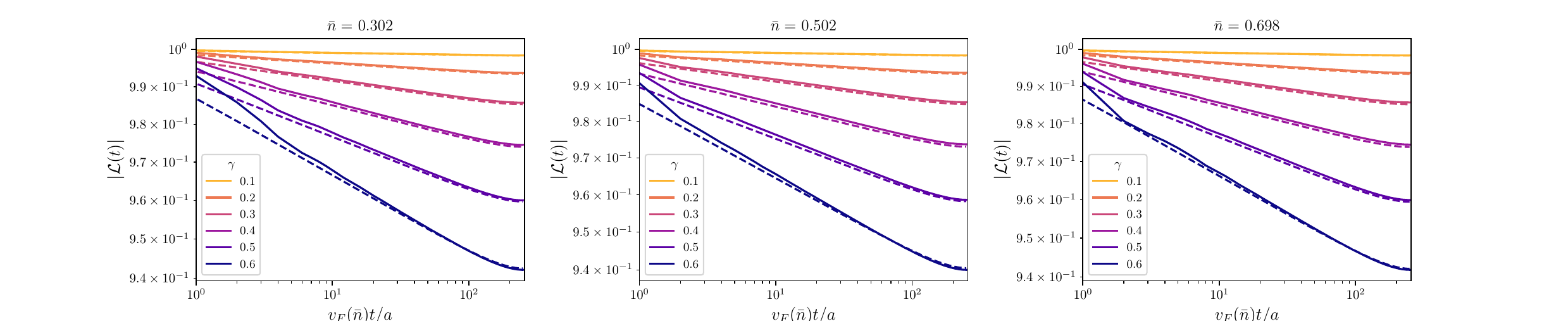}}\\
    \subfloat[]
    {\includegraphics[width=0.95\linewidth]{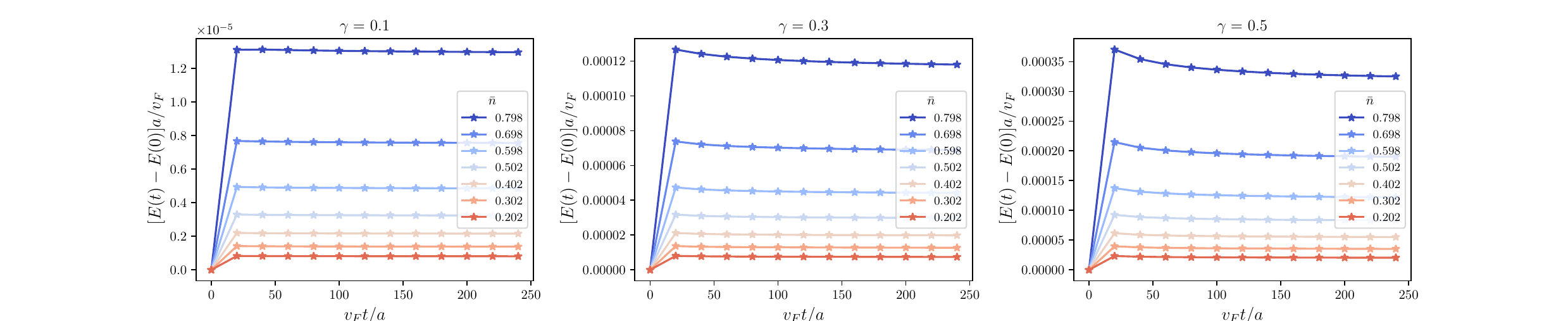}}
    \caption{Observables for the Tomonaga model with $L=500$ and $v_F=1$, showing agreement with bosonization. (a) The return amplitude (continuous lines) at various values of $\gamma$ and $\bar{n}$ compares with the prediction from finite-size bosonization (dashed lines). (b) The absorbed kinetic energy saturates, in agreement with bosonization.}
    \label{fig: results tomonaga}
\end{figure}
\par If our theory about the relation between a flat imaginary part of the spectrum and the validity of bosonization is valid, then we should observe that the various observables should be in agreement with the predictions from $K_{TLL}$. Indeed this prediction is confirmed, as shown in \figs~\ref{fig: results tomonaga}. In the upper panel, we show that the return amplitude $\mathcal{L}(t)$ is a pure power-law for various densities and values of $\gamma$, and that at late times it is in good agreement even with the finite-size prediction from bosonization $\mathcal{L}(t)=\exp(-F_L(t))$~, with
\begin{equation}
    F_L(t)\equiv\frac{\gamma_c^2}{L}\sum_{q>0}V_q^2\frac{1-\ii\ce_q t-\ee^{-\ii \ce_q t}}{\ce_q^2}=\beta_b\ln\frac{1-\ee^{-2\pi(\alpha+\ii v_F t)/L)}}{1-\ee^{-2\pi\alpha/L}},
\end{equation}
whose only free parameter is the UV cutoff $\alpha$. The lower panel shows that the kinetic energy absorbed by the system saturates, in agreement with the predictions from bosonization. Moreover, particle-hole symmetry (i.e. invariance of observables for reciprocal densities $\bar{n}$ and $1-\bar{n}$) is restored for the return amplitude. It is obvious that, since the imaginary part of the spectrum is not exactly flat (\fig~\ref{fig: im spectrum tomo}), bosonization will eventually break down. However, for $\gamma$ sufficiently far from the exceptional point, the timescale of the breakdown (i.e. what would be $\Gamma$, the rate of exponential decay of $\mathcal{L}(t)$) is pushed to very large values.
\par The conclusion from this discussion is that the effects that we observe beyond bosonization belong to the so-called beyond-Luttinger physics \cite{ImambekovSchmidtGlazman}, in the sense that they are corrections to the predictions of bosonization that stem from the inability of the latter to take into account the curvature of the fermionic band dispersion. Similarly to the standard beyond-Luttinger phenomena, such as the presence of X-ray edges in the dynamical structure factor, these effects are non-perturbative in nature. On the other hand, rather than being consequences of the band curvature in the vicinity of the Fermi surface, the new phenomena depend on the curvature far from it---actually, from the very edge of the band. 
  
\section{Crucial role of post-selection}
In this section we show how post-selection is crucial for the result we presented in the main text. First, we are going to prove how the nonlinear dynamics dictated by post-selection is responsible for making the effects beyond bosonization more apparent. Then, we are going to present the opposite case of complete absence of post-selection, comparing our results with the case of localized dephasing.
\subsection{Importance of nonlinear dynamics: cumulant expansion}
The nonlinear dynamics induced by measurements seems to be crucial to expose the effects beyond bosonization. Indeed, the return amplitude is given by the ratio 
\begin{equation}
    \mathcal{L}(t)=\frac{\sbraket{FS}{\tilde{\psi}(t)}}{\snorm{\tilde{\psi}(t)}^{1/2}}~,
\end{equation}
and it is easy to see that if we took only the numerator, $\tilde{\mathcal{L}}(t)=\sbraket{FS}{\tilde{\psi}(t)}=\ev{\ee^{-\ii K t}}{FS}$ (the “unrenormalized” return amplitude) we would obtain a leading exponential decay $\ee^{-\gamma\bar{n} t/2}$. This can be seen by expanding in $\gamma$:
\begin{equation*}
    \begin{aligned}
        \tilde{\mathcal{L}}(t)&=\ee^{-\ii E_{FS}t}\ev{\mathcal{T}\ee^{-\gamma/2\, \int_0^t \dd{t^\prime} n_0(t^\prime)}}{FS}=\\
        &=\ee^{-\ii E_{FS}t}\Big[1-\frac{\gamma}{2} \int_0^t \dd{t^\prime} \expval{n_0(t^\prime)}+\Big(\frac{\gamma}{2}\Big)^2\int_0^t\dd{t_1}\int_0^{t_1}\dd{t_2}\expval{n_0(t_1)n_0(t_2)}+\order{\gamma^3}\Big]=\\
        &=\ee^{-\ii E_{FS}t}\Big[1-\frac{\gamma}{2} \bar{n} t+\Big(\frac{\gamma}{2}\Big)^2\int_0^t\dd{t_1}\int_0^{t_1}\dd{t_2}\expval{n_0(t_1)n_0(t_2)}+\order{\gamma^3}\Big]
    \end{aligned}
\end{equation*}
where $E_{FS}=\ev{H}{FS}$, $\expval{\cdot}$ is the average on the ground state and $n_0(t)\equiv \ee^{\ii H t}n_0 \ee^{-\ii H t}$ is the number operator at the measured site in the interaction picture. Then, we assume $\tilde{\mathcal{L}}(t)=\exp[-\ii E_{FS} t +\tilde{L}_1+\tilde{L}_2+\order{\gamma^3}]=\exp[-\ii E_{FS} t][1+\tilde{L}_1+\tilde{L}_2+\order{\gamma^3}]$, where $\tilde{L}_k=\order{\gamma^k}$, and we determine $\tilde{L}_{1,2}$ by comparing with the direct expansion in $\gamma$, reported above: $\tilde{L}_1=-\gamma \bar{n} t/2$, $\tilde{L}_2=\big(\frac{\gamma}{2}\big)^2\int_0^t\dd{t_1}\int_0^{t_1}\dd{t_2}\expval{(n_0(t_1)-\bar{n})(n_0(t_2)-\bar{n})}$. By expressing $n_0(t)$ in terms of momentum modes, the latter integral can be evaluated to
\begin{equation*}
    \tilde{L}_2=\Big(\frac{\gamma}{2}\Big)^2\frac{1}{L^2}\sum_{\abs{p_1}<k_F}\sum_{k_F<\abs{p_2}<\pi}\frac{1+\ii(\ce_{p_1}-\ce_{p_2})t-\ee^{\ii (\ce_{p_1}-\ce_{p_2}) t}}{(\ce_{p_1}-\ce_{p_2})^2}~,
\end{equation*}
where $k_F=\pi\bar{n}$ is the Fermi momentum. At long times, the sum is dominated by the terms with $\ce_{p1}\approx\ce_{p_2}$ which due to the limits in the sums implies that both momenta have to be close to the Fermi surface. If we linearize the energy in this area, we find $\tilde{L}_2\approx F(t)\sim -\ii\Delta\ce\, t +\beta_b \ln (v_F t/\alpha)+\ii\zeta +\order{t^{-1}}$, where $\Delta\ce=\gamma^2/(2\pi v_F)^2\, v_F t/\alpha$ and $\zeta=\gamma^2/(8\pi v_F^2)$ are real constants. The important property of $\tilde{L}_2$ is that it does not contain any real term proportional to $t$, i.e. it does not contribute to the exponential decay, but only to the OC. Summing up, we have
\begin{equation}
    \tilde{\mathcal{L}}(t)=\ee^{-\ii E_{FS} t-\tfrac{\gamma}{2}\bar{n} t +\tilde{L}_2(t)+\order{\gamma^3}}\sim \ee^{\ii\zeta-\ii (E_{FS}+\Delta\ce) t-\tfrac{\gamma}{2}\bar{n} t+\beta_b\ln(v_F t/\alpha)+\order{\gamma^3,t^{-1}}}~,
\end{equation}
which decays exponentially, as advertised earlier, with a decay constant $\tilde{\Gamma}=\gamma\bar{n}/2+\order{\gamma^3}$.
A calculation along the same lines as the previous shows that 
\begin{equation}
    \snorm{\tilde{\psi}(t)}^{1/2}=\ee^{-\frac{\gamma}{2}\bar{n}t+2\Re\tilde{L}_2(t)+\order{\gamma^3}}\sim\ee^{-\frac{\gamma}{2}\bar{n}t+2\beta_b\ln(v_F t/\alpha)+\order{\gamma^3, t^{-1}}}~.
\end{equation}
These perturbative calculations show a number of relevant points. First, we see that both numerator and denominator decay exponentially, but with the \emph{same} exponent $-\gamma \bar{n}/2$, so that this leading behavior disappears from the result:
\begin{equation}
    \mathcal{L}(t)=\frac{\tilde{\mathcal{L}}(t)}{\snorm{\tilde{\psi}(t)}^{1/2}}\sim\ee^{\ii\zeta-\ii (E_{FS}+\Delta\ce) t-\beta_b\ln(v_F t/\alpha)+\order{\gamma^3,t^{-1}}}~,
\end{equation}
which agrees with the predictions from bosonization. Second, both $\tilde{\mathcal{L}}(t)$ and the state norm display an OC-like contribution in the form of a power-law growth, instead of a decay. An analogous behavior was observed in \cite{Tonielli} in a dissipative context. In our case, the usual algebraic decay is recovered once the state is normalized. The third aspect is that we find no sign of the decay rate $\Gamma$ that we measured in the numerics, which implies that it has to appear at a perturbative order larger than the second. Unfortunately, the cumulants of third and higher orders do not lend themselves easily to analytic calculations, so we have not been able to extract an expression for $\Gamma$. Moreover, we have argued that $\Gamma$ may not be an analytic function in $\gamma$, so trying to compute it from any finite number of cumulants might be doomed to failure\footnote{Cumulant expansions are also known for being possibly unstable beyond the lowest perturbative order.}. Nevertheless, in the main text we have found that $\Gamma$ is far smaller than $\gamma\bar{n}/2$, and thus it would represent only a small correction to the leading decay of the return amplitude if we did not normalize the state. We conclude that the setup of post-selected measurements, and the nonlinear state evolution that it entails, is crucial to reveal the beyond-TLL behavior. 
\subsection{Exact relations}
There exist exact (i.e. non-perturbative in the impurity potential) calculations of the return amplitude for the Hermitian OC problem \cite{NozieresDeDominicis,CombescotNozieres}, that prove its algebraic decay. In particular, the approach of \cite{CombescotNozieres} calculates $\ev{\ee^{-\ii (H+V n_0) t}}{FS}$ from the spectral properties of the (Hermitian) OC Hamiltonian $H+V n_0$, which are analogous to the non-Hermitian one in the $\gamma<4J$ regime. Since these calculations express everything in terms of the scattering phase $\delta_p$, which enters only through analytic functions, we can employ their results directly for our case with a complex $\delta_p$. The main result of \rref~\cite{CombescotNozieres} that we need is that the “unrenormalized” return amplitude $\tilde{\mathcal{L}}(t)\equiv\sbraket{FS}{\tilde{\psi}(t)}=\ev{\ee^{-\ii K t}}{FS}$ has the following asymptotic form at large times:
\begin{equation}
    \ln\tilde{\mathcal{L}}(t)\sim -\ii\Delta E t -\beta\ln(\alpha t)~,
\end{equation}
where $\alpha$ is a UV cutoff and the two constants $\Delta E$ and $\beta$ are completely determined by the scattering phase:
\begin{subequations}\label{eqs: fumi e beta}
    \begin{align}
        \Delta E&=\sum_{p\in FS}(\lambda_p-\ce_p)=-\frac{2\pi}{L}\sum_{p\in FS}v_p\frac{\delta_p}{\pi}=-\int_{-2J}^{\ce_F}\dd{\ce}\frac{\delta(\ce)}{\pi}~,\\
        \beta&=\frac{\delta_{k_F}^2}{\pi^2}=\frac{\delta(\ce_F^2)}{\pi^2}~,
    \end{align}
\end{subequations}
where $\delta(\ce)$ is the scattering phase expressed in terms of the energy $\ce=\ce_p$ in the continuum limit. In these calculations it is convenient to symmetrize the momentum states as in \eq~\eqref{eq: symmetrization}. In this way only the symmetric modes interact with the impurity and we can work in the half-Brillouin zone $p\in\{\frac{2\pi n}{L}\vert n=0,1,\dots,\frac{L}{2}\}$, where the dispersion relation $\ce=\ce_p$ is invertible. In the Hermitian scenario, the two constants $\Delta E$ and $\beta$ have very different roles---the former contributing to an overall phase while the latter gives the algebraic decay of the modulus of $\tilde{\mathcal{L}}(t)$. The key novelty in the present, non-Hermitian case, is that the scattering phase $\delta_p$ of the non-Hermitian Hamiltonian $K$ is generally \emph{complex} \footnote{This behavior of the scattering phase out of equilibrium has been associated to irreversibility \cite{PRB_Ng}.}, so that both the modulus and the phase of $\tilde{\mathcal{L}}(t)$ receive contributions from each of the terms. 
\par The OC exponent is $\beta=\delta^2(\ce_F)/\pi^2$, which depends only on the scattering phase at the Fermi energy $\ce_F$. If $\ce_F$ close to mid-band we find $\delta(\ce_F)\approx \ii\gamma/(2v_F)$, which gives $\beta$ in accord with bosonization, although with an extra minus sign (this phenomenon has been already noticed in \cite{Tonielli}---the correct sign is obtained when $\tilde{\mathcal{L}}(t)$ is divided by the norm of the state, as confirmed by the cumulant expansion described above). The most interesting part is the term linear in time, which is proportional to $\Delta E=-\int_{-2J}^{\ce_F}\dd{\ce}\delta(\ce)/\pi$ (known as Fumi's theorem \cite{Mahan,CombescotNozieres}). For an ordinary scattering problem, $\Delta E$ is the shift in the ground-state energy caused by the external potential, a quantity which is obviously non-universal and depending on the full dispersion, but that contributes only to an innocuous overall phase. However, in our model $\Delta E$ has an imaginary part, which means that it contributes to an exponential decay of $\tilde{\mathcal{L}}(t)$ with a decay constant that is sensitive to all occupied states---hence, with a nontrivial density dependence.
\par Here we report a more in-depth discussion of the properties of the exponential decay rate of $\tilde{\mathcal{L}}(t)$, $\tilde{\Gamma}\equiv-\Im\Delta E$.
\begin{figure}
    \centering
    \subfloat[ \label{fig: fumi n}]{\includegraphics[width=0.45\linewidth]{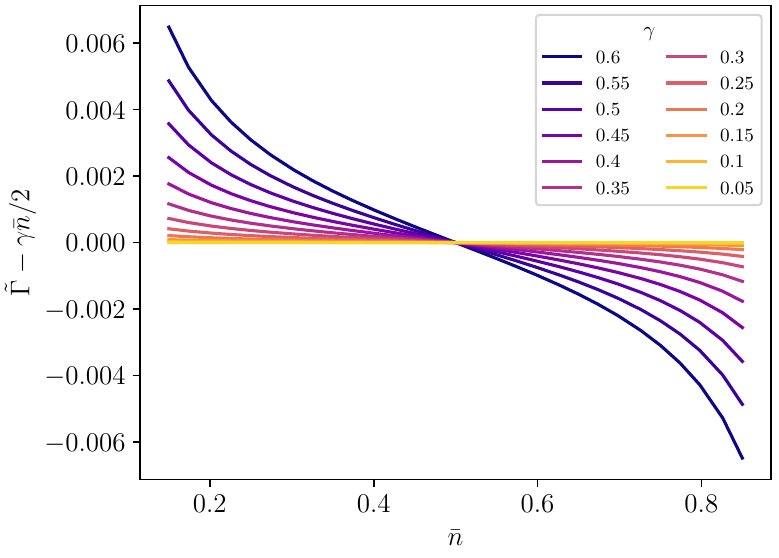}}
    \subfloat[\label{fig: fumi collapse}]{\includegraphics[width=0.45\linewidth]{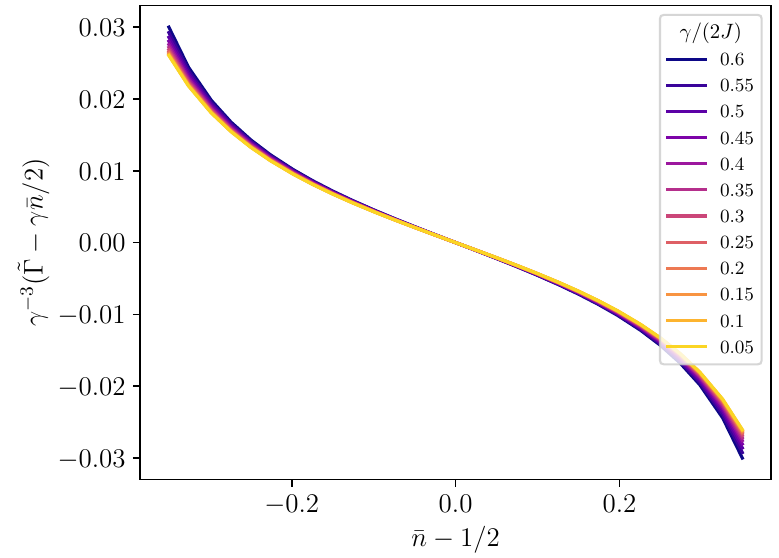}}
    \caption{Exponential decay rate of $\tilde{\mathcal{L}}(t)$, $\tilde{\Gamma}\equiv-\Im\Delta E$, computed from the numerical diagonalization of the single-particle version of $K$ for $L=500$ sites, with $2J=1$. (a) As a function of the density, showing the decreasing behavior once the average behavior $\gamma\bar{n}/2$ is subtracted. (b) Collapse of the data at different $\gamma$s, showing that $\tilde{\Gamma}-\gamma\bar{n}/2=\order{\gamma^3}$.}
    \label{fig: fumi}
\end{figure}
We show $\tilde{\Gamma}$ in \fig~\ref{fig: fumi}, computed from the numerical diagonalization of $K$. The left panel shows that if we subtract the leading term $\gamma\bar{n}/2$, we obtain a correction monotonically decreasing with the density. Indeed, since $\tilde{\Gamma}_b=\gamma\bar{n}/2$ is the prediction of bosonization, this correction is already a beyond-TLL effect. Although the qualitative behavior is the same as $\Gamma$ (i.e. the decay rate of the true return amplitude), the quantitative relation is different, since $\tilde{\Gamma}-\gamma\bar{n}/2$ can be fitted to an odd polynomial in $(\bar{n}-1/2)$. The left panel shows the collapse of the curves at different $\gamma$s by rescaling them by $\gamma^3$, which suggests an overall form
\begin{equation}
    \tilde{\Gamma}\approx\frac{\gamma}{2}\bar{n}+\gamma^3\Big[a_1 \Big(\bar{n}-\frac{1}{2}\Big)+a_3\Big(\bar{n}-\frac{1}{2}\Big)^3\Big]~,
\end{equation}
for certain coefficients $a_{1,3}$. This discussion is meant to illustrate concretely how the density can enter nontrivially in the return amplitude. It also shows that the behavior that we observed in $\Gamma$ for the full return amplitude ultimately depends on the interplay between $\tilde{\mathcal{L}}(t)$ and the norm of the state $\snorm{\tilde{\psi}}^{1/2}$.
\par Unfortunately, the results of \rref~\cite{CombescotNozieres} can not be used to compute the $\snorm{\tilde{\psi}}$, but we can still make some useful remarks on its behavior. It is easy to prove that
\begin{equation}\label{eq: norm vs n0}
    \snorm{\tilde{\psi}}^{1/2}=\ee^{-\frac{\gamma}{2}\int_0^t\dd{t^\prime}\ev{n_0}(t^\prime)}~,
\end{equation}
by direct differentiation. If, as observed in \fig~\ref{fig: n0}, $\ev{n_0}(t)$ tends to a constant\footnote{By comparing \eq~\eqref{eq: norm vs n0} with $\ln\abs{\tilde{\mathcal{L}}(t)}\sim -\tilde{\Gamma} t+\beta\ln t+\order{1}$ and $\ln\abs{\mathcal{L}(t)}\sim -\Gamma t-\beta\ln t+\order{1}$, we obtain $\ev{n_0}(t)\sim n_\infty+4\beta/(\gamma t)+\order{t^{-2}}$.}, $n_\infty$, then the full decay rate is $\Gamma=\tilde{\Gamma}-\gamma n_\infty/2$, and the observed exponential decrease of $\Gamma$ with the density is equivalent to say that $\gamma n_\infty/2$ tends to $\tilde{\Gamma}$ for increasing $\bar{n}$. This is indeed confirmed by the numerical data.  
\subsection{Comparison with localized dephasing}
We complement the previous discussion on the role of post-selection by comparing the result for the non-Hermitian OC with its closest dissipative analogue---the case of localized dephasing, defined by the master equation
\begin{equation}\label{eq: localized dephasing}
    \dv{}{t}\rho(t)=-\ii\comm{H}{\rho(t)}+\gamma\big(n_0\rho(t)n_0-\tfrac{1}{2}\{n_0,\rho(t)\}\big)~.
\end{equation}
This model has been already considered in \cite{Dolgirev,Tonielli}, and it can be obtained from the dynamics induced by the measurement operators \eqref{eq: dephasing M} if the state is averaged on the measurement outcomes, thus eliminating any post-selection. As noted at the beginning of this Supplementary Material, the master equation \eqref{eq: localized dephasing} shares the same non-Hermitian Hamiltonian $K=H-\ii\gamma n_0/2$ part of the dissipative dynamics as the nonlinear Schr\"odinger equation \eqref{eq: nlse}. We can simulate the dynamics of \eq~\eqref{eq: localized dephasing} efficiently with the same algorithm used in this work, because it is equivalent \cite{Dolgirev} to the stochastic dynamics induced by a noisy (real) impurity
\begin{equation}
    H_\xi(t)=H+\gamma^{1/2}\xi(t)n_0~,
\end{equation}
where $\xi(t)$ is a Gaussian white noise defined by the correlator $\overline{\xi(t)\xi(t^\prime)}=\delta(t-t^\prime)$ ($\overline{\mathcal{O}}$ is the noise average). Then, the state $\rho(t)$ evolving with \eq~\eqref{eq: localized dephasing} is obtained by averaging the state $\rho_\xi(t)=\ketbra{\psi_\xi(t)}$ evolved with $H_\xi(t)$ over the realizations of the noise. Since $H_\xi(t)$ is quadratic (unlike the time evolution induced by the master equation), we can apply the stochastic version of the algorithm in \rref~\cite{CaoTilloyDeLuca}. 
\begin{figure}
    \centering
    \subfloat[\label{fig:dephasing L}]{\includegraphics[width=0.45\linewidth]{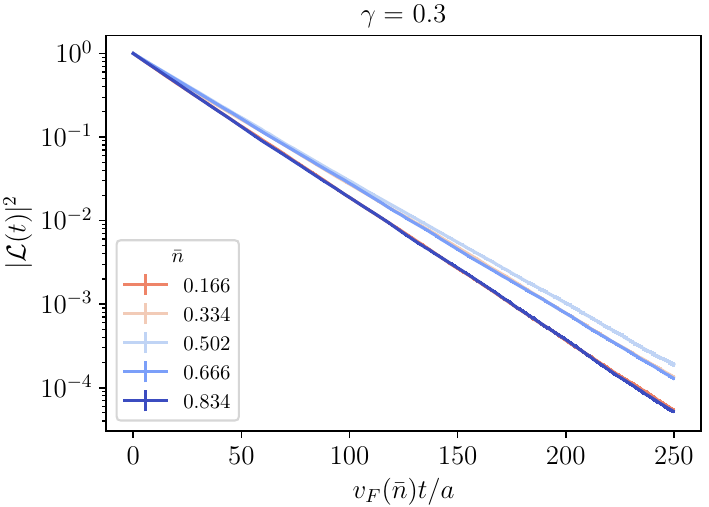}}
    \subfloat[\label{fig:dephasing E}]{\includegraphics[width=0.45\linewidth]{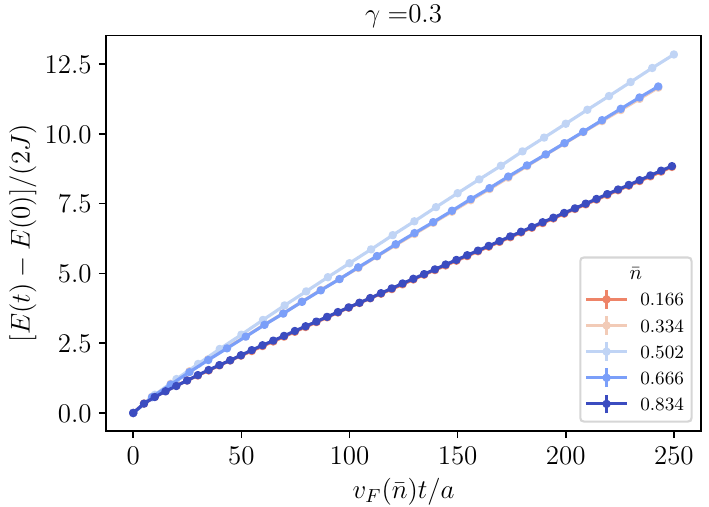}}\\
    \subfloat[\label{fig:dephasing nq}]{\includegraphics[width=0.45\linewidth]{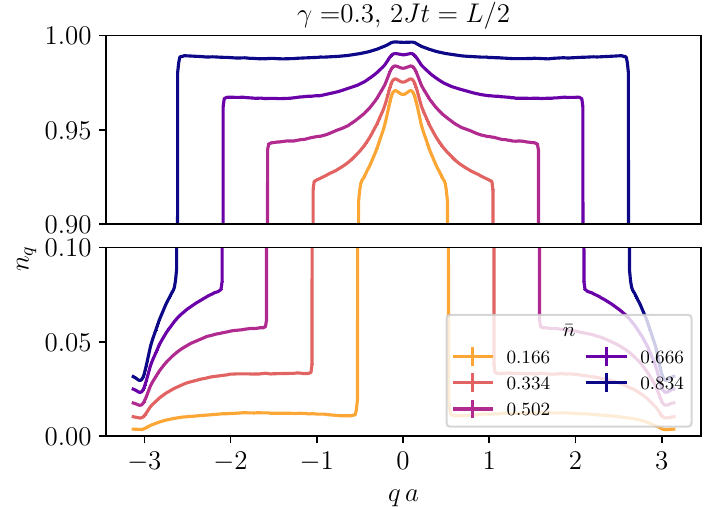}}
    \caption{Properties of a system with $L=500$ sites subjected to measurements \eqref{eq: dephasing M} without post-selections, namely local dephasing [\eq~\eqref{eq: localized dephasing}]. (a) Average square of the return amplitude, i.e. the Loschmidt echo $\ev{\rho(t)}{FS}$. (b) Average energy absorption. (c) Occupation of the momentum modes for varying density. The plots have been obtained by averaging $1024$ trajectories generated by $H_\xi(t)$. The error bars are calculated as the standard deviation of the empirical mean, which is comparable with the thickness of the lines.}
    \label{fig:dephasing}
\end{figure}
\par We can anticipate that we will not find any monotonic behavior in the density, since \eq~\eqref{eq: localized dephasing} is ph-symmetric---it is easy to verify that $\mathcal{C}\rho(t)\mathcal{C}^\dag$ satisfies the same equation as $\rho(t)$. Thus, we expect that all the quantities will be invariant under the exchange $\bar{n}\to 1-\bar{n}$. Indeed, this is what we observe from the numerics, as shown in \fig~\ref{fig:dephasing}. Figure \ref{fig:dephasing L} shows the Loschmidt echo, which is the average of the squared modulus of the return amplitude, $\overline{\abs{\braket{FS}{\psi_\xi(t)}}^2}=\ev{\rho(t)}{FS}$, while \fig~\ref{fig:dephasing E} shows the absorbed energy $E(t)=\Tr[\rho(t) H]$. In both Figures we observe a perfect ph symmetry, such that the data for conjugate densities $(\bar{n},\,1-\bar{n})$ coincides within the scale of the Figures (the differences are of the same order of magnitude as the residual randomness in the averaged data). On a qualitative level, we notice that the Loschmidt echo decays much faster than in the post-selected setting, owing to a decay rate which is linear in $\gamma$ to the leading order \cite{Tonielli} $\Gamma_\textup{deph}=\tfrac{1}{2}\gamma \bar{n}(1-\bar{n})+\order{\gamma^3}$ (which has a ph-symmetric dependence on the density, of course). This fast exponential decay overshadows the $\order{\gamma^2}$ algebraic behavior predicted in \cite{Tonielli}. The behavior of the energy (\fig~\ref{fig:dephasing E}) is qualitatively similar to the post-selected setup, except that the growth rate is two order of magnitude larger---in accord with the fast decay of the Loschmidt echo. In \fig~\ref{fig:dephasing nq} we show the average occupation in momentum space, for different densities. The behavior is quite different from that of the post-selected scenario, with a Fermi sea that is significantly depleted from the outside in. The possibility of a relation between the two momenta which seem more resistant to dissipation and the special modes $q^\ast$ requires further investigations.
\bibliography{biblioOC.bib}